**ISO/IEC JTC 1/SC 29/WG 1**

**(ITU-T SG16)**

# Coding of Still Pictures

| **JBIG** | **JPEG** |
|---|---|
| Joint Bi-level Image Experts Group | Joint Photographic Experts Group |

**TITLE:** **IT/IST/IPLeiria Response to the Call for Proposals on JPEG Pleno Point Cloud Coding**

**AUTHORS:** André F. R. Guarda[1], Nuno M. M. Rodrigues[2], Manuel Ruivo[1], Luís Coelho[1], Abdelrahman Seleem[1], Fernando Pereira[1]

**CONTACT EMAILS:** andre.guarda@lx.it.pt, nuno.rodrigues@co.it.pt, fp@lx.it.pt

**AFFILIATION:** [1] Instituto Superior Técnico - Universidade de Lisboa, and Instituto de Telecomunicações, Lisbon, Portugal; [2] ESTG, Politécnico de Leiria and Instituto de Telecomunicações, Leiria, Portugal

## Table of Contents

## Executive Summary

This document describes a deep learning (DL)-based point cloud (PC) geometry codec and a DL-based PC joint geometry and colour codec, submitted to the Call for Proposals on JPEG Pleno Point Cloud Coding issued in January 2022 [1].

These proposals have been originated by research developed at Instituto de Telecomunicações (IT), in the context of the project Deep-PCR entitled "Deep learning-based Point Cloud Representation" (PTDC/EEI-COM/1125/2021), financed by Fundação para a Ciência e Tecnologia (FCT).

In summary, this proposal includes:

1) **IT Deep Learning-based Point Cloud Codec for Geometry only (IT-DL-PCC-G) – Codec ID 01**: This codec represents the geometry as binary 3D blocks and includes two DL models, independently trained, one (end-to-end) for block-based coding and another (optional) for advanced block-based up-sampling (ABU). The DL coding model includes three parts addressing latents creation, entropy coding and latents decoding. Depending on the PC density, adaptive down-sampling may be applied to densify the PC being coded. The ABU model allows to increase the reconstructed PC density and quality at no rate cost, thus improving the overall rate-distortion (RD) performance.

2) **IT Deep Learning-based Point Cloud Codec for Geometry and Colour (IT-DL-PCC-GC) – Codec ID 02**: This codec builds on the IT-DL-PCC-G codec above, extending it to code not only the geometry but also three 8-bit colour components, notably R, G and B. IT-DL-PCC-GC uses for geometry and colour the same coding model/architecture adopted in IT-DL-PCC-G for geometry only, and jointly optimizes their coding; naturally, the (optional) ABU model is also extended to 4 channels to improve the geometry and (3) colour qualities at no additional rate cost.

In practice, the two PC codecs co-exist in a single PC coding framework, naturally with different trained DL models, running in a single software, controlled by a configuration parameter which signals geometry-only or joint geometry and colour coding.

The proposed codecs are based on recent developments in deep learning-based PC geometry coding and offer some of the key functionalities targeted by the Call for Proposals on JPEG Pleno Point Cloud Coding issued in January 2022.

The proposed IT-DL-PCC-G codec offers a compression efficiency that outperforms the MPEG G-PCC standard [2] and outperforms or is competitive with the V-PCC Intra standard [3] for the JPEG Call for Proposals test set; however, the same does not happen for the IT-DL-PCC-GC codec due to a quality saturation effect that needs to be overcome.

The proposed learning-based PC coding solutions are based on work already published by the proponents, notably [4-9].

# PART I – IT Deep Learning-based Point Cloud Codec for Geometry-only (IT-DL-PCC-G)

The first part of this document provides a description of the first proposed DL-based codec, IT-DL-PCC-G, targeting geometry-only coding.

## 1. Architecture and High-level Description

The proposed IT-DL-PCC-G codec consists of two main DL-based, neural network, modules, one focusing on compressing the PC geometry data, and the other (optional) focusing on post-processing with the goal of performing up-sampling/super-resolution to increase the quality at no rate cost.

The overall architecture of the IT-DL-PCC-G codec is presented in Fig. 1, with the various modules briefly described as follows:

- **Encoder**:
  - **PC Block Partitioning**: The voxelized PC is divided into disjoint 3D binary blocks (‘1’ and ‘0’ corresponding to filled and empty voxels, respectively) of a fixed target size, which are coded separately; the size of these block defines the random access granularity.
  - **Basic Block Down-sampling**: Depending on the PC characteristics, each partitioned block may be down-sampled to a lower grid precision; this is typically advantageous when the PC is sparse, since it increases the density of the blocks to be encoded.
  - **Deep Learning (DL)-based Block Encoding**: Each block is encoded with an end-to-end DL coding model, which uses the architecture in Fig. 2. This process can be compared to a typical transform coding approach, except that in this case a convolutional autoencoder (AE) with Inception-Residual [10] style Blocks (IRB) is used to learn a non-linear transform. The transform generates a set of coefficients, referred to as the *latent representation,* which are then explicitly quantized by scaling with a defined real-valued quantization step (QS), followed by rounding, and finally entropy coded. A learned adaptive entropy coding model is used, after being estimated via a variational autoencoder (VAE) [11].
  - **Binarization Optimization**: In order to provide adaptability to different PC densities, this module optimizes the binarization process that will be applied at the decoder. Since the output of the DL model decoder does not immediately correspond to a PC, but rather to the probabilities of voxels being occupied, this binarization process has the task to select the voxels and thus PC coordinates which will be ‘filled’, based on the generated probabilities. The optimization process at the encoder produces a binarization parameter *k*, which corresponds to the number of filled voxels that will be reconstructed at the decoder, thus it needs to be transmitted in the bitstream.

- **Decoder**:
    - **DL-based Block Decoding**: Blocks are decoded using the decoder counterpart of the DL-based block encoder mentioned before (see Fig. 2).
    - **Binarization**: The decoded voxel probabilities for each block are binarized to reconstruct the PC coordinates using the binarization parameter value resulting from the optimization performed at the encoder.
    - **Basic Block Up-sampling**: If down-sampling was performed at the encoder, each block is here up-sampled back to the original precision, without increasing the number of points, thus using a basic up-sampling solution.
    - **Advanced Block Up-sampling** (optional): This DL-based post-processing module can be used to perform an advanced set up-sampling of each block (since basic grid up-sampling has already been performed in the Basic Block Up-sampling module), in order to increase the number of points and, therefore, the density of the reconstructed PC at no rate cost, akin to a super-resolution approach.
    - **Binarization**: After the Advanced Block Up-sampling, it is necessary to once again apply the binarization process, since voxel filling probabilities are created by the ABU model.
    - **PC Block Merging**: The decoded blocks are merged to reconstruct the full PC.

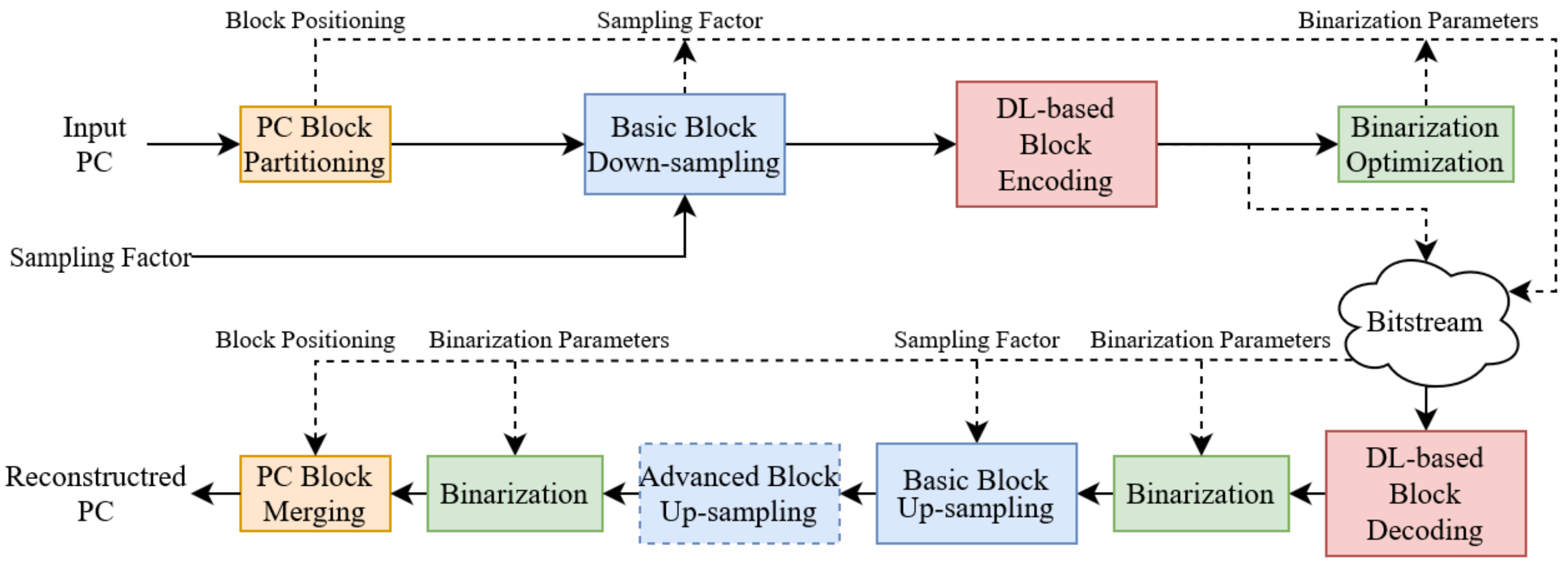

*Fig. 1. Overall architecture of the proposed IT-DL-PCC-G codec.*

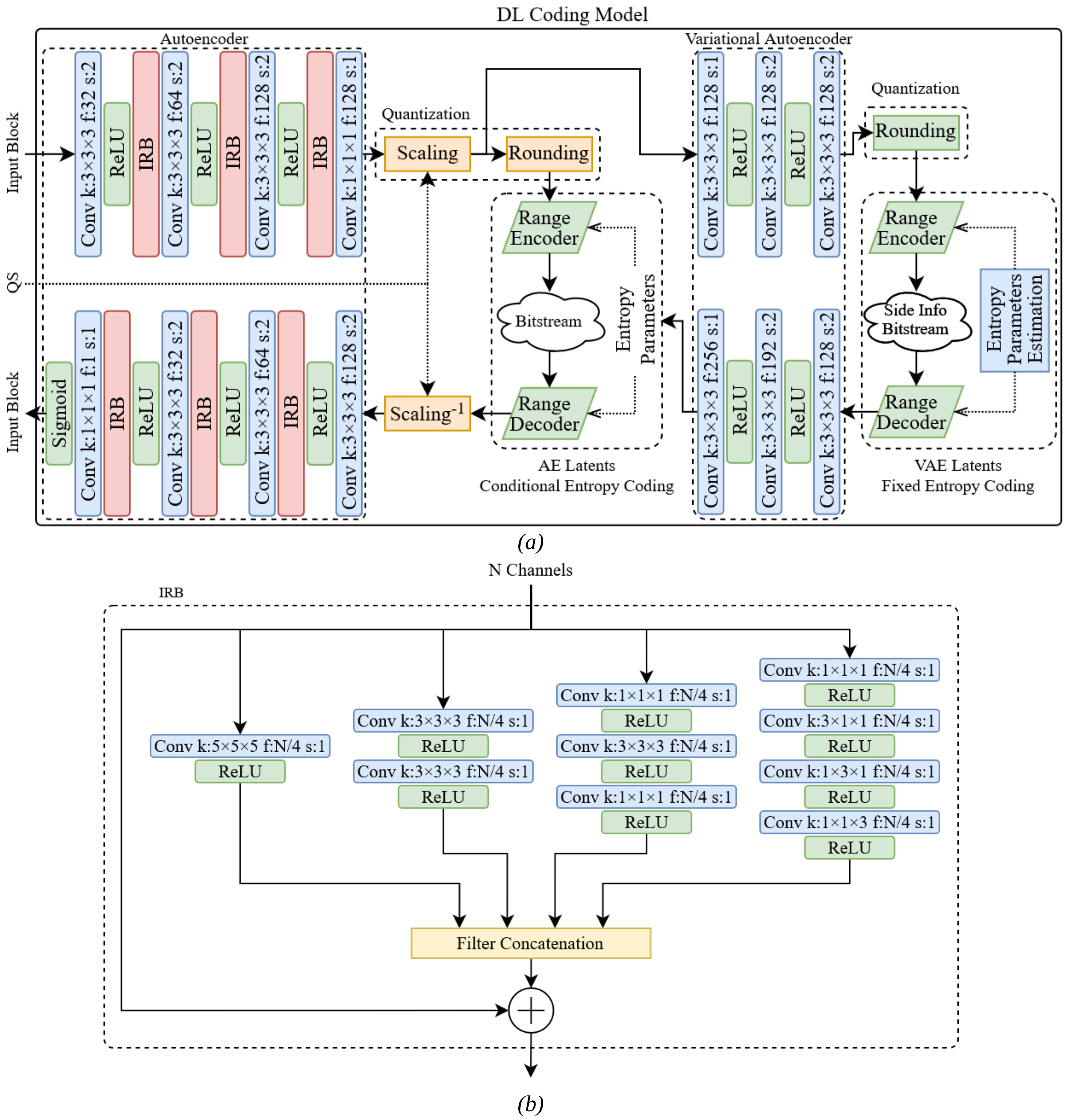

*Fig. 2. End-to-end DL-based coding model architecture. (a) Overall DL coding model. (b) Detail of the IRB block.*

## 2. Detailed Module Description

Each of the codec architecture modules presented in Fig. 1 is described here in more detail.

## 2.1 PC Block Partitioning/Merging

Before encoding, the PC geometry (3D coordinates) is converted into a binary, voxel-based 3D block representation, where voxels may be occupied or not; in practice, a ‘1’ signals a filled voxel while a ‘0’ signals an empty voxel. This voxel-based representation defines a regular structure that allows the use of convolutional neural networks (CNNs), similarly to image and video data; an example is shown in Fig. 3.

| *x* | *y* | *z* |
|---|---|---|
| 3 | 0 | 0 |
| 3 | 0 | 1 |
| 3 | 0 | 2 |
| 3 | 0 | 3 |
| 3 | 1 | 2 |
| 3 | 2 | 1 |
| 3 | 3 | 0 |
| 3 | 3 | 1 |
| 3 | 3 | 2 |
| 3 | 3 | 3 |

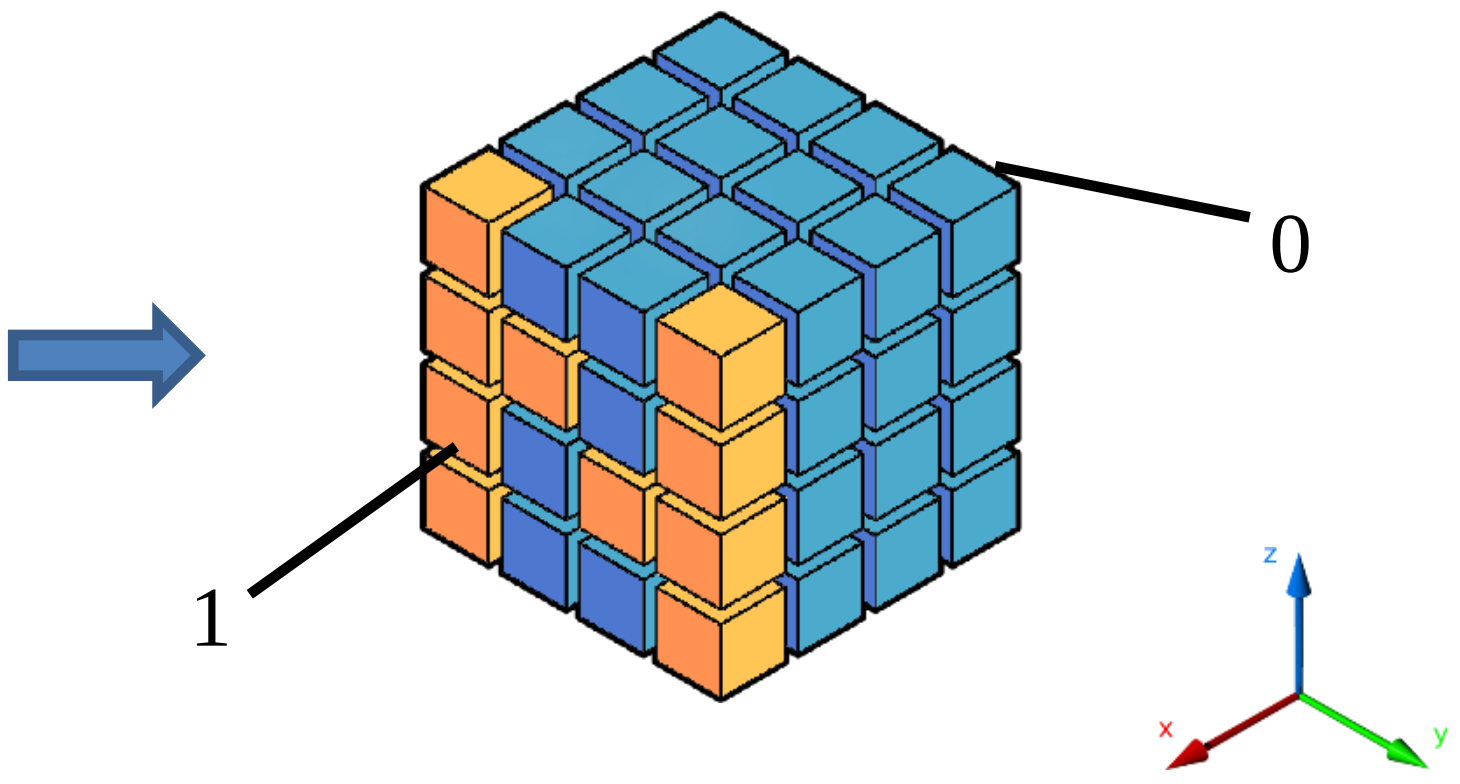

*Fig. 3. Example of conversion from 3D PC coordinates to a 3D block of binary voxels.*

Considering this new representation, a straightforward way to organize a PC is to divide it into disjoint blocks of a specific size, e.g., 64×64×64, which can then be coded separately with a DL coding model. The position of each single 3D block is transmitted to the decoder (*Block Positioning* in Fig. 1).

At the decoder side, given the decoded blocks and their position, the full PC is reconstructed by merging the blocks accordingly.

## 2.2 Basic Block Down/Up-sampling

This pair of modules is used when appropriate depending in the PC characteristics to reduce the PC coding precision (at encoder), allowing a more efficient compression, and then to restore the original precision (at decoder). This is achieved by simply scaling the input PC coordinates by a given sampling factor, followed by a rounding operation, which in turn induces a loss of points/information, but results in a denser surface (also larger voxels). At the decoder, the reconstructed PC coordinates are simply scaled back by the inverse sampling factor, which does not change the number of points or their relative positions.

This approach is particularly useful in two situations:

1. **Sparse PCs coding:** Since the DL coding model requires the conversion of the PC into a 3D block of binary voxels, the entire 3D space is represented including the empty voxels. When dealing with sparse PCs, the number of actual points in each block is much smaller than for dense PCs, meaning that the expended bits per input point (filled voxels) tends to be significantly higher. In addition, the DL coding model tends to have more difficulty to correctly reconstruct sparser surfaces. As such, by reducing the coding precision, the coding blocks are densified, giving an easier task to the DL coding model, and bringing the coding rate to a more reasonable range.

2. **Low rate coding:** It is common to experience some limitations when trying to train a DL coding model to reach low rates, even for dense PCs. Furthermore, the quality of the reconstructed PCs at low rates can be considerably degraded. This approach provides a simple solution to reach low rates with fewer severe coding artifacts by down-sampling the PC and coding the blocks with DL coding models trained for higher qualities/rates.

### 2.3 DL-based Block Encoding and Decoding

This section presents the adopted DL-based PC geometry coding model acting at block-level. Based on successful CNN architectures for image coding [12], the adopted end-to-end DL coding model is presented in Fig. 2. The full architecture can be divided into five main coding stages as follows:

1. **Autoencoder** – The convolutional autoencoder (AE) transforms the input binary 3D block into a latent representation with lower dimensionality, in a way comparable to the transform coding stage in traditional image coding. This latent representation can be regarded as the coefficients of the transform, and consists of multiple feature maps, which number depends on the chosen number of filters for the convolutional layers. The AE consists of a combination of 3D convolutional layers and Inception-Residual Blocks (IRB). The IRB is inspired in the Inception-ResNet [10], a popular neural network used for diverse image processing tasks; it contains several convolutional layers in parallel with different filter support sizes, which allow to extract different types of features from varying neighbouring contexts (from 5×5×5 to 1×1×1); in addition, a residual skip connection allows to propagate the features along the network, which also facilitates the training of deeper models. The number of filters starts from 32 in the first layer, and progressively increases to 128 at the final encoder layer, resulting in a rich latent representation. The AE contains a total of 2842016 trainable parameters, with 1208432 at the encoder side, and 1633584 at the decoder side.

2. **Quantization** – The AE latent representation is explicitly quantized before entropy coding. Considering a given quantization step (QS), which can be any positive real value, the latents are first scaled by QS, and then rounded to the closest integer. This explicit quantization approach allows to fine tune the target rate at coding time for a single trained DL coding model. At training time, an implicit quantization approach is considered (i.e., QS=1), and the rounding is replaced by a differentiable approximation, which consists in adding uniform noise to simulate the quantization error [12].

3. **AE Latents Conditional Entropy Coding –** A conditional entropy bottleneck layer from the Tensorflow compression library [12] is used to entropy code the block latent representation. This entropy coding approach uses a Gaussian mixture model conditioned on a hyperprior as the entropy coding model. During training, this layer estimates the entropy of the latent representation according to the entropy coding model, which is used for the rate-distortion (RD) optimization process. At coding time, a range encoder is used to create the block coding bitstream.

4. **Variational Autoencoder –** A variational autoencoder (VAE) is used to capture possible structure information still present in the block latent representation, which is then used as a hyperprior for the conditional entropy bottleneck. This way, the entropy coding model parameters can be more accurately estimated and adapted for each coded block. The mean-scale hyperprior as proposed in [11] was adopted. In this process, the VAE generates its own latent representation, which must also be coded and transmitted in the bitstream as additional side information to the decoder, so that the entropy coding model parameters can be replicated at the decoder. The VAE has a similar but simpler design to the AE, with only 3 convolutional layers at the encoder and another 3 at the decoder. The VAE contains a total of 3760960 trainable parameters.

5. **VAE Latents Fixed Entropy Coding –** Similar to the conditional entropy coding, this module entropy codes the VAE latent representation. However, it uses a fixed entropy coding model for all blocks instead of an adaptive one, which is learned during training. As all the components of the end-to-end DL coding model are jointly trained, the additional side information rate is compensated by reducing the rate associated with the latents, thus optimizing the overall RD performance.

The total number of trainable parameters in the full DL coding model (AE + VAE) is 6602976.

At the decoder side, each block is decoded with the DL coding model shown in Fig. 2. The “Side Info Bitstream” containing entropy coding related metadata is decoded to generate the entropy coding model parameters used for the current block, so that its “Bitstream” can finally be decoded.

### 2.4 Advanced Block Up-sampling

This section presents the optional Advanced Block Up-sampling (ABU) module. It is important to notice that the ABU receives the output of the Basic Block Up-sampling module, which means that the PC is already in the original precision, although sparser. The ABU goal is to densify the PC, increasing the reconstructed quality at no rate cost; naturally, some complexity cost is involved. In practice, the ABU model expresses how a surface may be densified given a sparser surface, in this case within a PC block.

The ABU module can offer significant RD performance gains, especially for originally dense and uniform PCs. However, this is not always the case, depending on the PC content characteristics (e.g., sparsity), as well as the reconstruction quality of the DL-based codec (whether it contains many coding artifacts or not). For this reason, the ABU is an optional post-processing module which may be activated via the configuration used to run the software (refer to Part III of this document for more information).

The ABU architecture is based on the solution proposed in [13], and consists on a 3D CNN shaped as a U-net [14], as shown in Fig. 4. The full architecture can be divided into two main processing stages as follows:

1. **Contracting Path** – The first path of the U-net is responsible for extracting features at different scales. Similarly to the AE in the DL coding model, it combines 3D down-sampling convolutional layers and IRB blocks, inspired in Inception-ResNet [10]. Compared to the DL coding model, the IRB is much simpler and lighter, with fewer and smaller filter supports (maximum of 3×3×3). On the other hand, it is a much deeper network, with many more convolutional layers and IRB’s. The number of filters/channels starts from C=16 in the first layer, and progressively increases.

2. **Expanding Path** – The second path is nearly symmetrical to the first path, now successively up-sampling the features, but with an additional task of aggregating the multiscale features extracted by the contracting path. By considering the features obtained at different scales, this path is able to accurately predict the occupation of the voxels which were lost due to the down-sampling process performed at encoder.

The total number of trainable parameters in the full ABU model is 7288893.

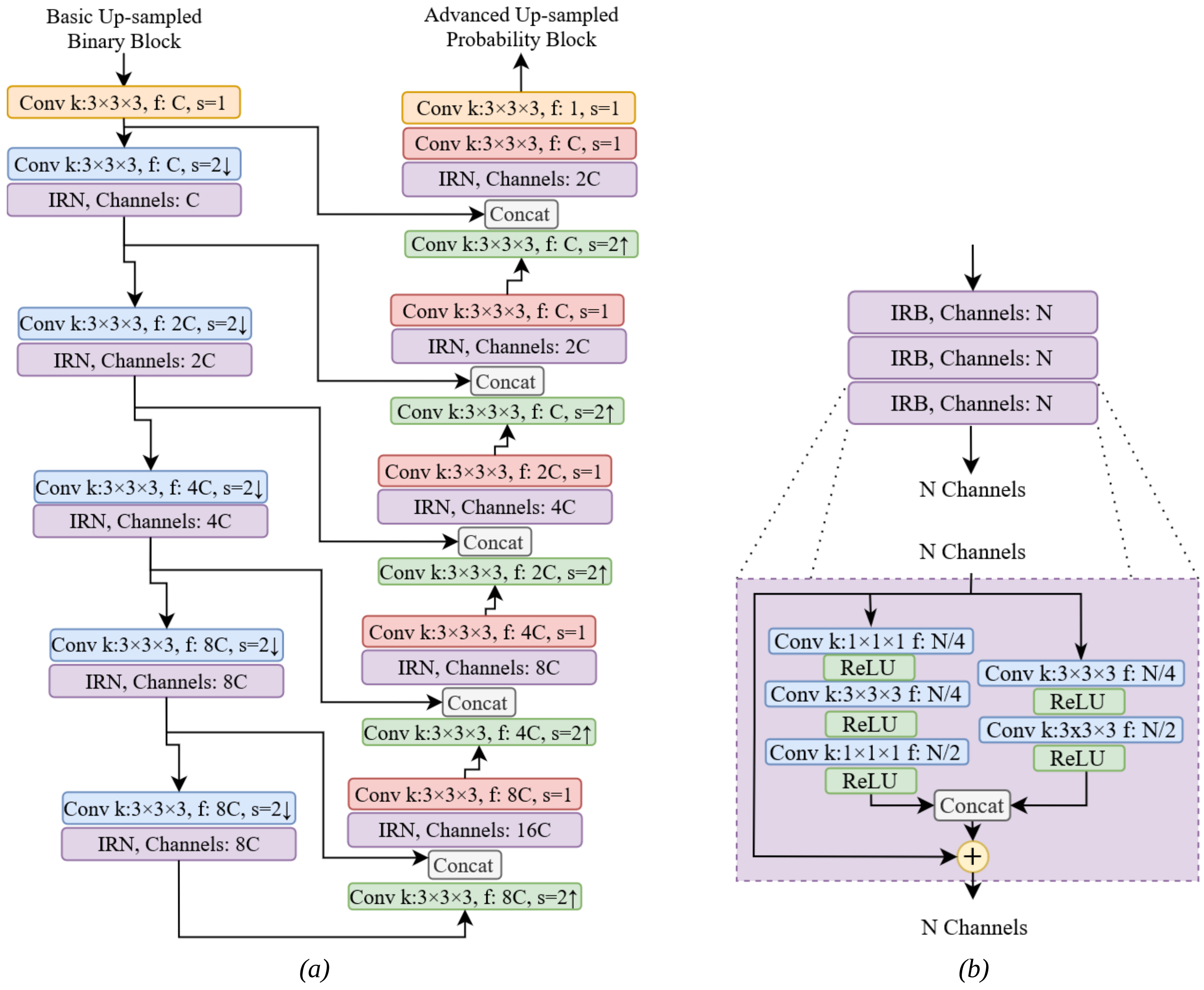

*Fig. 4. DL-based ABU model architecture. (a) Overall ABU model. (b) Detail of the IRN block.*

## 2.5 Binarization

The output of the DL-based decoder is a block with values in the interval [0, 1] for each voxel where each value represents the probability of a given voxel being filled. As such, it is necessary to transform these probabilities into binary values which can directly correspond to the final reconstructed points. A straightforward approach would be to use a simple fixed 0.5 threshold; however, this rigid solution presents a major drawback: by using a fixed threshold, the final reconstructed block becomes highly dependent on the block density assumed during training.

As such, a so-called *optimized Top-k* binarization approach was adopted for selecting the filled voxels, in which only the *k* voxels with the largest probabilities are selected as points, with *k* being defined as:

$$k_{Codec} = N_{input} \times \beta, \tag{1}$$

where $N_{input}$ is the number of input points of the original block (known), and β is a factor selected at the encoder. This β factor is optimized at the encoder by determining the value which results in the best reconstructed quality, using a quality metric such as PSNR-D1 or PSNR-D2; naturally, the best value of $k_{Codec}$ needs to be coded in the bitstream to be available to the decoder.

Furthermore, the occupation of each octant of the block is determined and transmitted to the decoder so that only voxels inside originally occupied octants can be selected as filled voxels.

Similarly to the DL coding model, the output of the ABU model is a block of voxel filling probabilities, thus also requiring some binarization process. To maximize the RD performance, two binarization approaches may be selected with different implications on the encoder complexity:

1. The same optimized Top-*k* approach used after the DL-based coding model may be applied; however, this requires ABU optimization during the encoding process, i.e. performing ABU at the encoder side, which can considerably increase the encoding time and complexity.

2. Alternatively, the same binarization multiplying factor β determined for the DL coding model optimization can be applied, at no increased encoding complexity cost. While it can still bring a good RD performance, notably for dense PCs, this approach is not optimal.

In both these approaches, only a binarization parameter $k_{ABU}$, computed as in Equation (1), needs to be transmitted in the bitstream together with the $k_{Codec}$ parameter, as well as the octant occupation code (*Binarization Parameters* in Fig. 1).

## 3. DL Model Training

This section describes the important training processes for the DL coding and ABU models.

### 3.1 Coding Model Training

In order to achieve efficient compression performance, the DL coding model from Fig. 2 was trained by minimizing a loss function that considers both the distortion of each decoded block, compared to the input block, as well as its estimated coding rate. For this purpose, the loss function follows a traditional formulation involving a Lagrangian multiplier, λ, and is given by:

$$Loss\ Function = Distortion + \lambda \times Coding\ rate. \tag{2}$$

DL-based codecs typically require training a different DL coding model for each target RD point, which is accomplished by varying the $\lambda$ parameter in Equation (2). A total of 6 models were trained, using $\lambda$ = 0.00025, 0.0005, 0.001, 0.0025, 0.005, and 0.01.

### Training Distortion Metric

As described in Section 2.1, a voxel-based representation was adopted to process the PCs. Thus, for the DL coding model, the input data is a block of binary voxels, and the decoded data represents a probability score between '0' and '1' for each voxel, i.e. the probability of each voxel being filled, since binarization cannot be performed during training as this is not a differentiable operation. Considering this, the block distortion is measured as the average voxel level distortion, computed as a binary classification error using the so-called *Focal Loss* (FL) [15], defined as follows:

$$FL(v,u) = \begin{cases} -\alpha(1-v)^{\gamma}\log v\,, & u=1 \\ -(1-\alpha)v^{\gamma}\log(1-v)\,, & u=0 \end{cases}, \quad (3)$$

where $u$ is the original voxel binary value and $v$ is the corresponding decoded voxel probability value. A weight parameter, α, is used to control the class imbalance effect since the number of '0' valued voxels in a block is vastly superior to the number of '1' valued voxels. The parameter γ allows increasing the importance of correcting misclassified voxels in relation to improving the classification score of already correct voxels. For the used models, the values α=0.7 and γ=2 for these two parameters were found to be appropriate.

### Training Coding Rate

The coding rate is estimated during training as the entropy of the AE and VAE latent representations, considering the computed conditional and fixed entropy coding models, respectively.

### Trained Models

The DL coding model is trained using *a selection of the static PCs* listed in the JPEG Pleno PCC Common Training and Testing Conditions (CTTC) [17]. As detailed in Table I, the selected PCs were down-sampled to a lower precision according to their sparsity (lighter colour signals the PCs which were down-sampled before training, and darker colour signals the PCs which were coded during training with the original precision), and then partitioned into blocks of size 64×64×64, as described in Section 2.1. The blocks with less than 500 'filled' voxels have been removed to avoid the negative effect on the training due to the increased class imbalance caused by such low point count blocks. In total, 35861 blocks were used for training and 3822 blocks for validation.

The PCs were split into training and validation sets, with the validation set being used for early stopping of the training process, in order to prevent overfitting. For early stopping, a patience of 5 epochs was defined, meaning that the training only stopped when the validation loss did not decrease further after 5 epochs.

*Table I. Dataset for training and validation of the DL coding model.*

| | Point Cloud | Frame | Original Precision | Original Points | Training Precision | Training Points | Blocks ($64^3$) |
|---|---|---|---|---|---|---|---|
| Training | Loot | 1200 | 10 | 805285 | 10 | 805285 | 192 |
| | Redandblack | 1550 | 10 | 757691 | 10 | 757691 | 166 |
| | Soldier | 690 | 10 | 1089091 | 10 | 1089091 | 235 |
| | Thaidancer | viewdep | 12 | 3130215 | 11 | 1007956 | 222 |
| | Andrew10 | 1 | 10 | 1276312 | 10 | 1276312 | 224 |
| | David10 | 1 | 10 | 1492780 | 10 | 1492780 | 277 |
| | Sarah10 | 1 | 10 | 1355867 | 10 | 1355867 | 258 |
| | The20sMaria | 600 | Not voxelized | 10383094 | 11 | 3681165 | 669 |
| | UlliWegner | 1400 | Not voxelized | 879709 | 10 | 537042 | 150 |
| | Basketball_player | 200 | 11 | 2925514 | 11 | 2925514 | 682 |
| | Exercise | 1 | 11 | 2391718 | 11 | 2391718 | 530 |
| | Model | 1 | 11 | 2458429 | 11 | 2458429 | 561 |
| | Mitch | 1 | 11 | 2289640 | 11 | 2289640 | 821 |
| | ThomasSenic | 170 | 11 | 2277443 | 11 | 2277443 | 749 |
| | Football | 1365600 | 11 | 1021107 | 11 | 1021107 | 160 |
| | Façade 15 | - | 14 | 8907880 | 12 | 6834258 | 2517 |
| | Façade 64 | - | 14 | 19702134 | 12 | 12755151 | 3539 |
| | Egyptian_mask | - | 12 | 272684 | 9 | 269739 | 141 |
| | Head_00039 | - | 12 | 13903516 | 12 | 13903516 | 9218 |
| | Shiva_00035 | - | 12 | 1009132 | 10 | 901081 | 282 |
| | ULB_Unicorn | - | 13 | 1995189 | 10 | 1588315 | 462 |
| | Landscape_00014 | - | 14 | 71948094 | 12 | 15270319 | 3069 |
| | Stanford_Area_2 | - | 16 | 47062002 | 12 | 19848824 | 4178 |
| | Stanford_Area_4 | - | 16 | 43399204 | 12 | 24236048 | 6559 |
| **Validation** | Boxer | viewdep | 12 | 3493085 | 11 | 3056129 | 915 |
| | Dancer | 1 | 11 | 2592758 | 11 | 2592758 | 604 |
| | Façade 09 | - | 12 | 1596085 | 11 | 1560834 | 836 |
| | Frog_00067 | - | 12 | 3614251 | 11 | 3321097 | 1467 |

Implementation and training were done in Tensorflow version 2.8, using the Tensorflow Compression library [12] version 2.8. For training, the Adam algorithm [16] was used with a learning rate of $10^{-4}$ and minibatches of 16 blocks.

### 3.2 ABU Model Training

Regarding the ABU, one model has to be trained for each relevant up-sampling factor, using the same training and validation datasets as described in the previous section; in this case, models have been trained for sampling factors 2 and 4. To generate the training input data, the previously generated blocks were down-sampled and up-sampled with the basic up-sampling approach described in Section 2.2, thus resulting into sparser blocks although at the same precision. Note that, just for training purposes, there was no coding involved.

Given that the ABU module has no impact on the rate, the ABU model was trained simply considering the distortion as loss function, using the same Focal Loss as presented in Equation (3), with the same parameter values. However, while the DL coding model measures the distortion between the decoded block and the block that is received as input, the ABU model measures the distortion between the original block and the basic up-sampled block after down-sampling.

At testing time, a single ABU model is used for all the rates, unlike the DL coding model. As such, the only dependency is the sampling factor, with two ABU models being trained in total, one considering a sampling factor of 2 and another considering a sampling factor of 4.

Implementation and training were done in Tensorflow version 2.8. For training, the Adam algorithm [16] was used with a learning rate of $10^{-4}$, and minibatches of 1 and 8 blocks for sampling factor of 2 and 4, respectively.

## 4. Performance Assessment for Geometry Codec

To assess the performance of the proposed geometry-only coding solution, IT-DL-PCC-G, the JPEG test PCs were coded following the JPEG CTTC [17].

### 4.1 Test Material

The test material used for the assessment of the proposed codecs consists of 20 PCs, including people and inanimate objects. This dataset was made available in the context of the JPEG Pleno Call for Proposals on Point Cloud Coding [1], and is presented in Fig. 5 and Table II. The sparsity of each PC was measured as the average distance between each point and its 20 nearest neighbours, averaged across the entire PC.

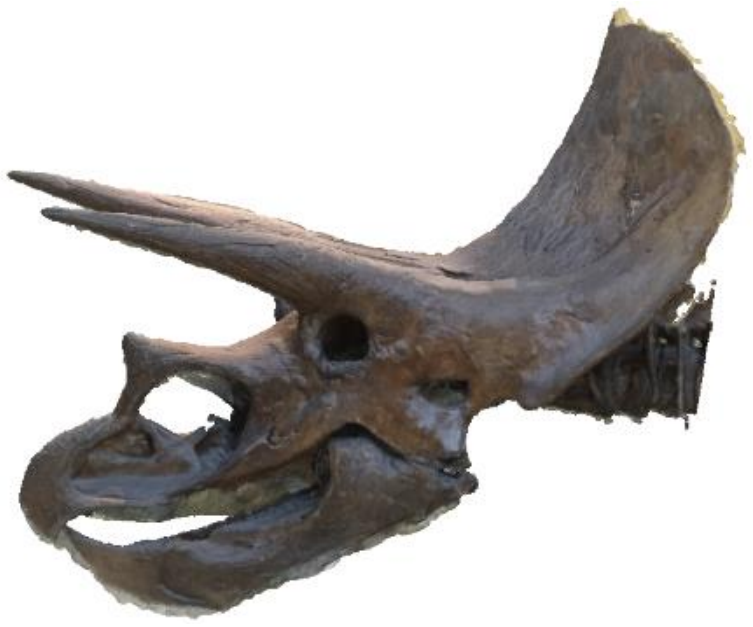

*(a) RWT2 (DinoSkull)*

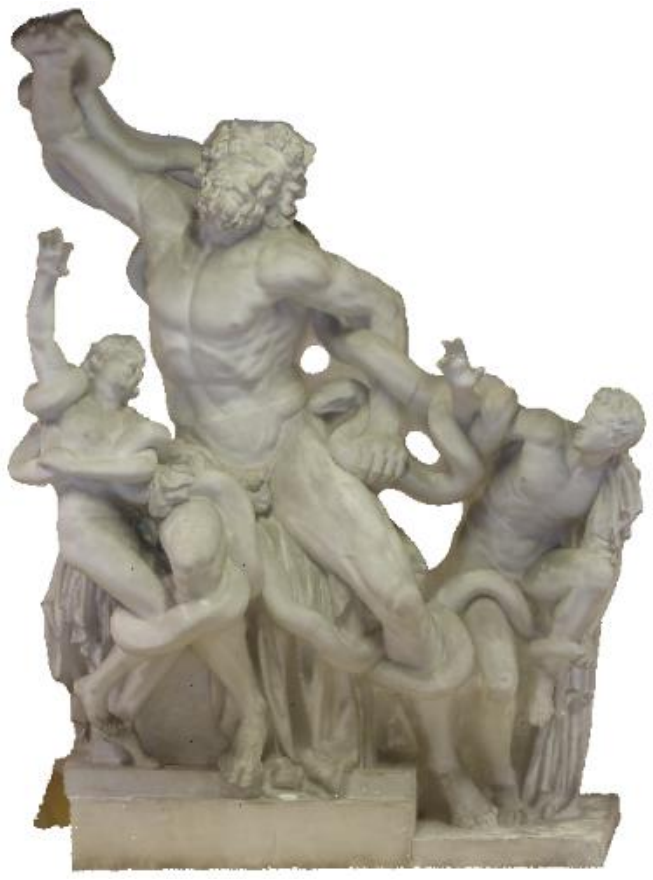

*(b) RWT34 (Lakoon)*

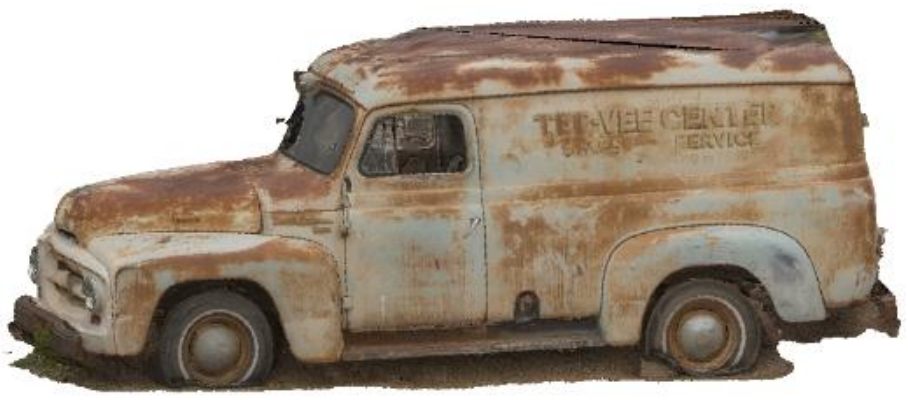

*(c) RWT53 (Van)*

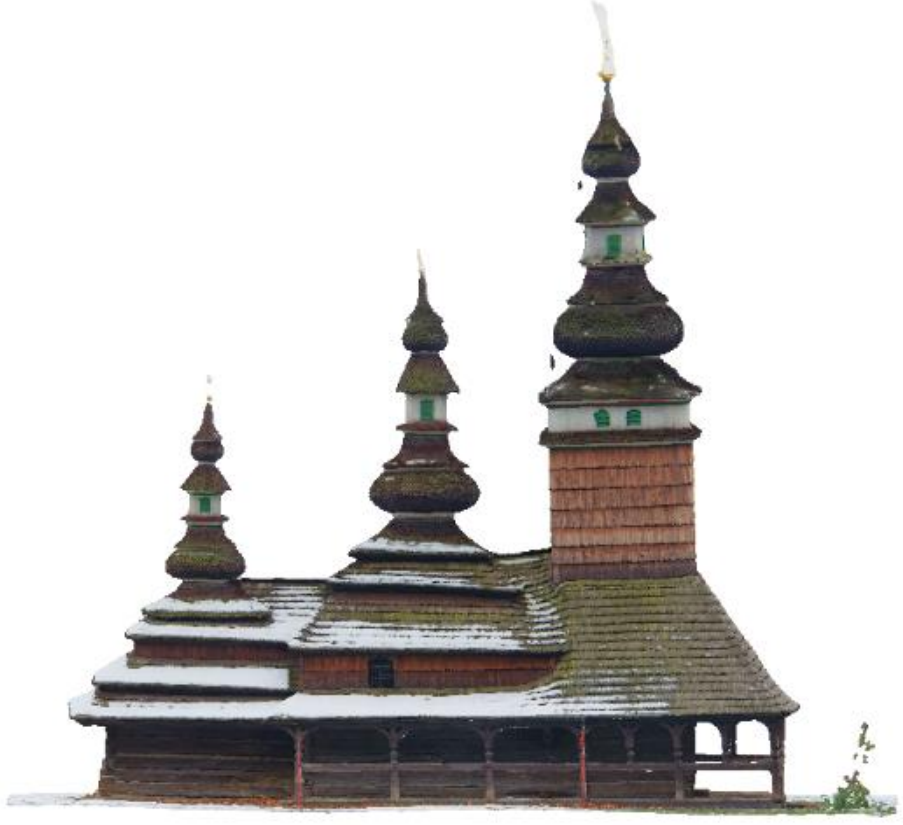

*(d) RWT70 (StMichael)*

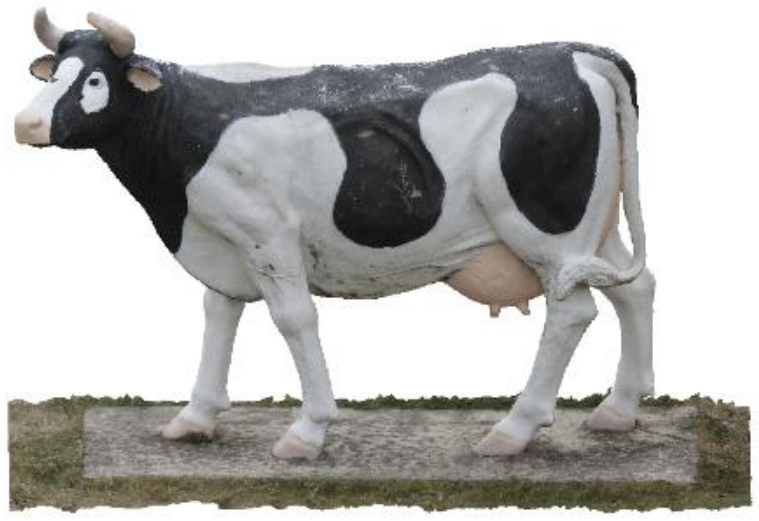

*(e) RWT120 (CowStatue)*

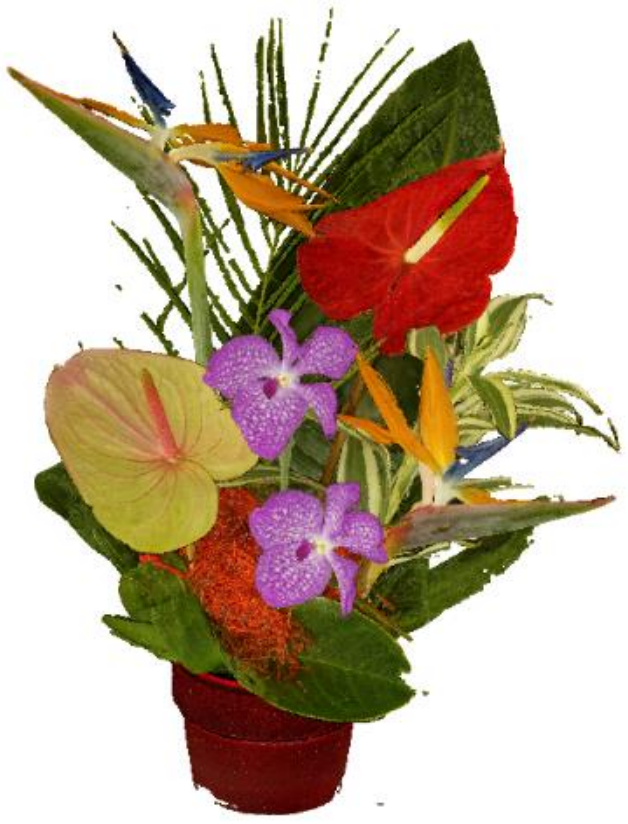

*(f) RWT130 (Bouquet)*

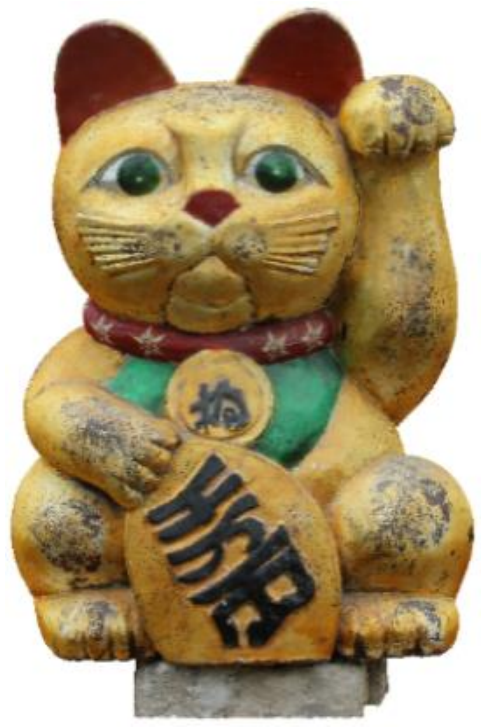
(g) RWT134 (CatStatue)

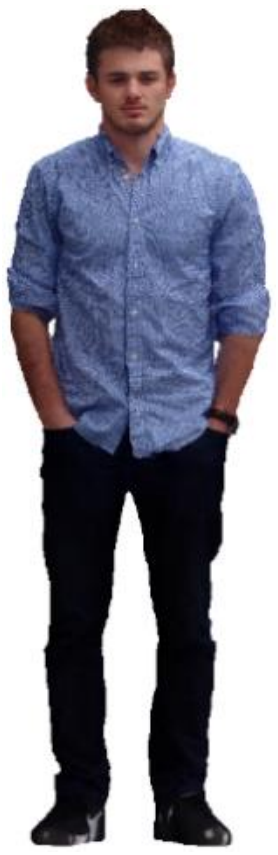
(h) RWT136 (BodyScanBlueShirt)

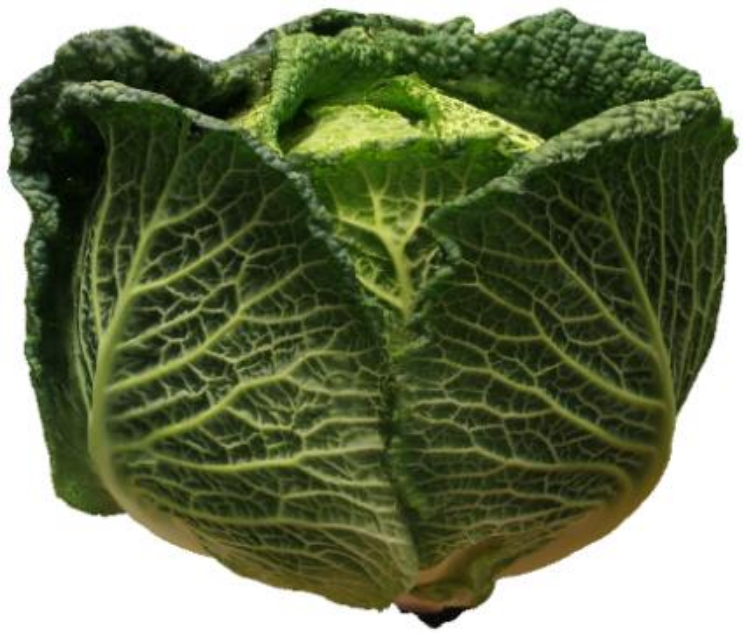
(i) RWT144 (Cabbage)

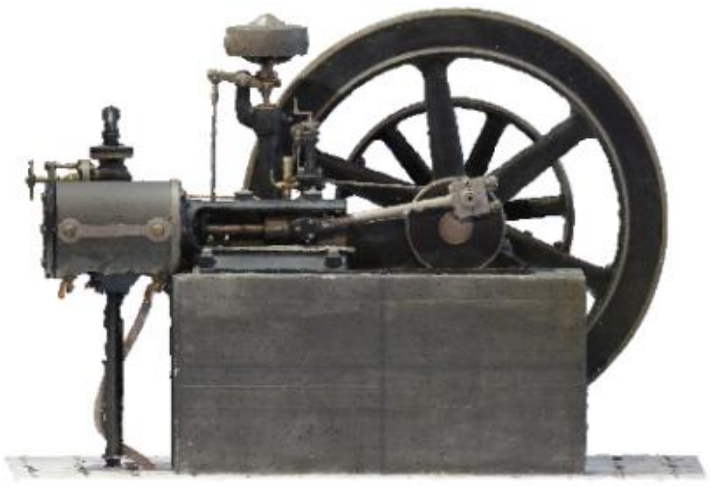
(j) RWT152 (SteamEngine)

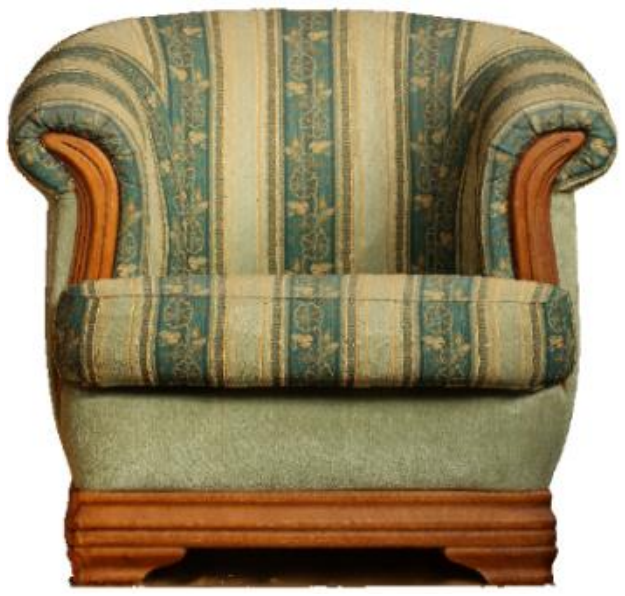
(k) RWT246 (ArmChair)

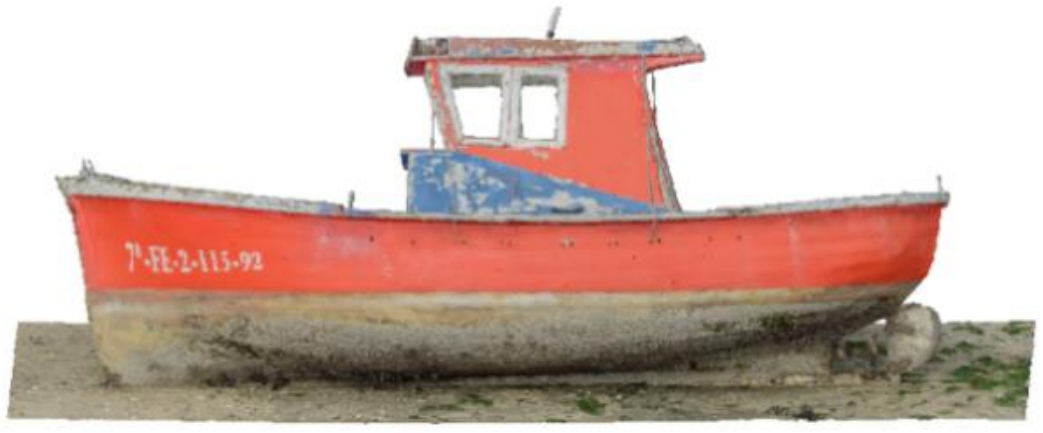
(l) RWT305 (BoatJosefa)

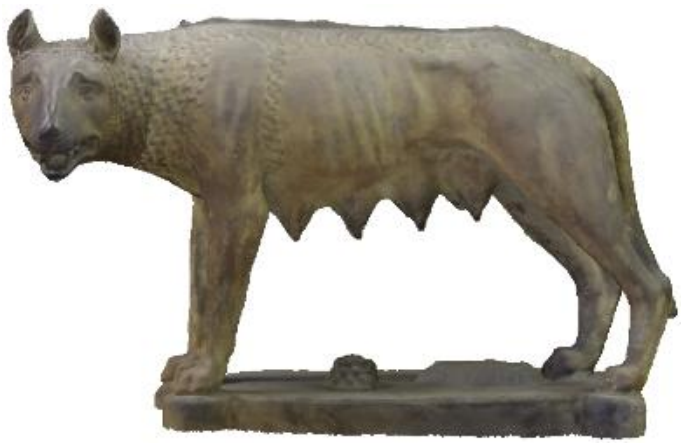
(m) RWT374 (CapitolineWolf)

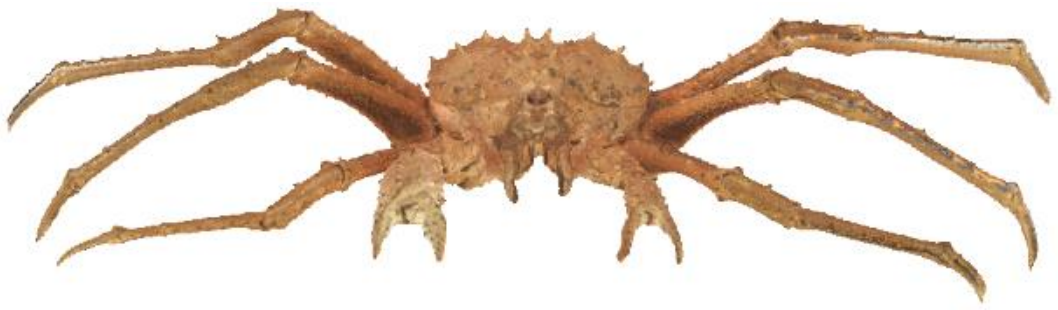
(n) RWT395 (KingCrab)

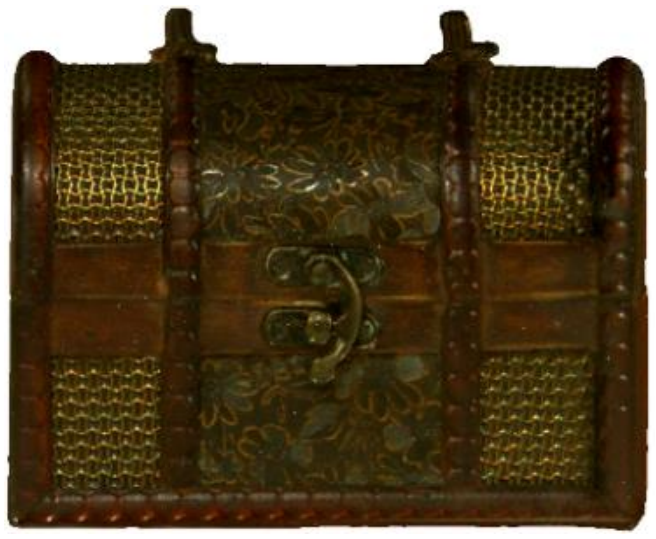
(o) RWT430 (WoodenChest)

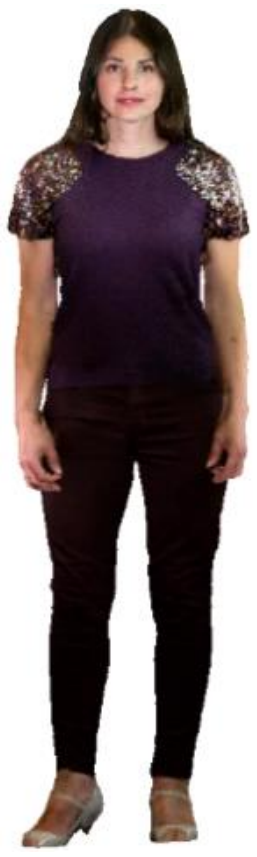
(p) RWT462 (BodyScanOlia)

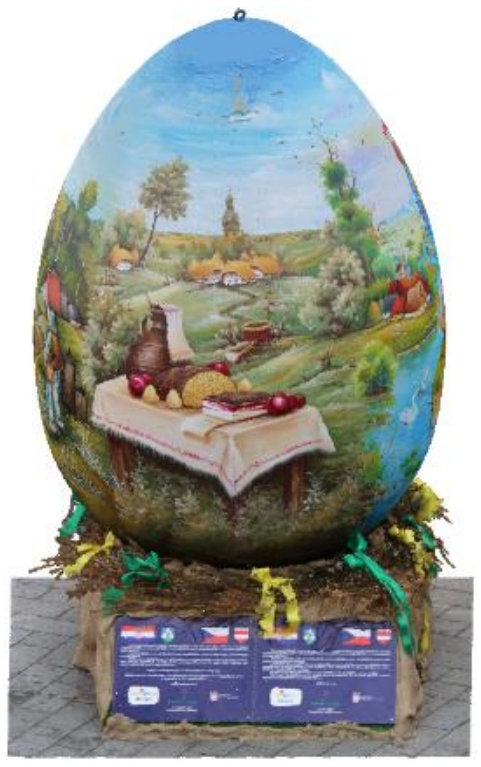
(q) RWT473 (PaintedEgg)

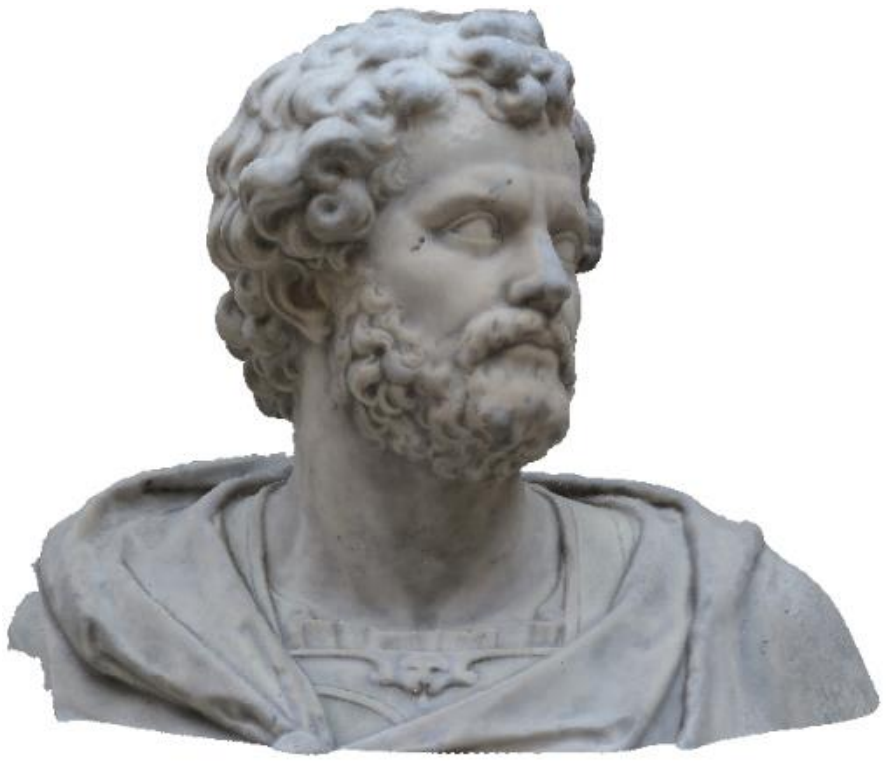
(r) RWT501 (Annibal)

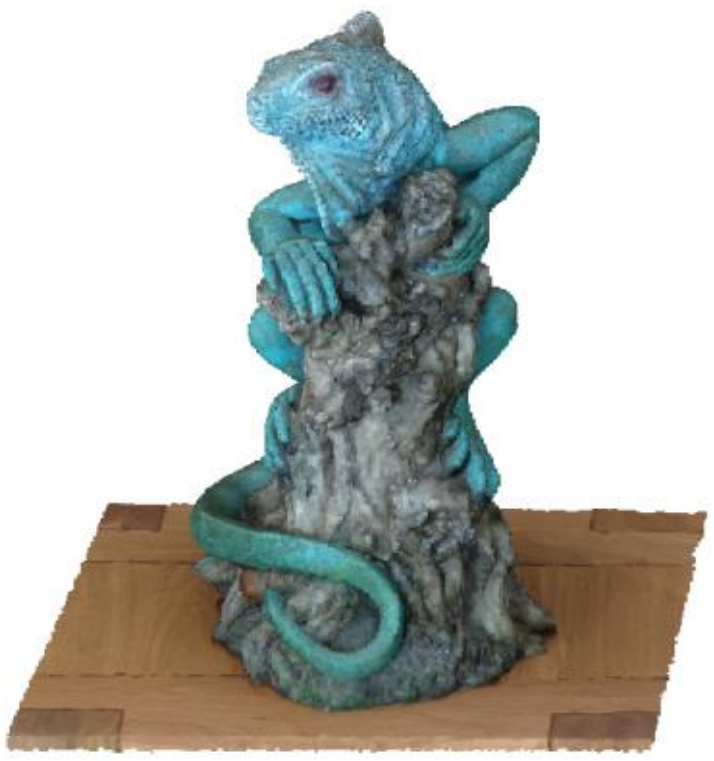

(s) RWT503b (Iguana)

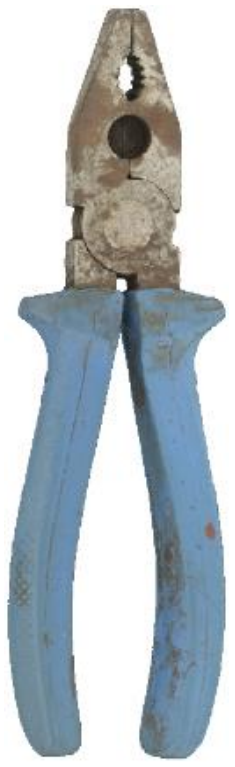

(t) RWT529 (Pliers)

*Fig. 5. Example rendering for the test PCs.*

*Table II. Test PCs and their descriptive names, number of points, and sparsity.*

| PC File | PC Name | Points | Sparsity |
|---|---|---|---|
| RWT2 | DinoSkull | 2 308 312 | 1.610369 |
| RWT34 | Lakoon | 2 212 435 | 1.614183 |
| RWT53 | Van | 1 563 866 | 1.60725 |
| RWT70 | StMichael | 1 871 158 | 1.60316 |
| RWT120 | CowStatue | 1 880 897 | 1.614311 |
| RWT130 | Bouquet | 3 150 249 | 1.602903 |
| RWT134 | CatStatue | 2 350 417 | 1.61968 |
| RWT136 | BodyScanBlueShirt | 835 264 | 1.571434 |
| RWT144 | Cabbage | 4 143 911 | 1.719278 |
| RWT152 | SteamEngine | 3 402 890 | 1.683609 |
| RWT246 | ArmChair | 3 486 192 | 1.663971 |
| RWT305 | BoatJosefa | 1 859 311 | 1.621199 |
| RWT374 | CapitolineWolf | 1 658 288 | 1.60332 |
| RWT395 | KingCrab | 655 412 | 1.545843 |
| RWT430 | WoodenChest | 3 839 604 | 1.716197 |
| RWT462 | BodyScanOlia | 813 808 | 1.570685 |
| RWT473 | PaintedEgg | 2 790 723 | 1.634813 |
| RWT501 | Annibal | 2 404 242 | 1.614124 |
| RWT503b | Iguana | 291 955 | 1.549672 |
| RWT529 | Pliers | 663 367 | 1.600387 |

## 4.2 Selecting the Coding Configurations

In DL-based coding solutions, it is typical to train a DL coding model for a specific target RD trade-off. However, such approach becomes impractical when aiming to achieve a given bitrate at testing time, without the possibility of training new models. As such, the proposed IT-DL-PCC-G solution avoids this issue by allowing a more flexible coding configuration at testing time, containing several parameters, namely:

- **DL coding model**: Six DL coding models are available, which were trained for different RD trade-offs, spanning a wide range of rates.
- **Quantization step (QS)**: The quantization step parameter applied to the latents can be used for finetuning the target rate for each PC after selecting a specific DL coding model.
- **Coding block size (BS):** The size of the 3D block coding units can be selected, allowing not only a finetuning of the rate, but also a trade-off between performance and random access granularity.
- **Sampling factor (SF):** The sampling factor can be defined to allow reaching lower rates (by increasing its value) or to improve coding performance for sparser PCs.
- **ABU:** The ABU module can be activated optionally, with the goal to improve the reconstruction quality when adopting a sampling factor larger than 1.

Considering these coding parameters and the target rates defined in the JPEG CTTC (0.05, 0.15, 0.5, 1.5 bpp), several coding configurations were adopted to code each test PC in order to select the coding configurations which allow reaching the target rates for each PC.

In this way, a "cloud" of RD points was generated for each PC (each point corresponding to a different configuration). Then, four RD points that correspond to the JPEG target rates (within a 10% tolerance margin) were selected, forming the convex hull of this "cloud" for each PC. An example of such a "cloud" and its convex hull is presented in Fig. 6.

With this approach, it was possible to reach the JPEG target rates for all test PCs. However, the last target rate (1.5 bpp) is considerably higher than what was possible to reach for all PCs. Nonetheless, the selected RD point is able to offer near-lossless geometry quality.

Table III shows the coding configurations determined to achieve the JPEG defined target rates for each PC.

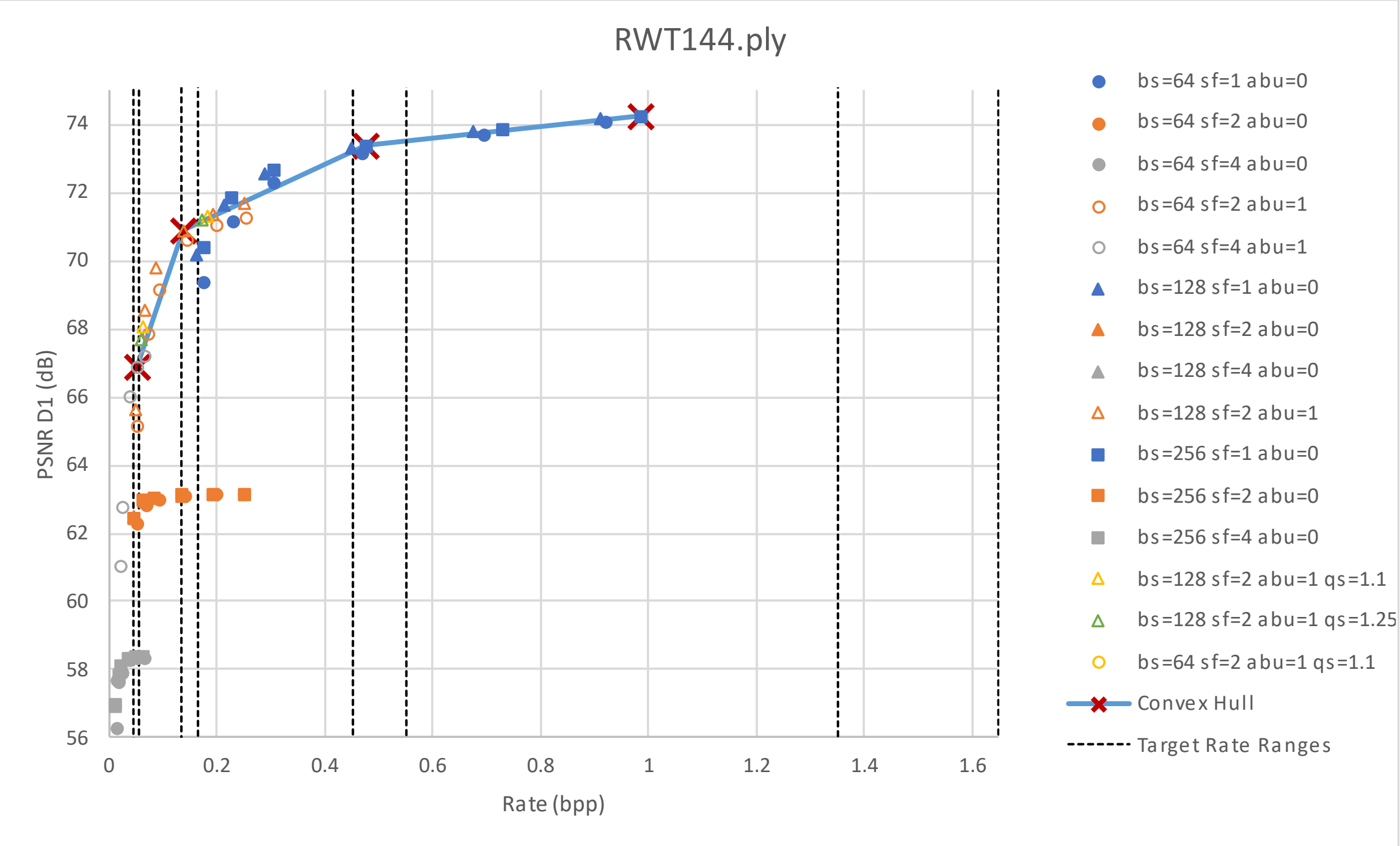

*Fig. 6. Example of a "cloud" of RD points for multiple coding configurations, for the RWT144 PC (bs means block size and sf means sampling factor).*

*Table III. IT-DL-PCC-G coding configurations to achieve the JPEG target rates.*

| PC | R1 | R2 | R3 | R4 |
|---|---|---|---|---|
| ***RWT2*** | bs=128 sf=2 abu=1 qs=1.1 λ=0.005 | bs=128 sf=2 abu=1 qs=1.25 λ=0.0005 | bs=256 sf=1 abu=0 qs=1 λ=0.001 | bs=256 sf=1 abu=0 qs=1 λ=0.00025 |
| ***RWT34*** | bs=128 sf=2 abu=1 qs=1.25 λ=0.005 | bs=128 sf=2 abu=1 qs=1.25 λ=0.0005 | bs=256 sf=1 abu=0 qs=1 λ=0.001 | bs=256 sf=1 abu=0 qs=1 λ=0.00025 |
| ***RWT53*** | bs=128 sf=2 abu=1 qs=1 λ=0.005 | bs=128 sf=2 abu=1 qs=1.1 λ=0.0005 | bs=256 sf=1 abu=0 qs=1 λ=0.001 | bs=256 sf=1 abu=0 qs=1 λ=0.00025 |
| ***RWT70*** | bs=128 sf=2 abu=1 qs=1.1 λ=0.005 | bs=128 sf=2 abu=1 qs=1.1 λ=0.0005 | bs=256 sf=1 abu=0 qs=1 λ=0.001 | bs=256 sf=1 abu=0 qs=1 λ=0.00025 |
| ***RWT120*** | bs=64 sf=2 abu=1 qs=1.1 λ=0.005 | bs=128 sf=2 abu=1 qs=1.1 λ=0.0005 | bs=256 sf=1 abu=0 qs=1 λ=0.001 | bs=256 sf=1 abu=0 qs=1 λ=0.00025 |
| ***RWT130*** | bs=128 sf=2 abu=1 qs=1.1 λ=0.005 | bs=128 sf=2 abu=1 qs=1.1 λ=0.0005 | bs=256 sf=1 abu=0 qs=1 λ=0.001 | bs=256 sf=1 abu=0 qs=1 λ=0.00025 |
| ***RWT134*** | bs=128 sf=2 abu=1 qs=1 λ=0.005 | bs=128 sf=2 abu=1 qs=1 λ=0.0005 | bs=256 sf=1 abu=0 qs=1 λ=0.001 | bs=256 sf=1 abu=0 qs=1 λ=0.00025 |
| ***RWT136*** | bs=128 sf=2 abu=1 qs=1 λ=0.005 | bs=128 sf=2 abu=1 qs=1 λ=0.0005 | bs=256 sf=1 abu=0 qs=1.3 λ=0.0005 | bs=256 sf=1 abu=0 qs=1 λ=0.00025 |
| ***RWT144*** | bs=64 sf=4 abu=1 qs=1 λ=0.0005 | bs=128 sf=2 abu=1 qs=1 λ=0.001 | bs=256 sf=1 abu=0 qs=1 λ=0.001 | bs=256 sf=1 abu=0 qs=1 λ=0.00025 |
| ***RWT152*** | bs=128 sf=2 abu=1 qs=1 λ=0.005 | bs=256 sf=1 abu=0 qs=1 λ=0.01 | bs=256 sf=1 abu=0 qs=1 λ=0.001 | bs=256 sf=1 abu=0 qs=1 λ=0.00025 |
| ***RWT246*** | bs=128 sf=2 abu=1 qs=1 λ=0.005 | bs=128 sf=2 abu=1 qs=1 λ=0.0005 | bs=256 sf=1 abu=0 qs=1 λ=0.001 | bs=256 sf=1 abu=0 qs=1 λ=0.00025 |
| ***RWT305*** | bs=128 sf=2 abu=1 qs=1 λ=0.005 | bs=256 sf=1 abu=0 qs=1.1 λ=0.005 | bs=256 sf=1 abu=0 qs=1 λ=0.0005 | bs=256 sf=1 abu=0 qs=1 λ=0.00025 |
| ***RWT374*** | bs=128 sf=2 abu=1 qs=1.1 λ=0.005 | bs=128 sf=2 abu=1 qs=1 λ=0.0005 | bs=256 sf=1 abu=0 qs=1 λ=0.001 | bs=256 sf=1 abu=0 qs=1 λ=0.00025 |
| ***RWT395*** | bs=128 sf=2 abu=1 qs=1 λ=0.01 | bs=128 sf=2 abu=1 qs=1 λ=0.001 | bs=256 sf=1 abu=0 qs=1 λ=0.001 | bs=256 sf=1 abu=0 qs=1 λ=0.00025 |
| ***RWT430*** | bs=64 sf=2 abu=1 qs=1 λ=0.005 | bs=128 sf=1 abu=0 qs=1 λ=0.005 | bs=128 sf=1 abu=0 qs=1.1 λ=0.0005 | bs=256 sf=1 abu=0 qs=1 λ=0.00025 |
| ***RWT462*** | bs=128 sf=2 abu=1 qs=1 λ=0.005 | bs=128 sf=2 abu=1 qs=1 λ=0.0005 | bs=256 sf=1 abu=0 qs=1.25 λ=0.0005 | bs=256 sf=1 abu=0 qs=1 λ=0.00025 |
| ***RWT473*** | bs=128 sf=2 abu=1 qs=1 λ=0.005 | bs=128 sf=2 abu=1 qs=1 λ=0.0005 | bs=256 sf=1 abu=0 qs=1 λ=0.001 | bs=256 sf=1 abu=0 qs=1 λ=0.00025 |
| ***RWT501*** | bs=64 sf=4 abu=1 qs=1 λ=0.0005 | bs=128 sf=2 abu=1 qs=1 λ=0.001 | bs=256 sf=1 abu=0 qs=1 λ=0.001 | bs=256 sf=1 abu=0 qs=1 λ=0.00025 |
| ***RWT503b*** | bs=128 sf=2 abu=1 qs=1 λ=0.01 | bs=128 sf=2 abu=1 qs=1 λ=0.001 | bs=128 sf=1 abu=0 qs=1 λ=0.001 | bs=256 sf=1 abu=0 qs=1 λ=0.00025 |
| ***RWT529*** | bs=128 sf=2 abu=1 qs=1 λ=0.005 | bs=128 sf=1 abu=0 qs=1 λ=0.005 | bs=128 sf=1 abu=0 qs=1 λ=0.0005 | bs=256 sf=1 abu=0 qs=1 λ=0.00025 |

### 4.3 RD Performance Results: RD Charts and BD-PSNR

The RD performance results for the IT-DL-PCC-G codec are plotted in RD charts and compared with the G-PCC Octree and V-PCC Intra anchors in Fig. 7, for both the PSNR-D1 and PSNR-D2 geometry objective quality metrics. For the *RWT152* PC, V-PCC Intra anchor results were only provided for the last two target rates. To summarize the results, Table IV shows the Bjontegaard-Delta PSNR gains (BD-PSNR) and Bjontegaard-Delta rate savings (BD-Rate) for the proposed IT-DL-PCC-G coding solution, over the G-PCC Octree and V-PCC Intra anchors.

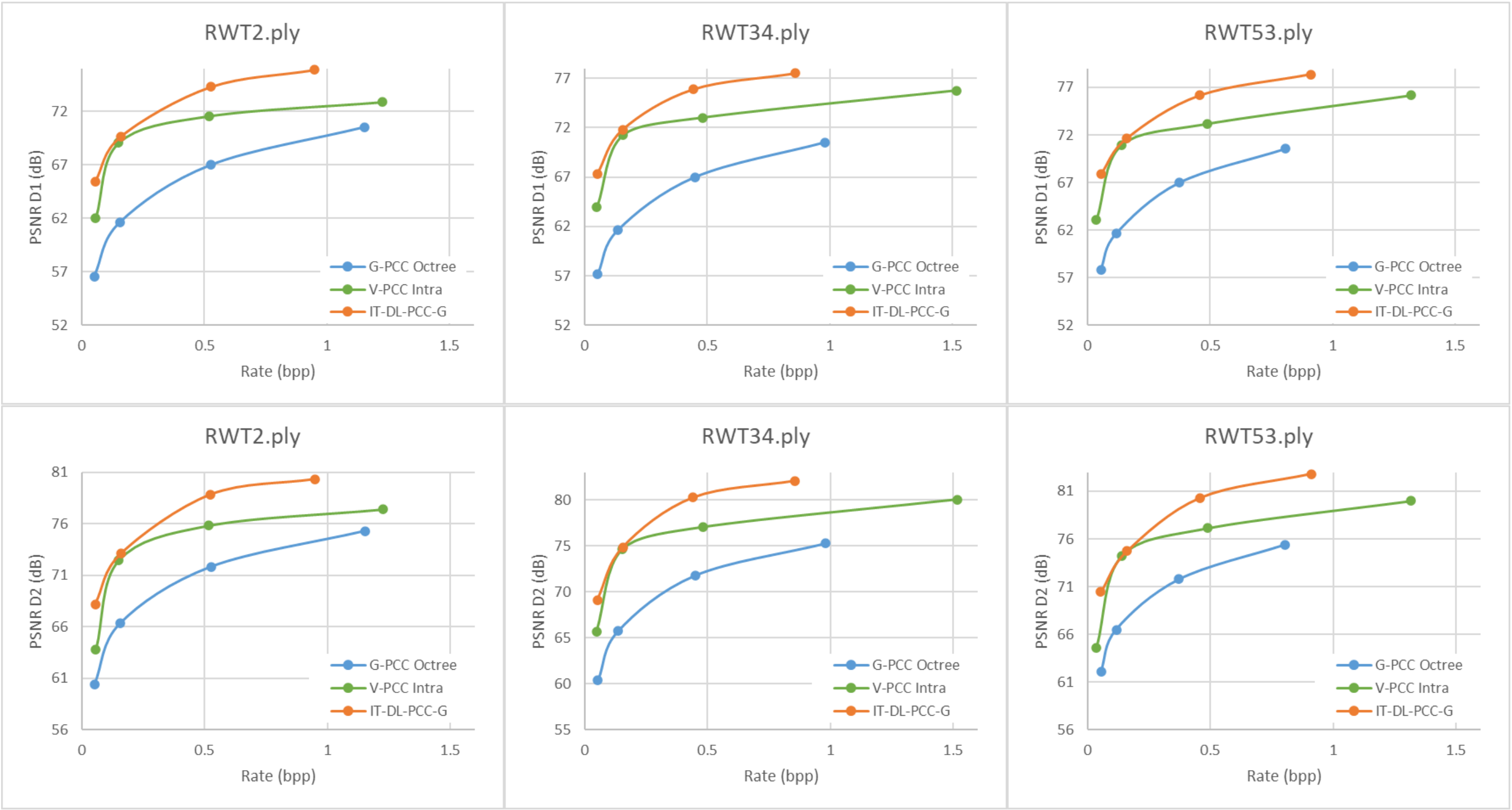

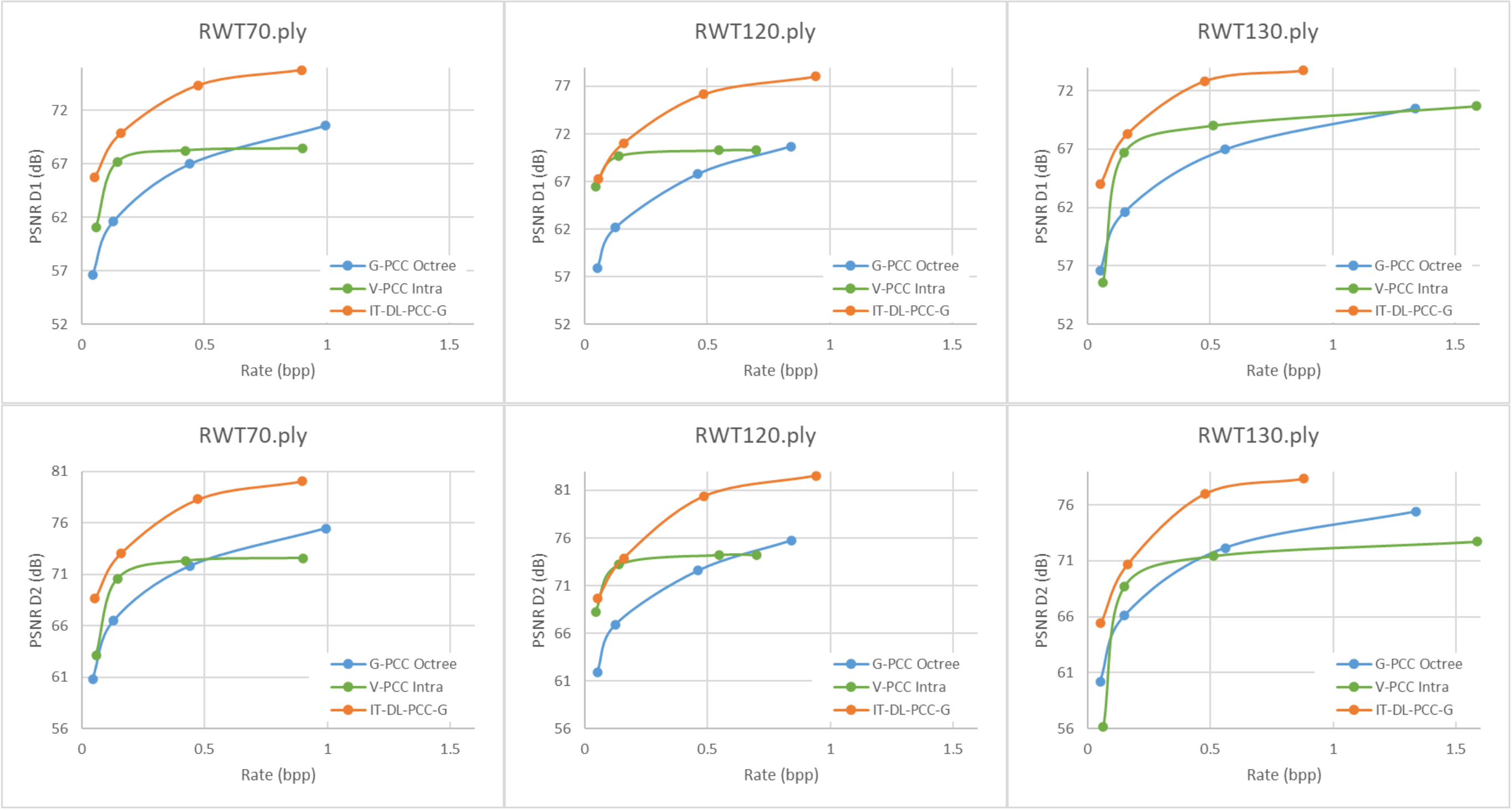

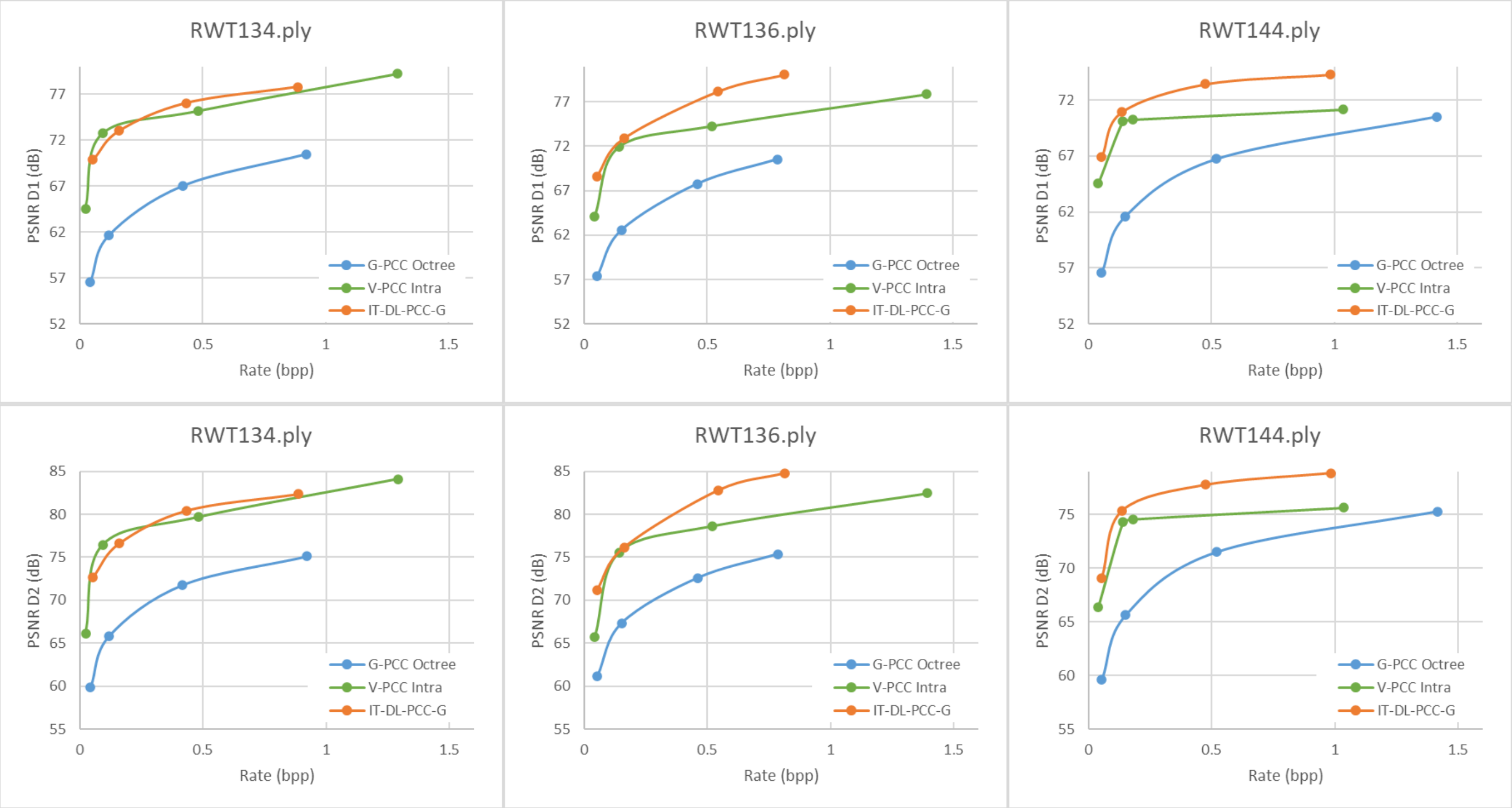

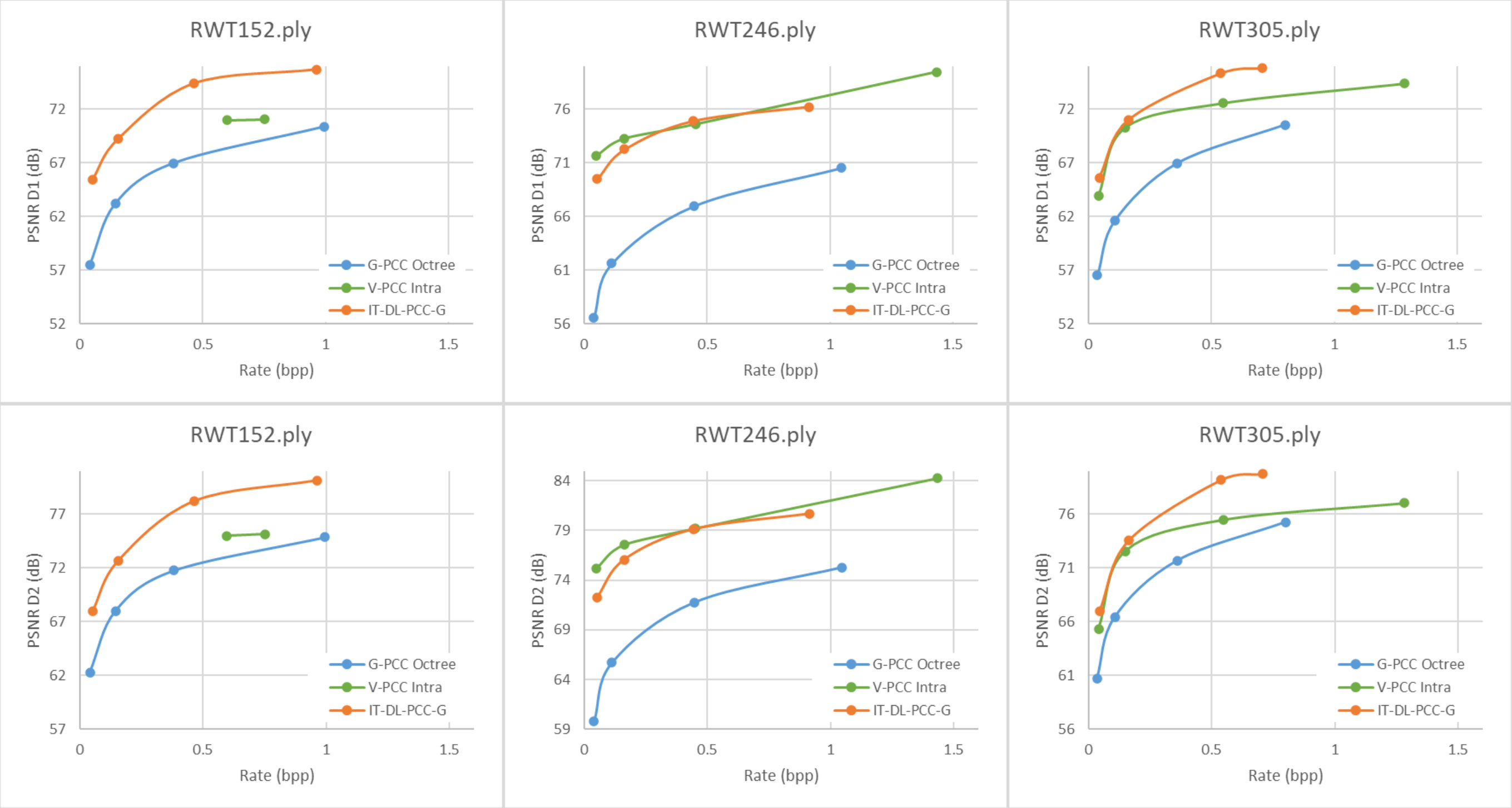

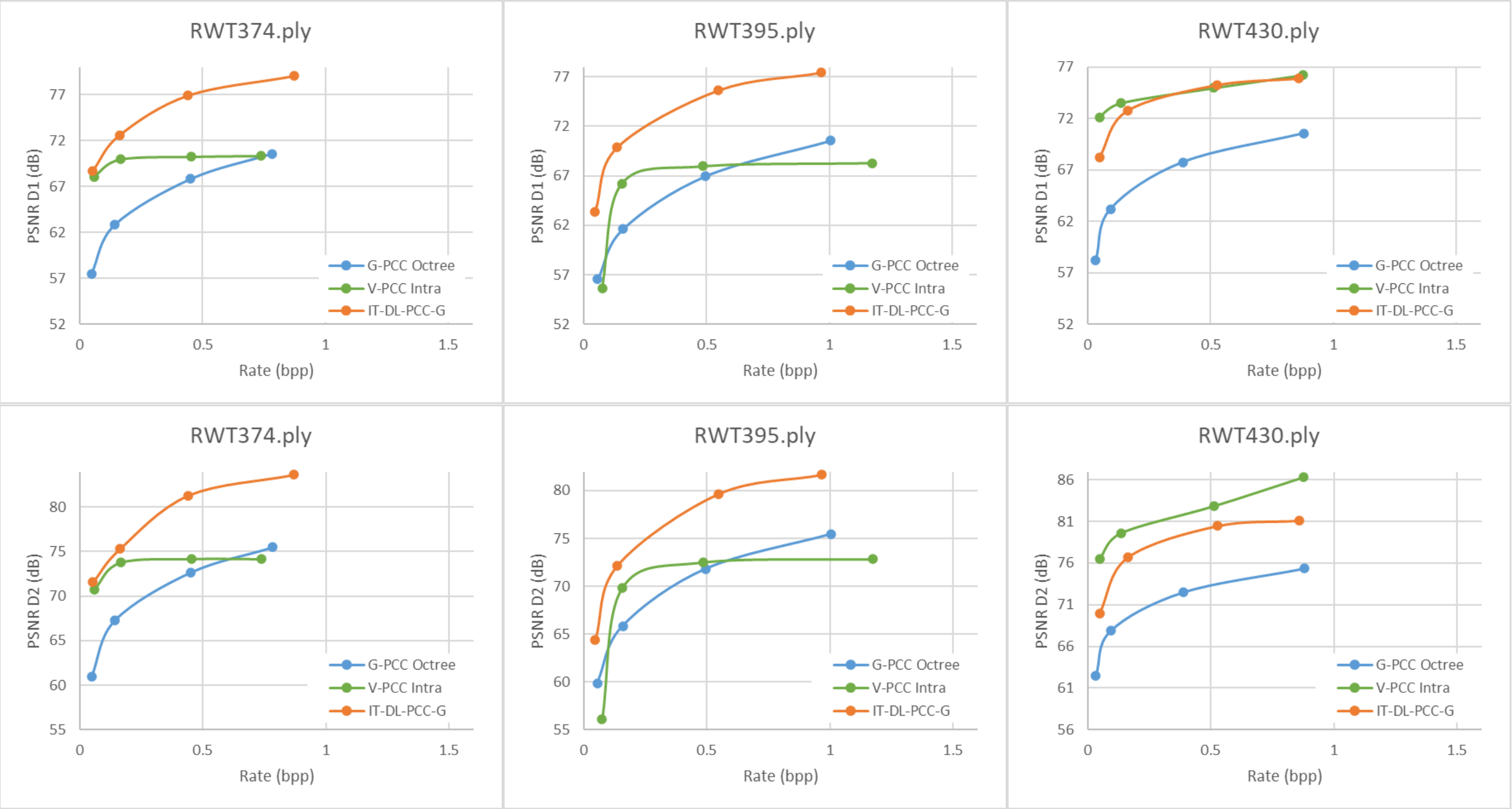

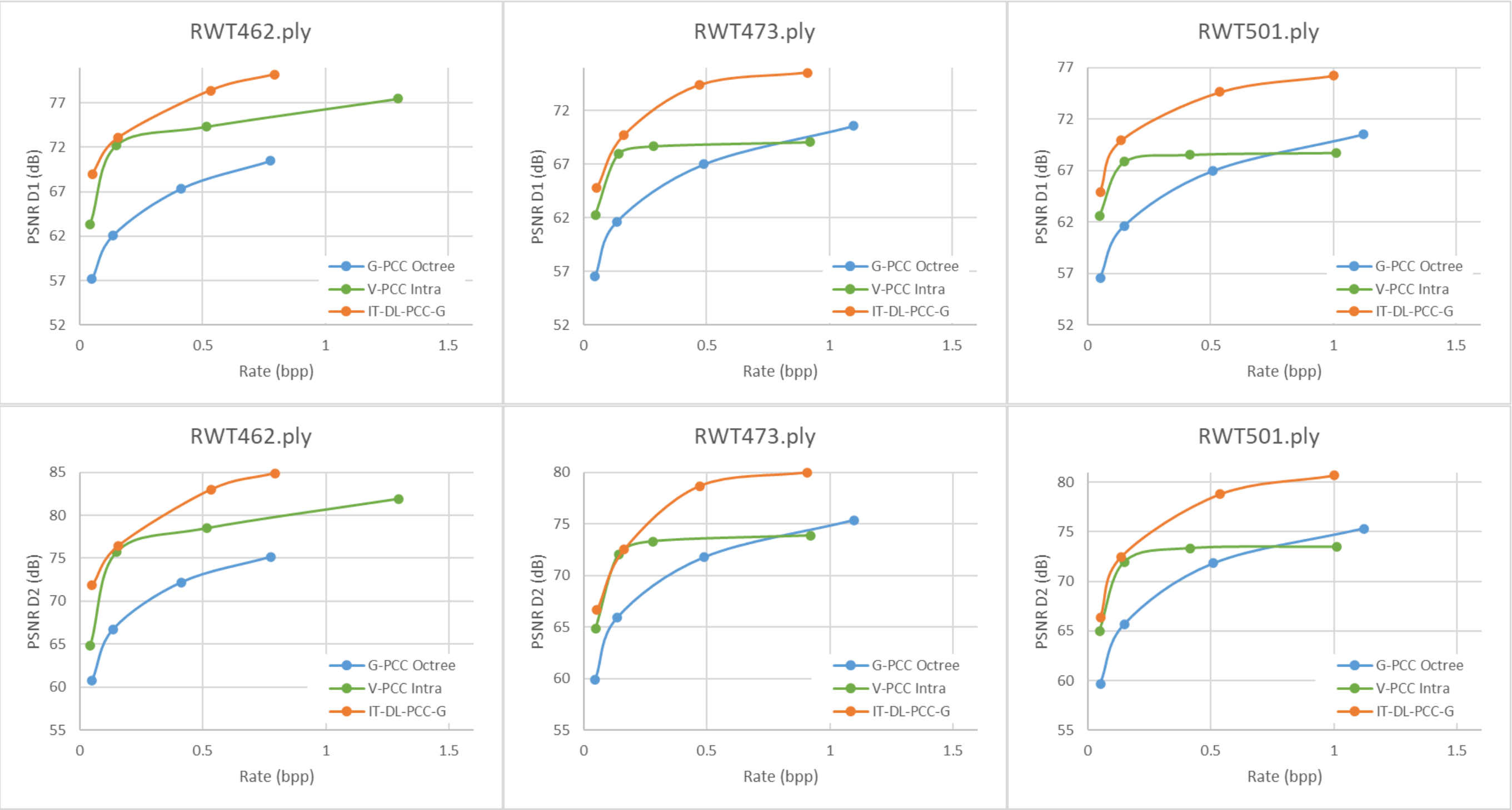

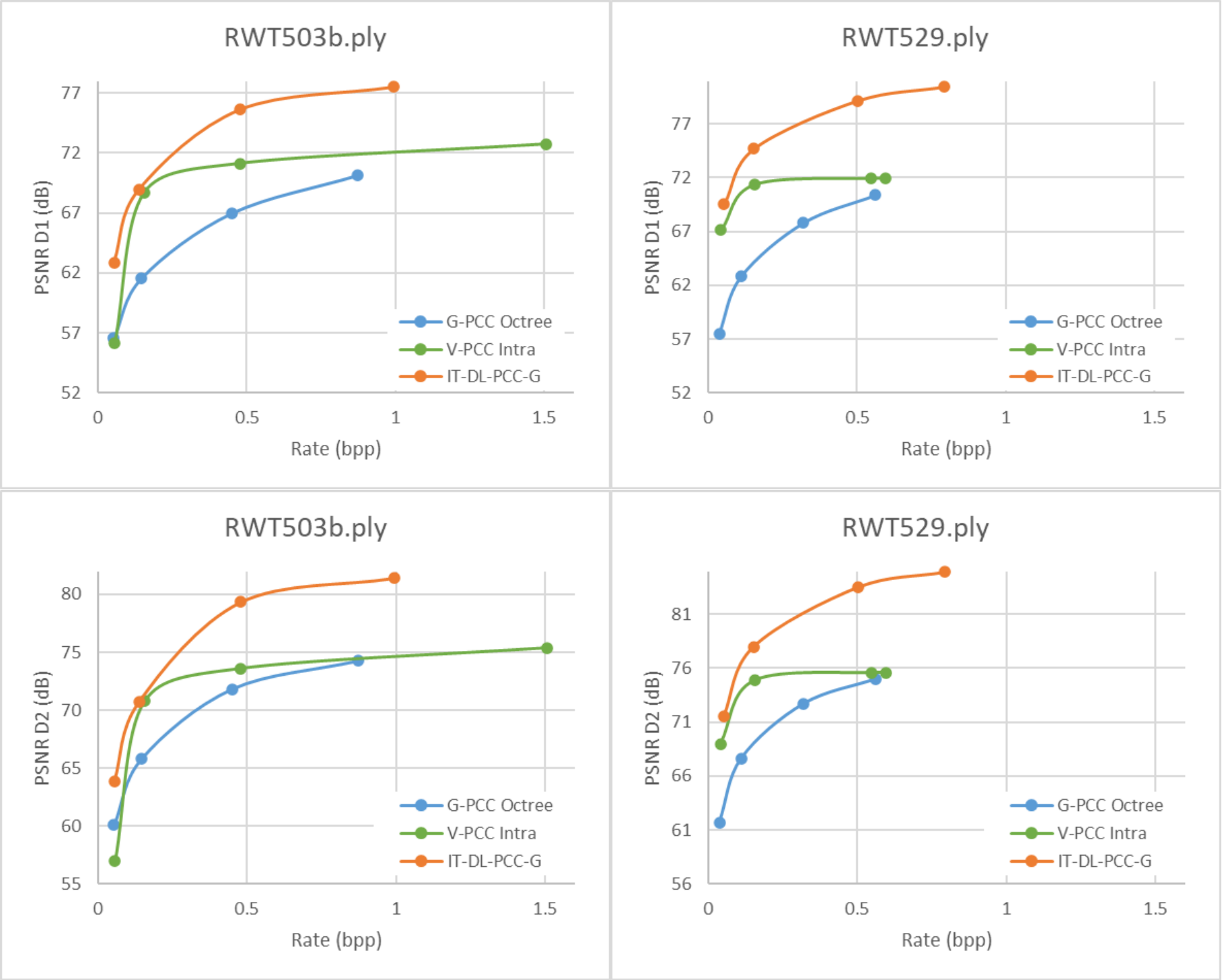

*Fig. 7. RD performance for the proposed IT-DL-PCC-G solution, in comparison with the V-PCC and G-PCC anchors. Results are presented in pairs for each test PC, with PSNR-D1 in the top row and PSNR-D2 in the bottom row.*

*Table IV. BD-PSNR and BD-Rate values for the proposed IT-DL-PCC-G solution over the two anchors.*

| | **G-PCC Octree Reference** | | | | **V-PCC Intra Reference** | | | |
|---|---|---|---|---|---|---|---|---|
| | PSNR-D1 | | PSNR-D2 | | PSNR-D1 | | PSNR-D2 | |
| | BD-Rate | BD-PSNR | BD-Rate | BD-PSNR | BD-Rate | BD-PSNR | BD-Rate | BD-PSNR |
| ***RWT2*** | -84.2% | 7.73 | -77.2% | 6.88 | -30.2% | 1.77 | -28.8% | 2.00 |
| ***RWT34*** | -89.0% | 9.32 | -81.2% | 8.43 | -40.5% | 1.87 | -32.0% | 1.88 |
| ***RWT53*** | -86.7% | 8.62 | -78.4% | 7.26 | -38.0% | 1.68 | -31.0% | 1.70 |
| ***RWT70*** | -82.5% | 7.28 | -73.8% | 5.98 | -55.2% | 4.39 | -49.5% | 4.35 |
| ***RWT120*** | -85.0% | 8.15 | -74.8% | 6.71 | -8.8% | 2.60 | 5.5% | 2.32 |
| ***RWT130*** | -79.8% | 6.48 | -62.9% | 4.79 | -51.7% | 3.38 | -48.5% | 4.44 |
| ***RWT134*** | -93.3% | 9.77 | -89.2% | 9.15 | 0.7% | -0.04 | 11.6% | -0.36 |
| ***RWT136*** | -89.8% | 10.08 | -83.8% | 9.03 | -42.9% | 2.04 | -34.1% | 1.93 |
| ***RWT144*** | -91.6% | 8.43 | -88.3% | 8.35 | -26.7% | 1.64 | -21.3% | 1.65 |
| ***RWT152*** | -78.4% | 6.19 | -67.7% | 4.95 | -66.9% | 4.22 | -63.9% | 4.26 |
| ***RWT246*** | -93.0% | 8.82 | -88.7% | 8.28 | 27.5% | -0.75 | 50.7% | -1.27 |
| ***RWT305*** | -83.0% | 7.38 | -68.3% | 5.38 | -28.1% | 1.24 | -25.2% | 1.59 |
| ***RWT374*** | -89.1% | 9.35 | -81.3% | 8.23 | -44.5% | 3.94 | -21.3% | 3.64 |
| ***RWT395*** | -83.5% | 8.50 | -71.5% | 7.10 | -62.6% | 6.63 | -48.9% | 6.02 |
| ***RWT430*** | -90.4% | 7.23 | -81.8% | 6.55 | 31.8% | -1.01 | 187.7% | -3.70 |
| ***RWT462*** | -90.7% | 10.32 | -85.8% | 9.45 | -45.5% | 2.32 | -41.7% | 2.57 |
| ***RWT473*** | -82.1% | 7.29 | -70.8% | 6.15 | -25.6% | 3.16 | 0.3% | 2.21 |
| ***RWT501*** | -85.8% | 8.07 | -75.4% | 6.94 | -38.1% | 3.90 | -9.4% | 2.84 |
| ***RWT503b*** | -76.9% | 7.73 | -59.5% | 5.89 | -36.2% | 3.39 | -33.2% | 3.93 |
| ***RWT529*** | -89.0% | 10.13 | -81.6% | 8.70 | -44.9% | 4.06 | -24.6% | 4.10 |
| **Average** | -86.2% | 8.34 | -77.1% | 7.21 | -31.3% | 2.52 | -12.9% | 2.31 |

### 4.4 Analysis of the Results

From the full set of performance results included in the previous subsection, it is possible to observe that:

**IT-DL-PCC-G versus G-PCC Octree**

- IT-DL-PCC-G achieves very significant compression gains when compared with G-PCC Octree for all rates.
- Results are consistent for all test PCs, with average rate savings of 86.2% and 77.1% and average BD-PSNR gains of 8.34 dB and 7.21 dB, for PSNR-D1 and PSNR-D2 metrics, respectively.

**IT-DL-PCC-G versus V-PCC Intra**

- IT-DL-PCC-G also outperforms V-PCC Intra for almost all test PCs.
- Results show substantial gains, with average rate savings of 31.3% and 12.9% and average BD-PSNR gains of 2.52 dB and 2.31 dB, for PSNR-D1 and PSNR-D2 metrics, respectively.
- The PCs for which IT-DL-PCC-G does not outperform V-PCC Intra are *RWT134* (CatStatue), *RWT246* (ArmChair) and *RWT430* (WoodenChest). These PCs represent objects with simpler shapes, smooth surfaces, and without obstructions, which V-PCC Intra can code very efficiently.

### 4.5 Complexity Evaluation

Regarding the complexity of the proposed IT-DL-PCC-G solution, the resource footprint and the computational complexity are both reported, following the JPEG CTTC recommendations [17].

To measure the resource footprint, the number of parameters of the DL models are stated, as well as their precision. For each DL coding model, the total number of parameters is 6602976, represented with 32-bit floats. In addition, for each ABU model, the total number of parameters is 7288893, also represented with 32-bit floats.

As for the computational complexity, the encoding and decoding times are reported in Table V and Table VI, with and without GPU. Tests were run on an Intel Core i5-12600k CPU @ 3.70 GHz computer with NVidia GeForce RTX 3080 Ti 12 GB GPU and 32 GB of RAM, running Ubuntu 20.04. **Due to time constraints,** **only one complete run over the entire test dataset was performed (and not the average of 10 runs as requested by the JPEG CTTC).**

*Table V. Real encoding times of the IT-DL-PCC-G codec, with and without GPU, for each target rate and each PC.*

| | CPU | | | | GPU | | | |
|---|---|---|---|---|---|---|---|---|
| | R1 | R2 | R3 | R4 | R1 | R2 | R3 | R4 |
| ***RWT2*** | 10m32.8s | 9m46.9s | 7m52.6s | 7m40.1s | 5m15.9s | 4m2.4s | 3m22.9s | 2m48.1s |
| ***RWT34*** | 9m51.5s | 8m48.8s | 7m34.8s | 6m46.4s | 3m55.7s | 3m39.0s | 3m26.3s | 2m28.2s |
| ***RWT53*** | 5m37.9s | 5m47.9s | 4m29.1s | 4m48.4s | 2m40.8s | 2m53.7s | 2m7.1s | 1m36.1s |
| ***RWT70*** | 8m52.3s | 8m36.2s | 6m59.1s | 5m47.5s | 4m12.6s | 3m26.8s | 3m9.2s | 2m11.6s |
| ***RWT120*** | 5m18.9s | 9m14.1s | 7m26.5s | 7m9.0s | 2m36.1s | 3m58.0s | 3m19.1s | 2m43.7s |
| ***RWT130*** | 13m25.5s | 14m25.9s | 9m16.4s | 9m8.6s | 6m17.9s | 5m52.2s | 5m28.1s | 4m45.8s |
| ***RWT134*** | 11m4.0s | 11m42.5s | 9m0.5s | 9m9.5s | 4m50.3s | 4m43.4s | 4m15.8s | 3m14.2s |
| ***RWT136*** | 3m51.3s | 3m22.1s | 2m13.5s | 2m9.0s | 1m58.5s | 2m13.9s | 2m32.7s | 1m23.4s |
| ***RWT144*** | 19m8.7s | 20m50.2s | 16m50.5s | 15m45.3s | 7m52.4s | 8m16.7s | 6m52.0s | 5m32.1s |
| ***RWT152*** | 16m47.9s | 11m1.8s | 10m14.3s | 10m29.1s | 6m24.8s | 5m47.4s | 5m39.0s | 3m49.7s |
| ***RWT246*** | 16m47.9s | 16m28.7s | 13m48.1s | 12m48.6s | 5m59.9s | 5m48.6s | 4m45.3s | 3m41.7s |
| ***RWT305*** | 5m5.6s | 3m54.5s | 3m37.3s | 3m35.5s | 2m54.5s | 2m6.8s | 1m48.0s | 1m44.5s |
| ***RWT374*** | 7m25.3s | 7m21.5s | 6m8.2s | 5m42.1s | 2m50.9s | 2m45.2s | 2m18.4s | 1m48.3s |
| ***RWT395*** | 2m57.6s | 2m41.7s | 2m11.6s | 2m1.9s | 1m30.1s | 1m14.7s | 1m0.9s | 0m52.3s |
| ***RWT430*** | 9m46.7s | 15m45.1s | 7m31.3s | 11m48.6s | 4m15.6s | 6m17.2s | 2m40.4s | 3m41.4s |
| ***RWT462*** | 2m37.1s | 2m34.8s | 1m56.1s | 1m55.9s | 1m22.9s | 1m21.5s | 0m55.1s | 0m54.3s |
| ***RWT473*** | 12m56.5s | 12m43.1s | 10m14.2s | 9m52.5s | 5m9.6s | 4m37.7s | 3m40.9s | 2m54.6s |
| ***RWT501*** | 11m14.1s | 12m5.9s | 9m59.2s | 9m17.3s | 4m19.6s | 4m22.6s | 3m29.0s | 2m39.8s |
| ***RWT503b*** | 1m59.4s | 1m54.4s | 0m41.1s | 1m25.4s | 0m55.8s | 0m51.1s | 0m22.4s | 0m34.2s |
| ***RWT529*** | 2m29.5s | 0m58.6s | 0m55.9s | 1m50.2s | 1m14.1s | 0m35.1s | 0m32.3s | 0m48.8s |

*Table VI. Real decoding times of the IT-DL-PCC-G codec, with and without GPU, for each target rate and each PC.*

| | CPU | | | | GPU | | | |
|---|---|---|---|---|---|---|---|---|
| | R1 | R2 | R3 | R4 | R1 | R2 | R3 | R4 |
| ***RWT2*** | 7m11.9s | 7m33.3s | 5m51.6s | 5m10.8s | 2m4.5s | 2m9.1s | 1m34.4s | 1m12.7s |
| ***RWT34*** | 6m42.1s | 6m46.4s | 4m57.0s | 4m31.2s | 2m37.9s | 1m47.9s | 1m52.8s | 0m57.7s |
| ***RWT53*** | 4m36.7s | 4m23.6s | 3m23.6s | 3m1.1s | 1m21.1s | 1m49.8s | 1m22.4s | 1m2.8s |
| ***RWT70*** | 6m21.2s | 6m22.7s | 4m40.9s | 4m19.4s | 2m14.5s | 2m33.1s | 1m48.4s | 0m54.8s |
| ***RWT120*** | 3m42.2s | 7m41.4s | 5m42.2s | 5m37.9s | 1m5.2s | 2m53.0s | 1m36.0s | 1m18.2s |
| ***RWT130*** | 8m44.9s | 9m4.5s | 6m49.7s | 6m0.3s | 2m39.7s | 2m16.5s | 2m11.7s | 1m12.1s |
| ***RWT134*** | 9m41.6s | 9m20.0s | 7m2.9s | 6m34.4s | 2m28.1s | 2m44.4s | 1m57.4s | 1m32.1s |
| ***RWT136*** | 2m28.7s | 2m15.5s | 1m40.0s | 1m32.6s | 1m15.3s | 1m2.7s | 0m58.9s | 0m38.8s |
| ***RWT144*** | 14m56.8s | 17m7.0s | 12m35.6s | 11m28.2s | 4m24.2s | 4m54.9s | 3m50.3s | 2m9.5s |
| ***RWT152*** | 10m10.7s | 8m5.8s | 7m56.4s | 7m9.2s | 3m23.6s | 2m41.1s | 2m2.9s | 1m44.9s |
| ***RWT246*** | 14m17.0s | 14m13.3s | 10m59.2s | 9m58.7s | 3m45.8s | 3m38.7s | 2m57.0s | 1m50.7s |
| ***RWT305*** | 3m15.7s | 2m28.5s | 2m17.1s | 2m16.2s | 1m3.4s | 0m51.1s | 0m36.9s | 0m35.9s |
| ***RWT374*** | 6m19.5s | 6m16.5s | 4m49.7s | 4m23.4s | 1m47.8s | 1m44.7s | 1m24.5s | 0m55.5s |
| ***RWT395*** | 2m14.9s | 2m14.8s | 1m42.3s | 1m32.0s | 0m49.1s | 0m49.1s | 0m38.8s | 0m28.8s |
| ***RWT430*** | 7m26.3s | 12m45.0s | 5m25.3s | 8m54.3s | 1m56.8s | 3m23.3s | 1m4.9s | 1m41.9s |
| ***RWT462*** | 2m1.6s | 2m1.8s | 1m21.4s | 1m20.9s | 0m48.2s | 0m47.5s | 0m27.2s | 0m26.8s |
| ***RWT473*** | 10m27.1s | 10m21.9s | 8m0.3s | 7m21.6s | 2m48.9s | 2m43.4s | 2m13.6s | 1m24.7s |
| ***RWT501*** | 9m29.6s | 10m27.1s | 7m58.8s | 7m16.7s | 2m35.4s | 2m49.1s | 2m11.6s | 1m23.9s |
| ***RWT503b*** | 1m42.2s | 1m42.2s | 0m30.6s | 1m8.7s | 0m39.8s | 0m39.7s | 0m14.0s | 0m23.6s |
| ***RWT529*** | 1m57.8s | 0m36.8s | 0m34.3s | 1m20.0s | 0m43.5s | 0m15.9s | 0m13.2s | 0m25.6s |

## 5. JPEG Pleno PCC Requirements Checking

Considering the JPEG Pleno PCC requirements [18], the proposed IT-DL-PCC-G solution offers the following properties:

- **Compression Efficiency**: IT-DL-PCC-G offers a considerably better RD performance than the G-PCC Octree and V-PCC Intra standards adopted as anchors.
- **Random Access**: In IT-DL-PCC-G, the PC is divided into blocks which are coded independently. Since the positions of the coded blocks are transmitted to the decoder, this allows the user to choose decoding only selected regions of the PC, thus providing spatial random access. The granularity of the random access may be controlled by selecting the block size for coding.

# PART II – IT Deep Learning-based Point Cloud Codec for Geometry and Colour (IT-DL-PCC-GC)

The second part of this document describes the second proposed codec, IT-DL-PCC-GC, notably targeting joint geometry and colour coding.

## 6. Architecture and High-level Description

The proposed IT-DL-PCC-GC codec offers a joint geometry and colour coding approach, in which the same DL model processes geometry and colour simultaneously. For this purpose, the overall architecture is the same as for the IT-DL-PCC-G geometry-only codec, shown in Fig. 1, with only a few changes to allow extending it for joint coding.

The main differences can be summarized as follows:

- **Input PC data**: The input PC contains now not only the geometry but also the colour attributes, typically in RGB colour space. As such, when dividing the PC into 3D blocks, each voxel consists of four channels (instead of a single one as for geometry-only), with one channel containing the binary geometry information, i.e., whether the voxel is occupied or not, and the remaining three channels containing the 8-bit RGB colour information. In order for geometry and colour values to be in the same range, the RGB colour values were scaled from the original [0, 255] range to [0, 1]. For empty voxels, the colour values are assumed to be zero.

- **DL models**: Regarding the DL models, for both the DL coding model and the ABU model, the only change is on the last convolutional layer in the decoder and in the expanding path, respectively, where four filters are used instead of a single filter, matching the number of channels of the input 3D block. Everything else remains the same, including the number of filters and the size of the latent representation. With this small change, the DL coding model goes from 6602976 (in IT-DL-PCC-G) to 6605664 trainable parameters (in IT-DL-PCC-GC), whereas the ABU model goes from 7288893 (in IT-DL-PCC-G) to 7291488 trainable parameters (in IT-DL-PCC-GC).

- **Binarization optimization**: Finally, the last difference is related to the binarization optimization performed at the encoder, in which the binarization parameters are now optimized considering not only just a geometry quality metric but also a colour quality metric, such as PSNR-YUV or PSNR-RGB.

## 7. DL Model Training

This section describes the important training processes for the DL coding and ABU models for joint geometry and colour coding/processing.

The training process and dataset were the same as described in Section 3 for the proposed IT-DL-PCC-G geometry-only codec. The only difference regards the training loss function, which naturally needs to incorporate the colour rate and distortion components.

### 7.1 Coding Model Training

For the coding model, the loss function uses the same Lagrangian formulation presented in Equation (2). Since the latent representation now contains both the compressed geometry and colour representation without separation between them, only the total (geometry + colour) rate is considered. On the other hand, the distortion is measured separately, with the total distortion being given by:

$$Total\ Distortion = (1-\omega) \times Distortion_{Geometry} + (\omega) \times Distortion_{Colour}\ , \qquad (4)$$

where $\omega$ is a weight that leverages the importance of colour over the geometry. Although intensive studies have been performed on the best weight, tests have demonstrated that $\omega = 0.5$ provides a good balance in terms of final RD performance.

As previously stated, the distortion function must be differentiable and thus the distortion metrics used for testing (D1, D2, MSE-Y, MSE-RGB) cannot be used. The geometry distortion ($Distortion_{Geometry}$) was once again defined as the average Focal Loss over the entire block, using the same parameters as for the IT-DL-PCC-G codec. As for the colour distortion ($Distortion_{Colour}$), a voxel-wise mean squared error (MSE) was defined, as follows:

$$Distortion_{Colour} = \frac{1}{N_{input}} \sum_{i \in N_{input}} \frac{(R_i - R_i')^2 + (G_i - G_i')^2 + (B_i - B_i')^2}{3}\ , \qquad (5)$$

where $R_i, G_i, B_i$ are the colour values of the occupied voxel *i* in the input block, $R_i', G_i', B_i'$ are the colour values of the collocated voxel in the decoded block, and $N_{input}$ is the number of occupied voxels in the input block.

### 7.2 ABU Model Training

Regarding the ABU model, once again the training followed the same procedure as for the IT-DL-PCC-G codec, with the distortion corresponding now to the total distortion as defined in Equation (4), using the same weights as for the DL coding model.

## 8. Performance Assessment for Geometry and Colour Codec

To assess the performance of the proposed joint geometry-colour coding solution, IT-DL-PCC-GC, the JPEG test PCs were coded following the JPEG CTTC [17]. To reach the target rates, the coding configurations for each test PC were determined similarly to the geometry-only IT-DL-PCC-G solution.

### 8.1 Coding Configurations

Using the same approach as for the IT-DL-PCC-G solution, the coding configurations determined for the IT-DL-PCC-GC solution to reach the target rates defined in the JPEG CTTC (0.1, 0.3, 1, 3 bpp) are shown in

Table VII.

*Table VII. IT-DL-PCC-GC coding configurations to achieve the JPEG target rates.*

| PC | R1 | R2 | R3 | R4 |
|---|---|---|---|---|
| ***RWT2*** | bs=128 sf=2 abu=1 qs=1 λ=0.005 | bs=128 sf=2 abu=1 qs=1.25 λ=0.0005 | bs=256 sf=1 abu=0 qs=1 λ=0.001 | bs=256 sf=1 abu=0 qs=1 λ=0.00025 |
| ***RWT34*** | bs=128 sf=2 abu=1 qs=1 λ=0.005 | bs=128 sf=2 abu=1 qs=1.25 λ=0.0005 | bs=256 sf=1 abu=0 qs=1 λ=0.001 | bs=256 sf=1 abu=0 qs=1 λ=0.00025 |
| ***RWT53*** | bs=128 sf=2 abu=1 qs=1.25 λ=0.005 | bs=128 sf=2 abu=1 qs=1 λ=0.001 | bs=128 sf=1 abu=0 qs=1 λ=0.001 | bs=256 sf=1 abu=0 qs=1 λ=0.00025 |
| ***RWT70*** | bs=128 sf=2 abu=1 qs=1 λ=0.01 | bs=128 sf=2 abu=1 qs=1.1 λ=0.001 | bs=256 sf=1 abu=0 qs=1.3 λ=0.001 | bs=256 sf=1 abu=0 qs=1 λ=0.00025 |
| ***RWT120*** | bs=128 sf=2 abu=1 qs=1.25 λ=0.005 | bs=128 sf=2 abu=1 qs=1 λ=0.001 | bs=128 sf=1 abu=0 qs=1 λ=0.001 | bs=256 sf=1 abu=0 qs=1 λ=0.00025 |
| ***RWT130*** | bs=128 sf=2 abu=1 qs=1.1 λ=0.01 | bs=128 sf=2 abu=1 qs=1.25 λ=0.001 | bs=256 sf=1 abu=0 qs=1 λ=0.0025 | bs=256 sf=1 abu=0 qs=1 λ=0.00025 |
| ***RWT134*** | bs=128 sf=2 abu=1 qs=1.25 λ=0.005 | bs=128 sf=2 abu=1 qs=1.1 λ=0.001 | bs=256 sf=1 abu=0 qs=1.3 λ=0.001 | bs=256 sf=1 abu=0 qs=1 λ=0.00025 |
| ***RWT136*** | bs=128 sf=2 abu=1 qs=1 λ=0.005 | bs=128 sf=2 abu=1 qs=1.25 λ=0.0005 | bs=256 sf=1 abu=0 qs=1 λ=0.001 | bs=256 sf=1 abu=0 qs=1 λ=0.00025 |
| ***RWT144*** | bs=64 sf=4 abu=1 qs=1.1 λ=0.001 | bs=128 sf=2 abu=1 qs=1 λ=0.0025 | bs=256 sf=1 abu=0 qs=1.45 λ=0.001 | bs=256 sf=1 abu=0 qs=1 λ=0.00025 |
| ***RWT152*** | bs=128 sf=2 abu=1 qs=1.25 λ=0.005 | bs=128 sf=2 abu=1 qs=1 λ=0.001 | bs=256 sf=1 abu=0 qs=1 λ=0.001 | bs=256 sf=1 abu=0 qs=1 λ=0.00025 |
| ***RWT246*** | bs=128 sf=2 abu=1 qs=1.25 λ=0.005 | bs=128 sf=2 abu=1 qs=1 λ=0.001 | bs=256 sf=1 abu=0 qs=1.3 λ=0.001 | bs=256 sf=1 abu=0 qs=1 λ=0.00025 |
| ***RWT305*** | bs=128 sf=2 abu=1 qs=1.25 λ=0.005 | bs=128 sf=2 abu=1 qs=1.25 λ=0.0005 | bs=256 sf=1 abu=0 qs=1.3 λ=0.001 | bs=256 sf=1 abu=0 qs=1 λ=0.00025 |
| ***RWT374*** | bs=128 sf=2 abu=1 qs=1 λ=0.005 | bs=128 sf=2 abu=1 qs=1 λ=0.0005 | bs=128 sf=1 abu=0 qs=1 λ=0.0005 | bs=256 sf=1 abu=0 qs=1 λ=0.00025 |
| ***RWT395*** | bs=128 sf=2 abu=1 qs=1.3 λ=0.005 | bs=128 sf=2 abu=1 qs=1 λ=0.001 | bs=256 sf=1 abu=0 qs=1.3 λ=0.001 | bs=256 sf=1 abu=0 qs=1 λ=0.00025 |
| ***RWT430*** | bs=128 sf=2 abu=1 qs=1 λ=0.005 | bs=128 sf=2 abu=1 qs=1 λ=0.001 | bs=128 sf=1 abu=0 qs=1 λ=0.001 | bs=256 sf=1 abu=0 qs=1 λ=0.00025 |
| ***RWT462*** | bs=128 sf=2 abu=1 qs=1 λ=0.005 | bs=128 sf=2 abu=1 qs=1 λ=0.0005 | bs=256 sf=1 abu=0 qs=1.3 λ=0.0005 | bs=256 sf=1 abu=0 qs=1 λ=0.00025 |
| ***RWT473*** | bs=128 sf=2 abu=1 qs=1.4 λ=0.005 | bs=128 sf=2 abu=1 qs=1.1 λ=0.001 | bs=256 sf=1 abu=0 qs=1.3 λ=0.001 | bs=256 sf=1 abu=0 qs=1 λ=0.00025 |
| ***RWT501*** | bs=128 sf=2 abu=1 qs=1 λ=0.005 | bs=128 sf=2 abu=1 qs=1.25 λ=0.0005 | bs=256 sf=1 abu=0 qs=1 λ=0.001 | bs=256 sf=1 abu=0 qs=1 λ=0.00025 |
| ***RWT503b*** | bs=128 sf=2 abu=1 qs=1 λ=0.01 | bs=128 sf=2 abu=1 qs=1 λ=0.001 | bs=128 sf=1 abu=0 qs=1.1 λ=0.001 | bs=256 sf=1 abu=0 qs=1 λ=0.00025 |
| ***RWT529*** | bs=128 sf=2 abu=1 qs=1 λ=0.005 | bs=128 sf=2 abu=1 qs=1.25 λ=0.0005 | bs=256 sf=1 abu=0 qs=1 λ=0.001 | bs=256 sf=1 abu=0 qs=1 λ=0.00025 |

## 8.2 RD Performance Results: RD Charts and BD-PSNR

The RD performance results for IT-DL-PCC-GC are plotted in RD charts and compared with the G-PCC Octree and V-PCC Intra anchors in Fig. 8, for both the PSNR-D1 and PSNR-D2 geometry objective quality metrics, for both the PSNR-Y and PSNR-YUV colour objective quality metrics, and for the PCQM joint objective quality metric. For the *RWT152* PC, V-PCC Intra anchor results were only provided for the last two target rates.

To summarize the results, Table VIII, Table IX, and Table X show the Bjontegaard-Delta PSNR gains (BD-PSNR) and Bjontegaard-Delta rate savings (BD-Rate) for the proposed IT-DL-PCC-GC solution over the anchors G-PCC Octree and V-PCC Intra, for each set of objective quality metrics.

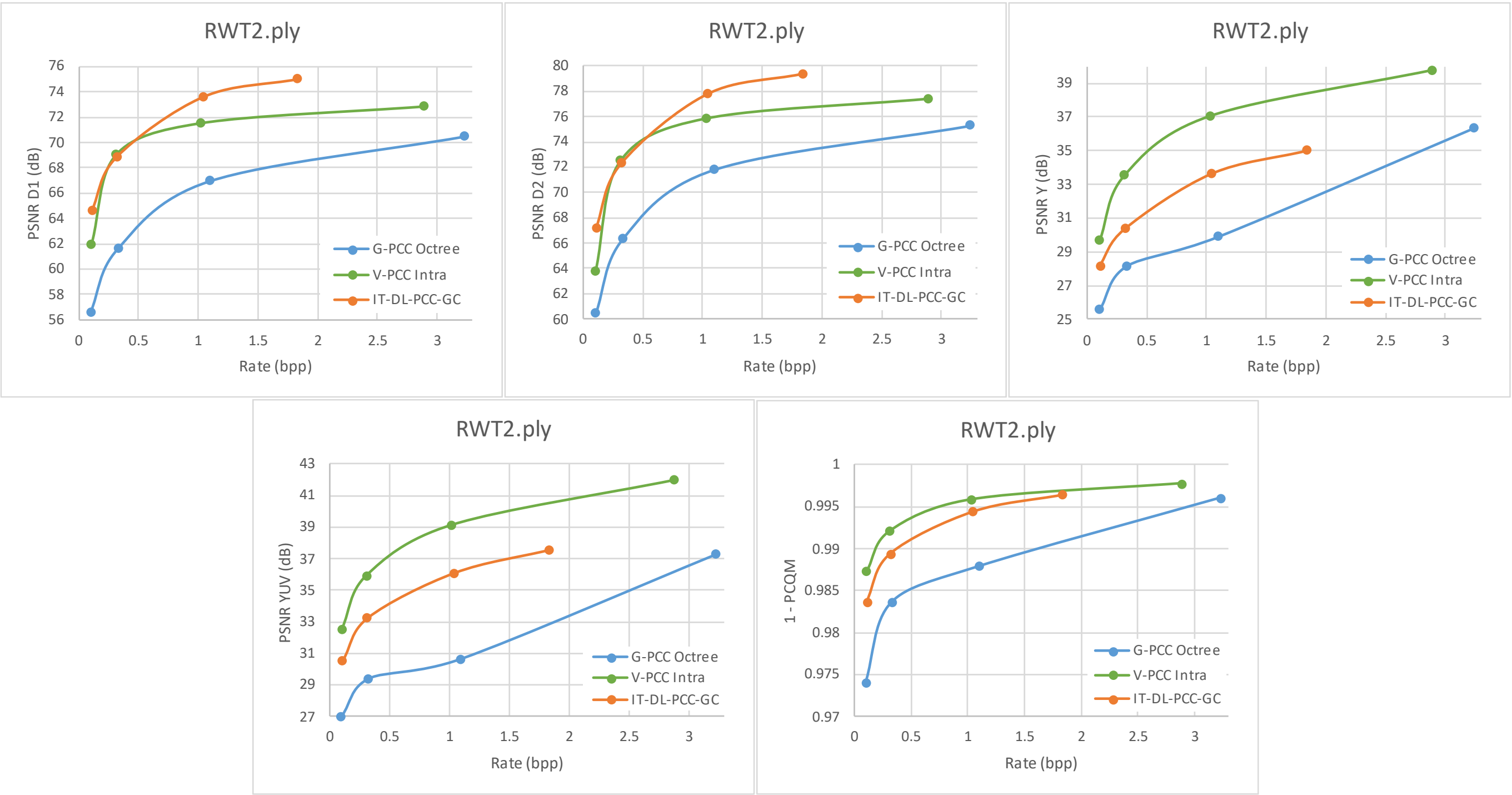

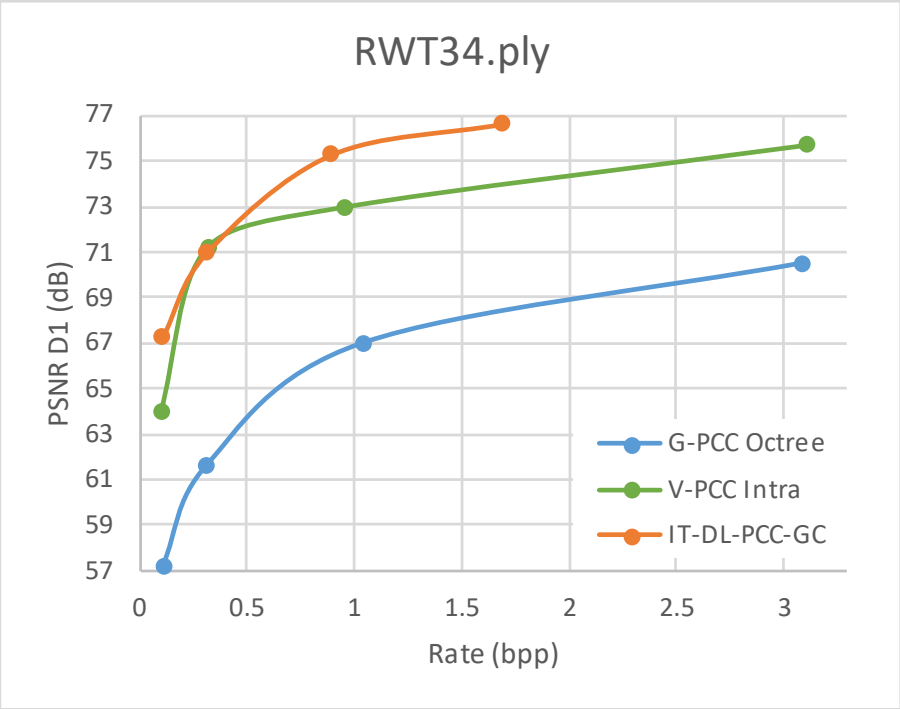

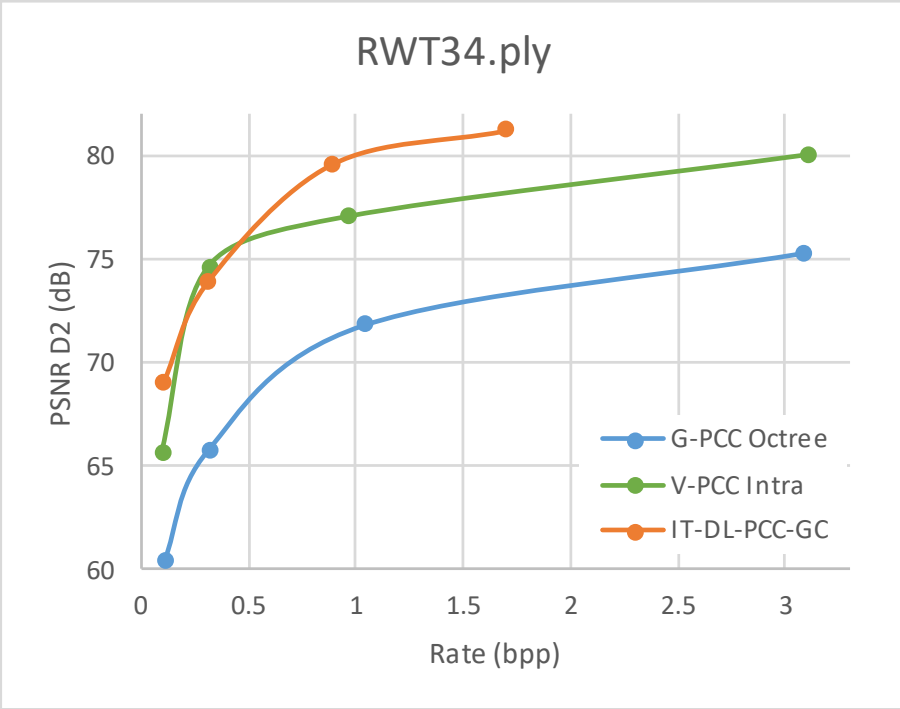

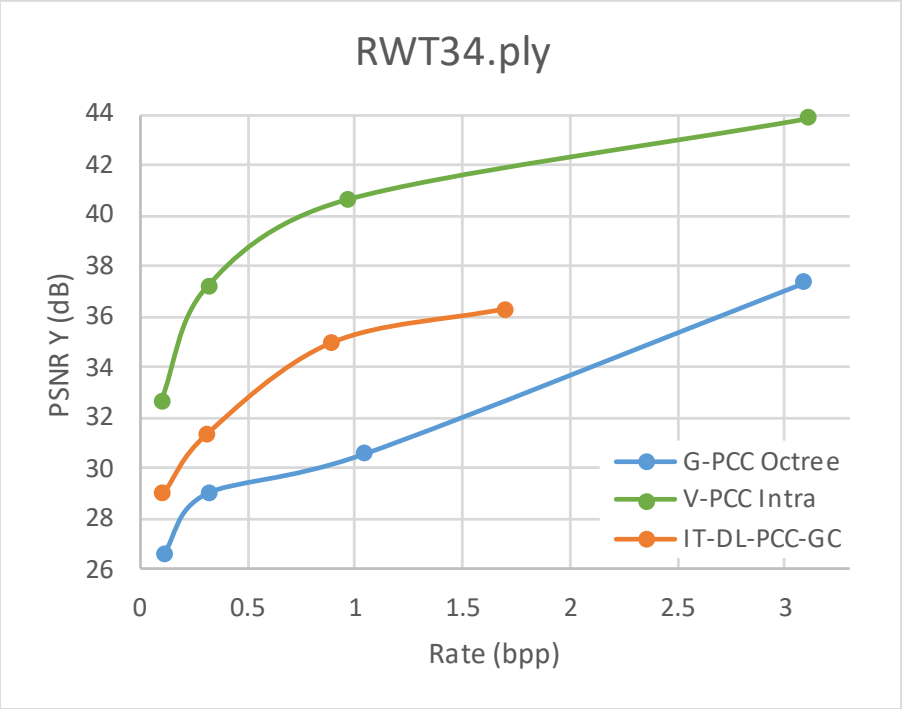

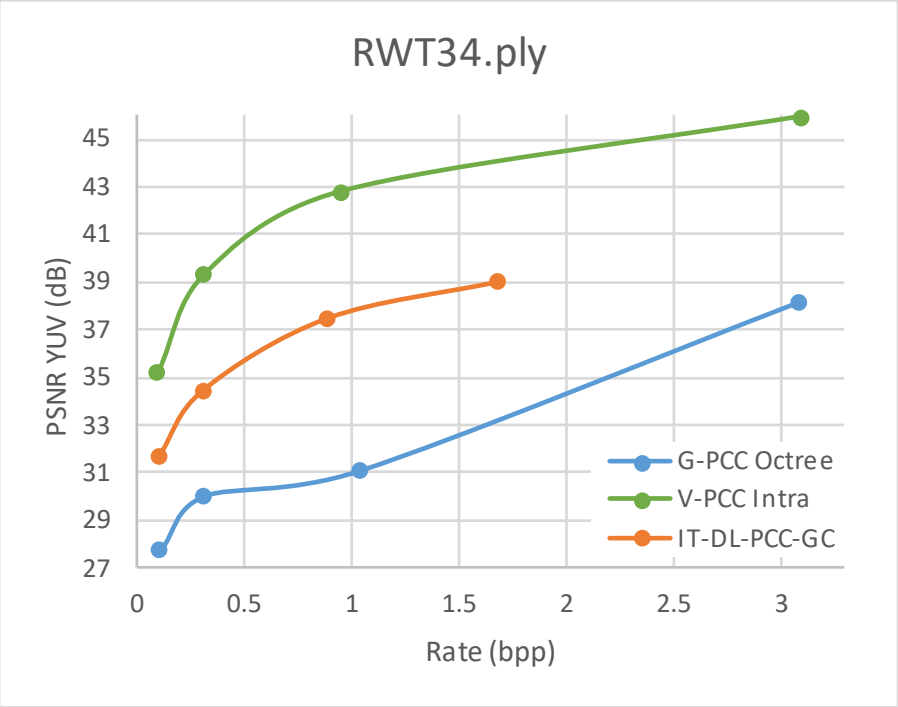

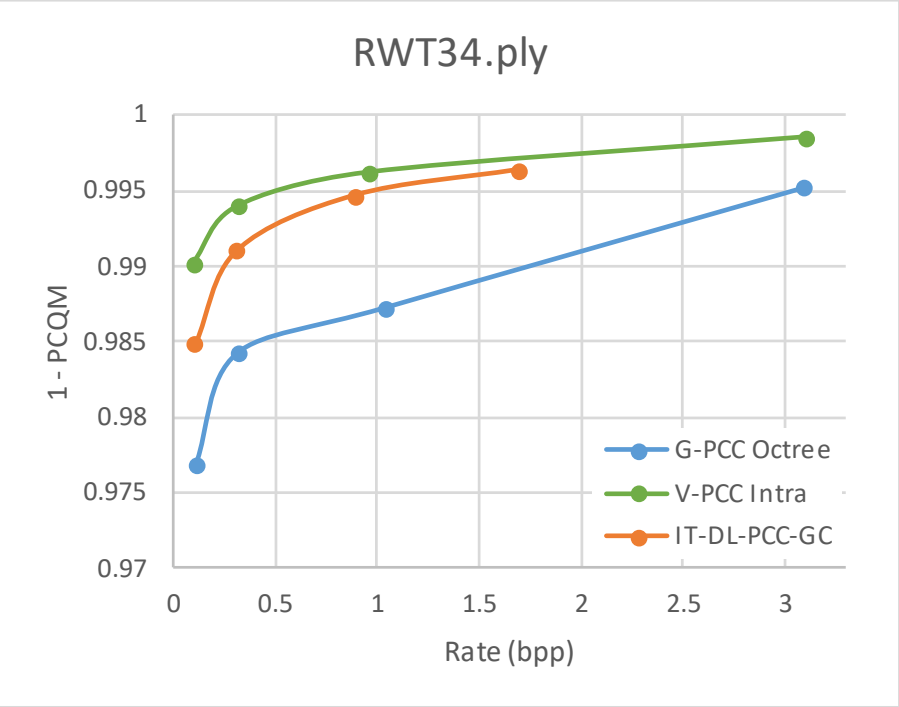

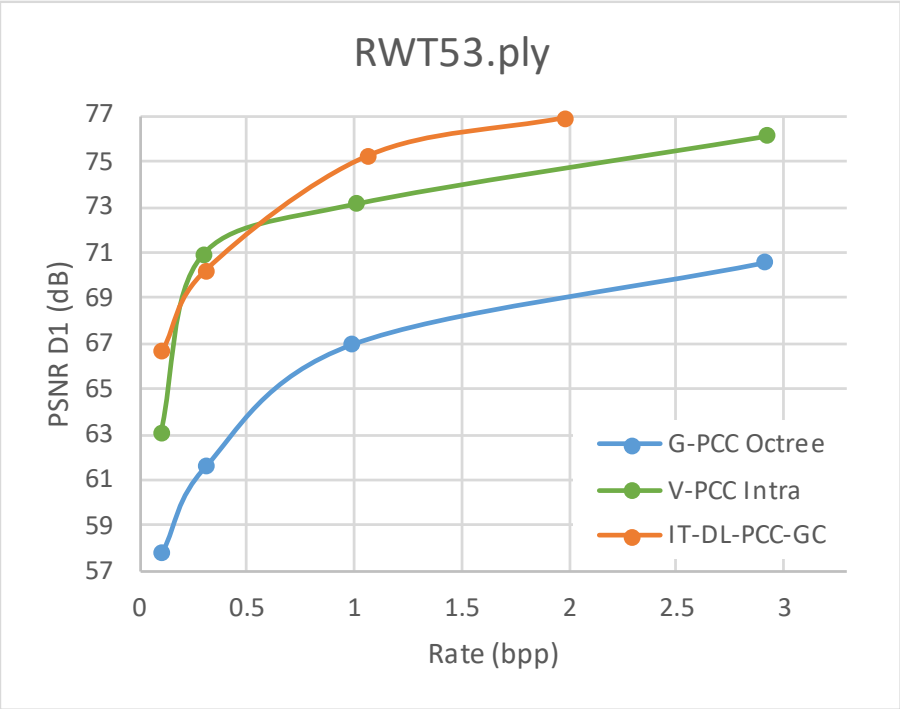

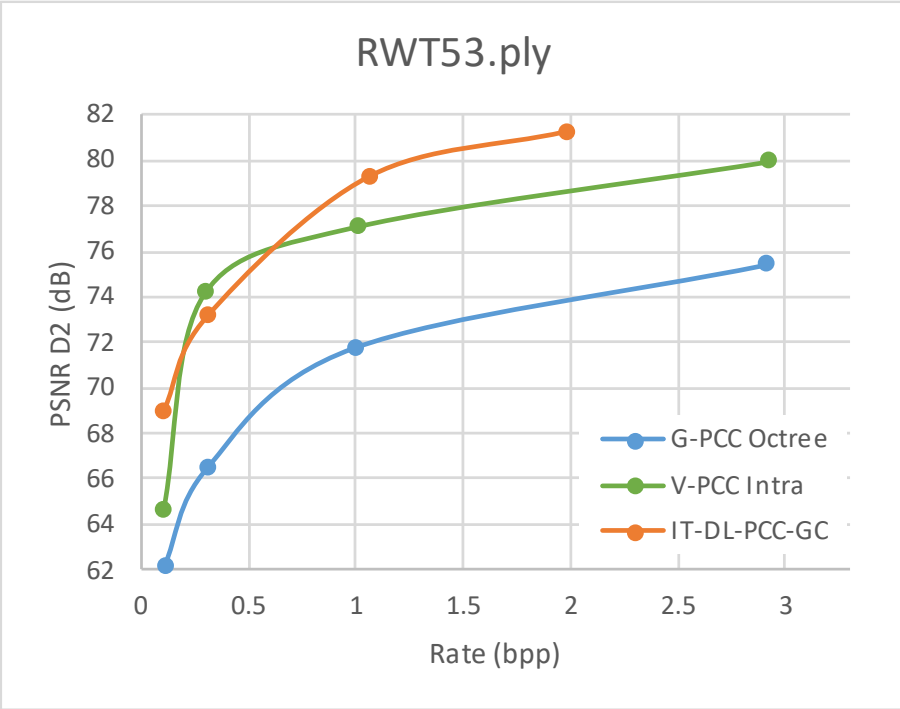

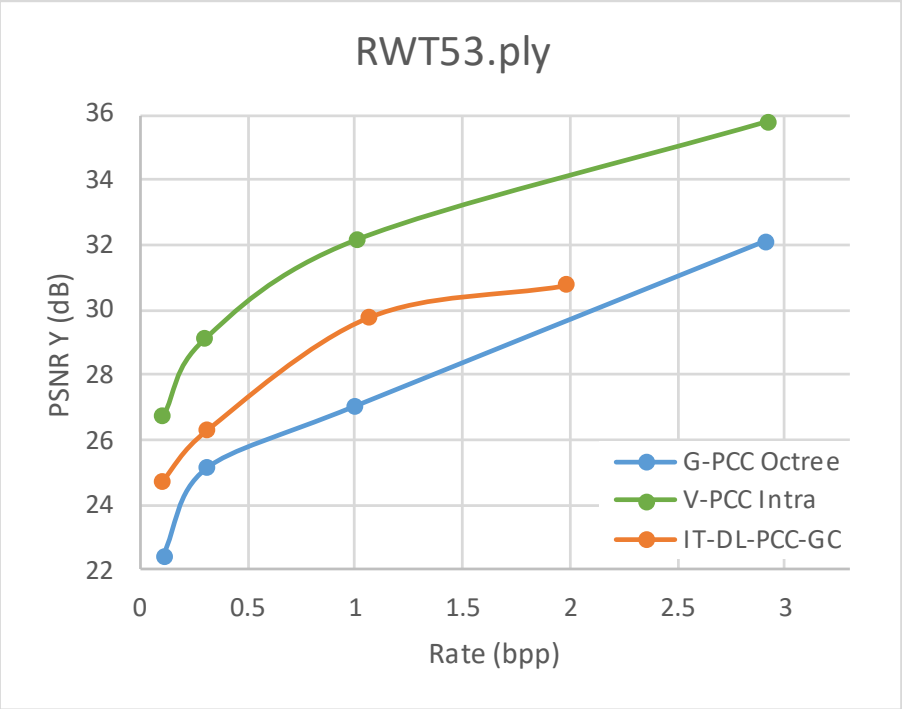

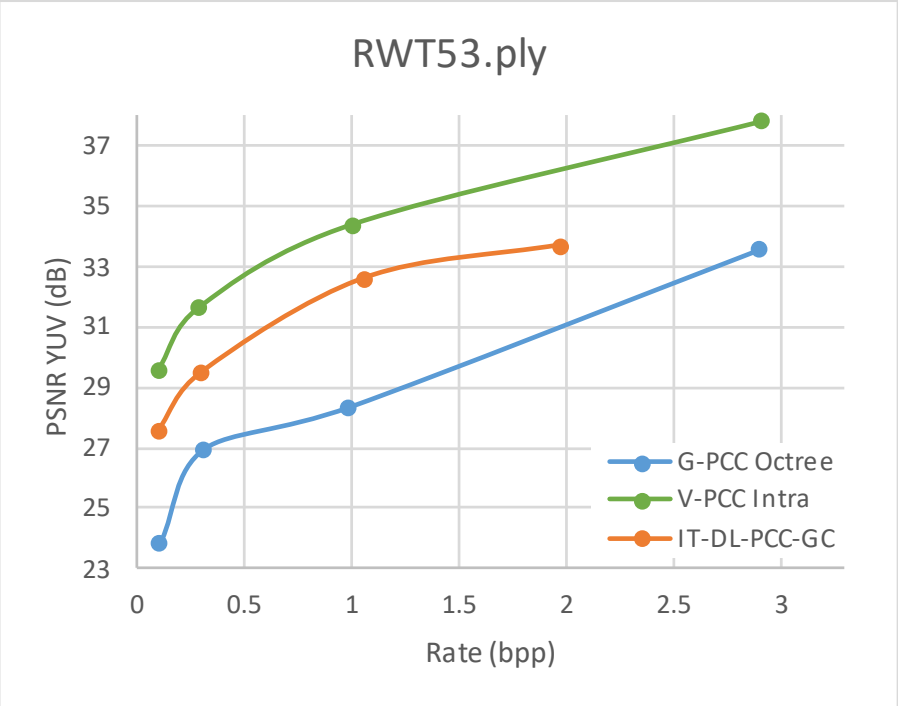

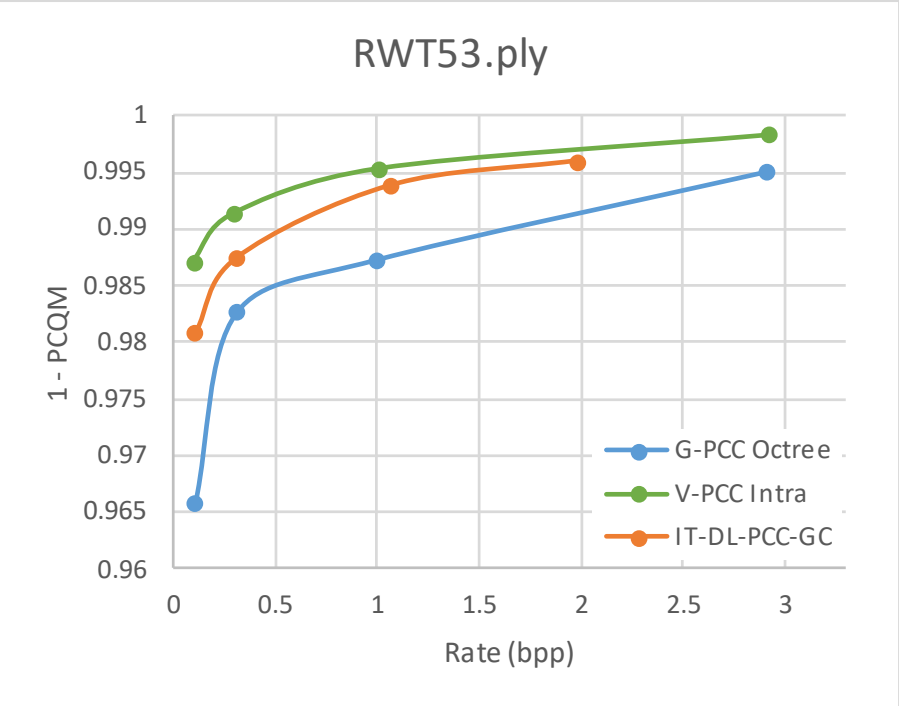

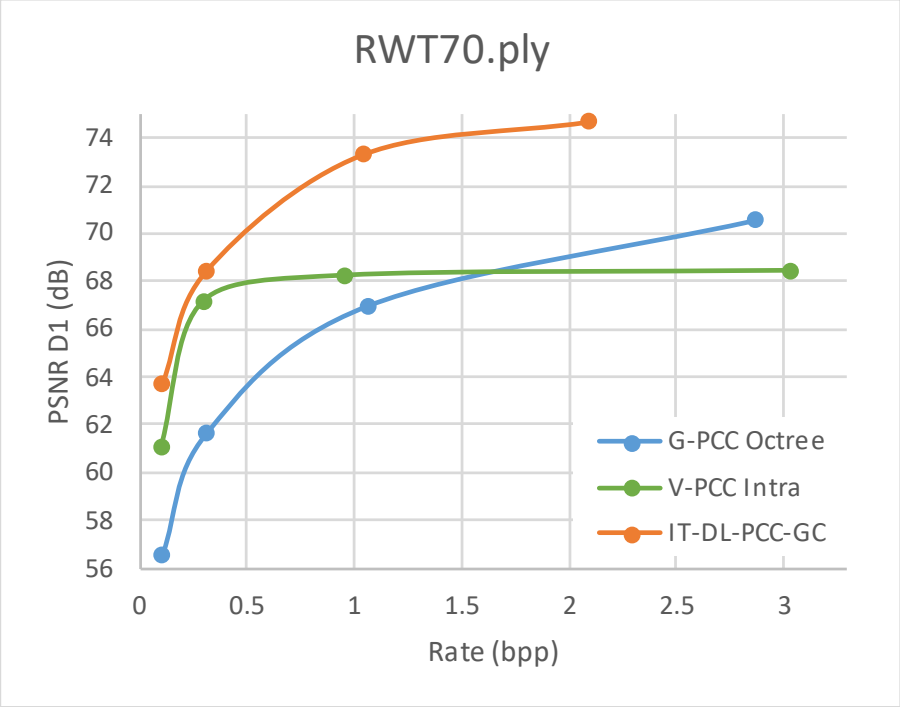
RWT70.ply
PSNR D1 (dB)
56 58 60 62 64 66 68 70 72 74
0 0.5 1 1.5 2 2.5 3
Rate (bpp)
G-PCC Octree
V-PCC Intra
IT-DL-PCC-GC

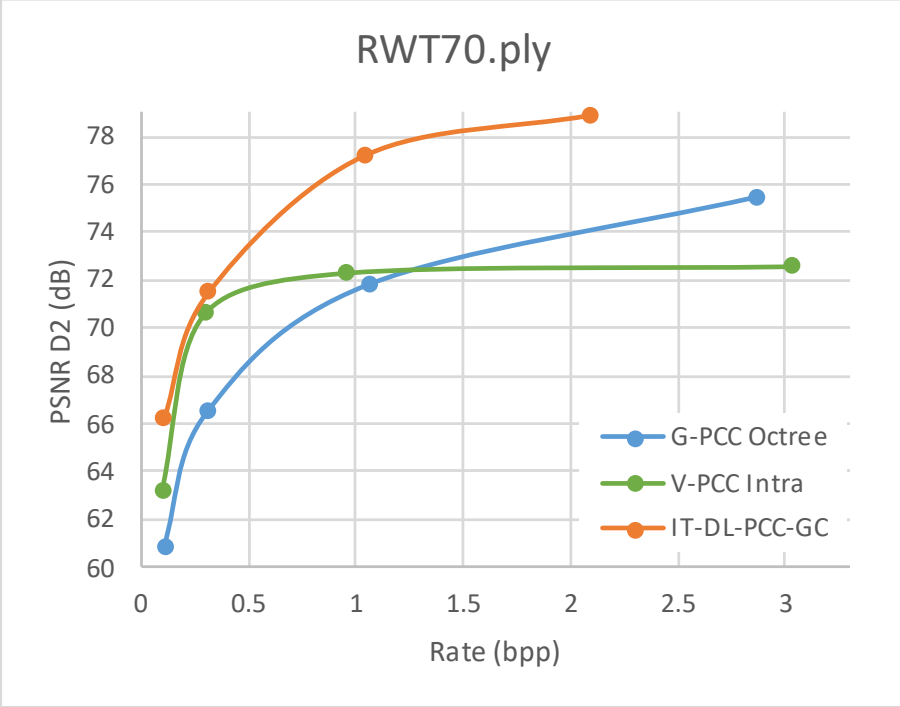
RWT70.ply
PSNR D2 (dB)
60 62 64 66 68 70 72 74 76 78
0 0.5 1 1.5 2 2.5 3
Rate (bpp)
G-PCC Octree
V-PCC Intra
IT-DL-PCC-GC

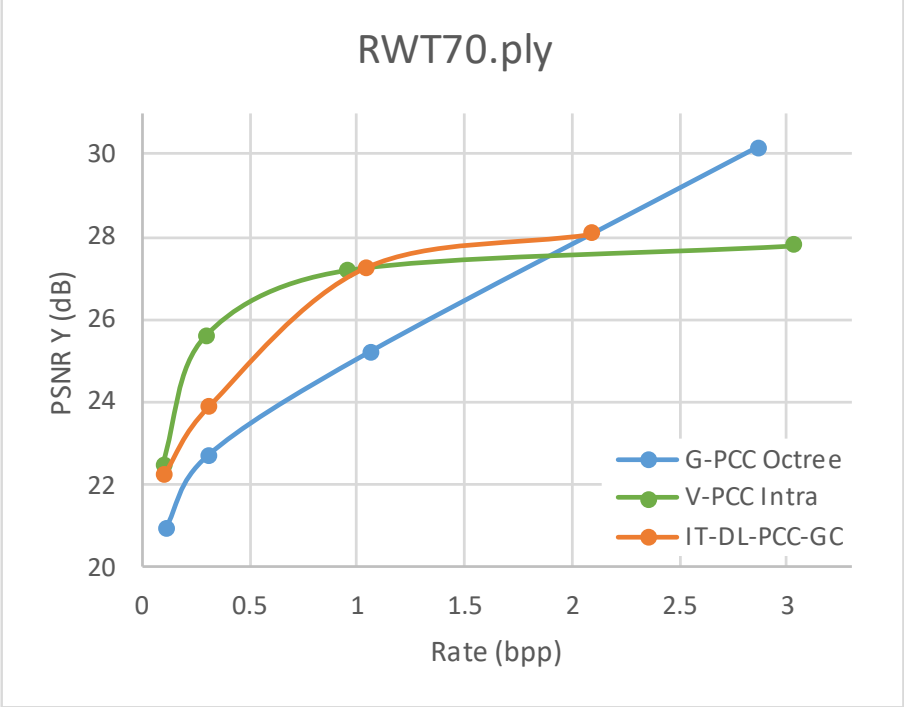
RWT70.ply
PSNR Y (dB)
20 22 24 26 28 30
0 0.5 1 1.5 2 2.5 3
Rate (bpp)
G-PCC Octree
V-PCC Intra
IT-DL-PCC-GC

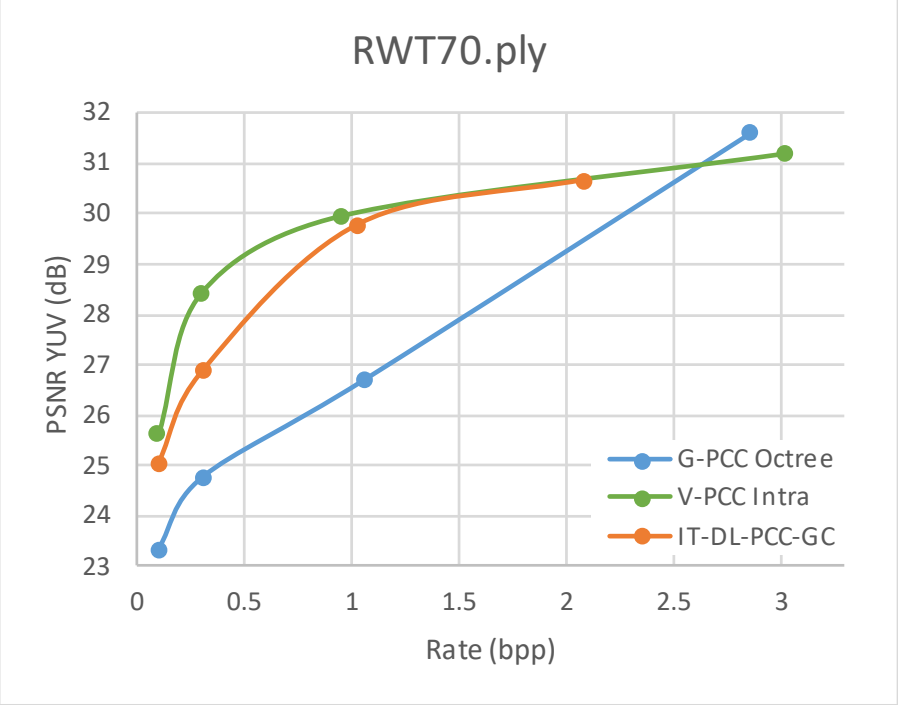
RWT70.ply
PSNR YUV (dB)
23 24 25 26 27 28 29 30 31 32
0 0.5 1 1.5 2 2.5 3
Rate (bpp)
G-PCC Octree
V-PCC Intra
IT-DL-PCC-GC

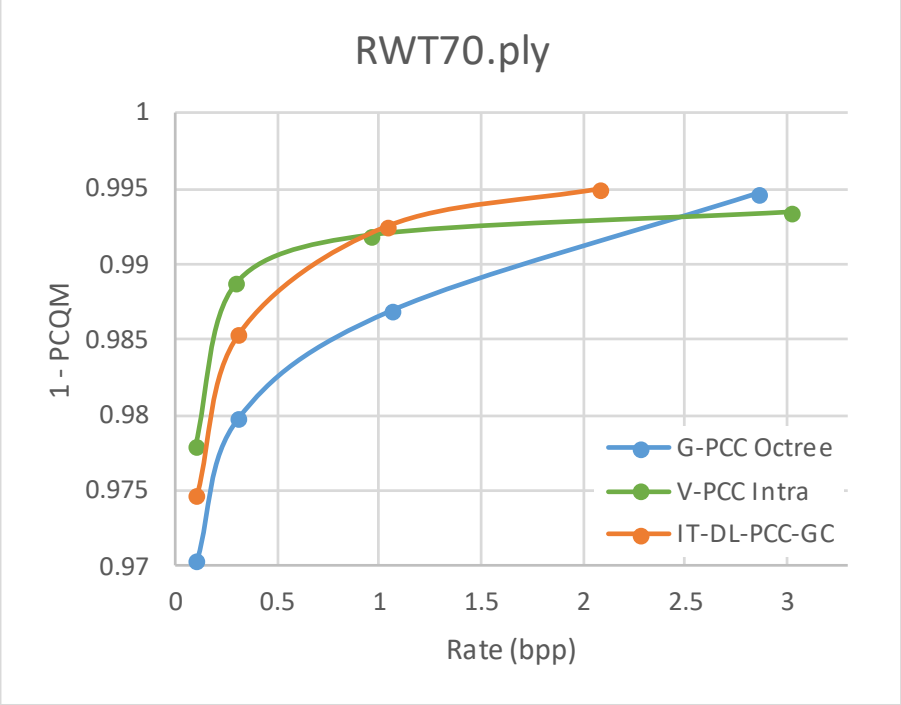
RWT70.ply
1 - PCQM
0.97 0.975 0.98 0.985 0.99 0.995 1
0 0.5 1 1.5 2 2.5 3
Rate (bpp)
G-PCC Octree
V-PCC Intra
IT-DL-PCC-GC

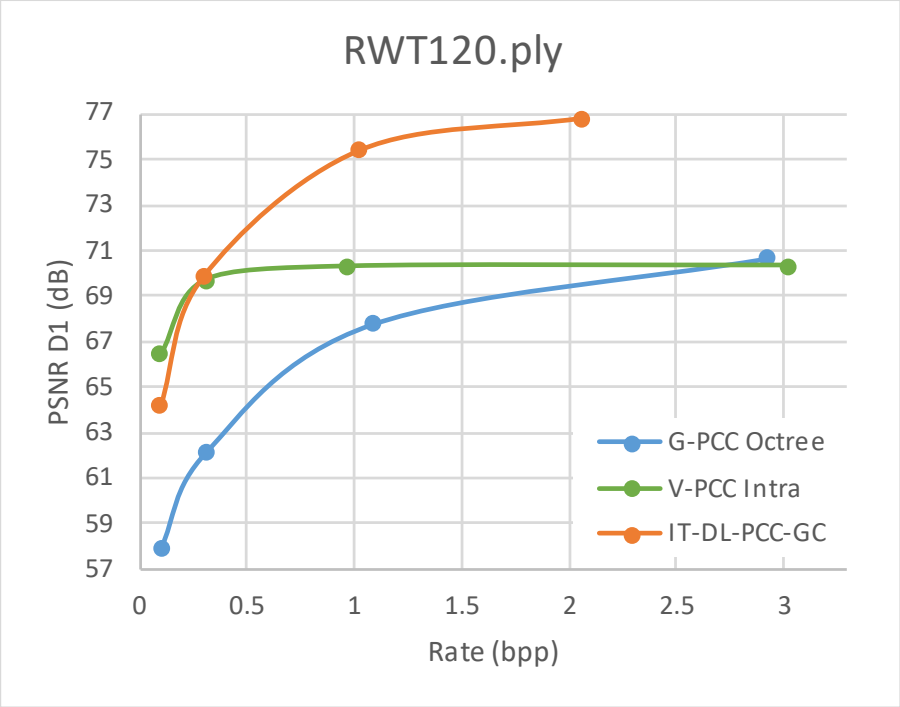
RWT120.ply
PSNR D1 (dB)
57 59 61 63 65 67 69 71 73 75 77
0 0.5 1 1.5 2 2.5 3
Rate (bpp)
G-PCC Octree
V-PCC Intra
IT-DL-PCC-GC

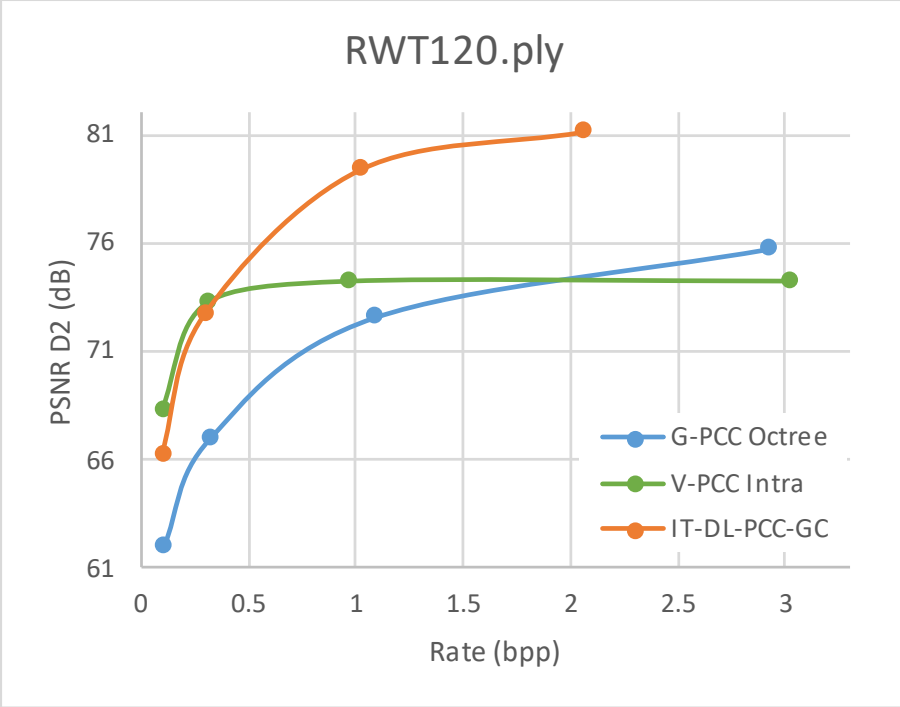
RWT120.ply
PSNR D2 (dB)
61 66 71 76 81
0 0.5 1 1.5 2 2.5 3
Rate (bpp)
G-PCC Octree
V-PCC Intra
IT-DL-PCC-GC

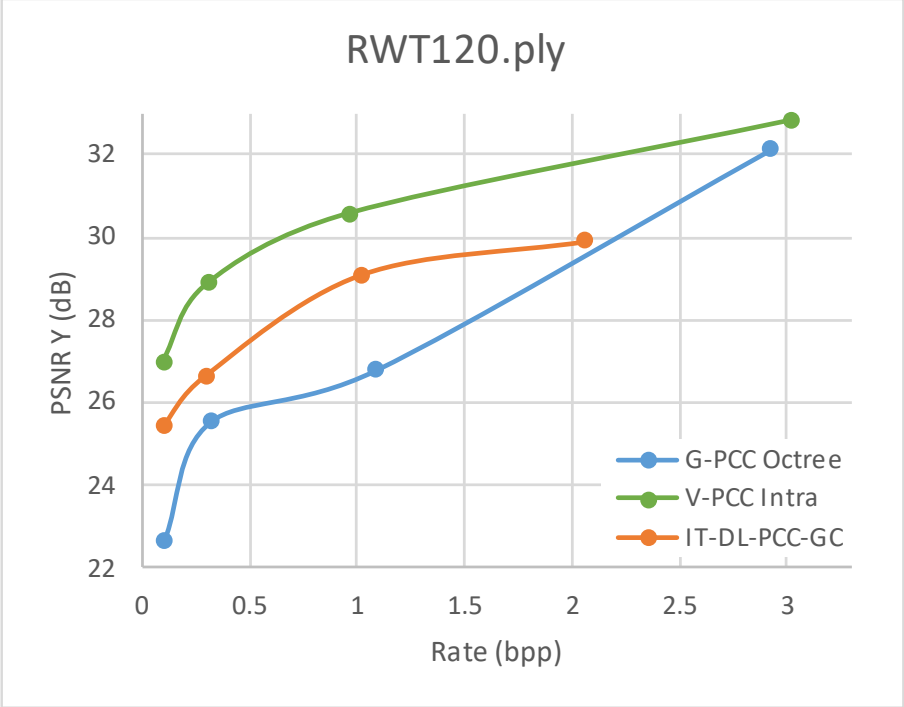
RWT120.ply
PSNR Y (dB)
22 24 26 28 30 32
0 0.5 1 1.5 2 2.5 3
Rate (bpp)
G-PCC Octree
V-PCC Intra
IT-DL-PCC-GC

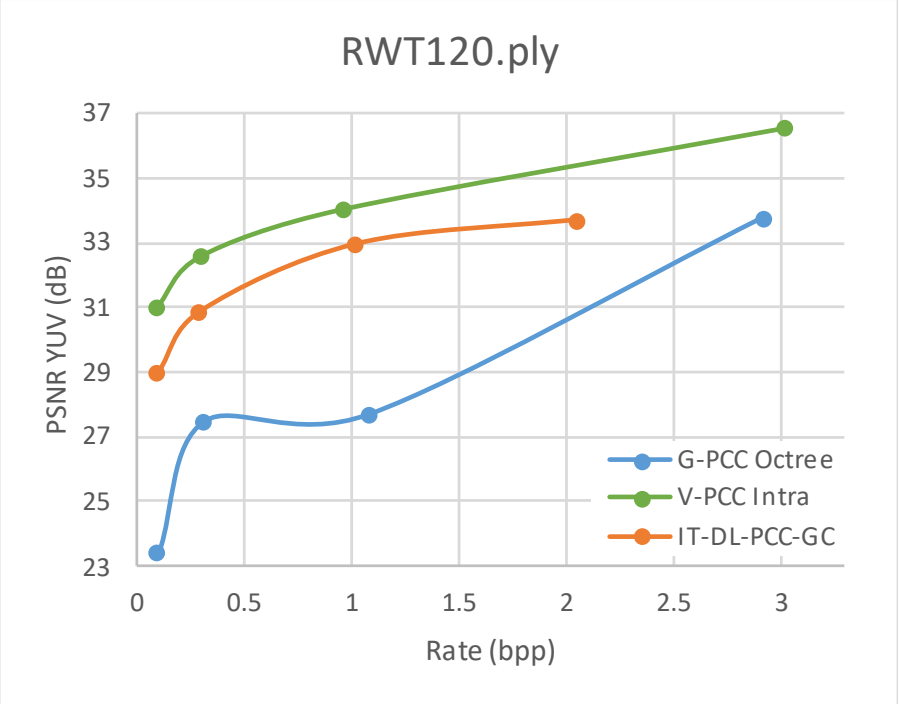
RWT120.ply
PSNR YUV (dB)
23 25 27 29 31 33 35 37
0 0.5 1 1.5 2 2.5 3
Rate (bpp)
G-PCC Octree
V-PCC Intra
IT-DL-PCC-GC

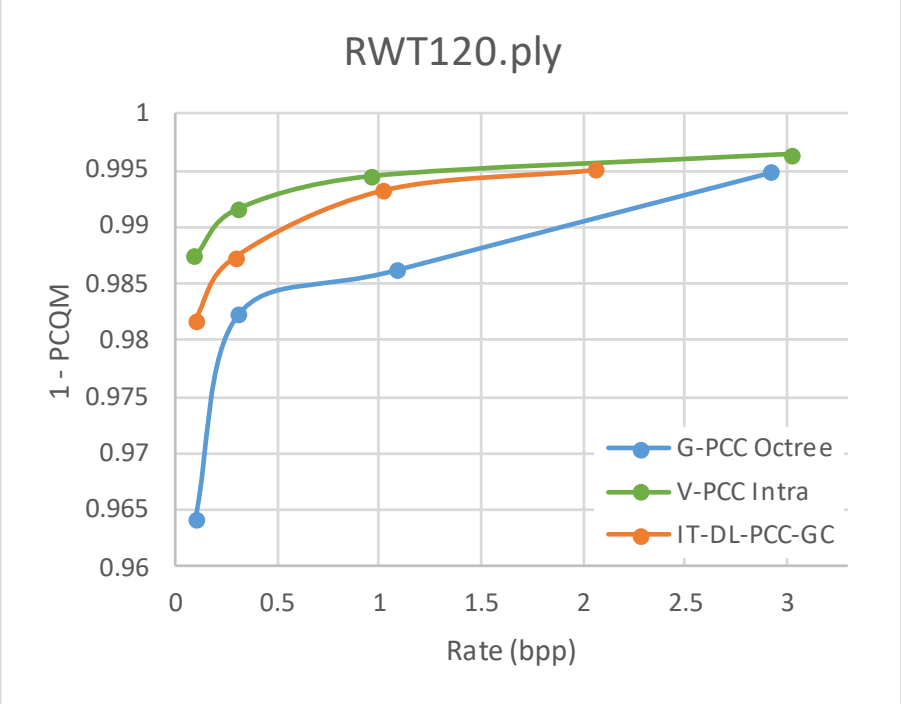
RWT120.ply
1 - PCQM
0.96 0.965 0.97 0.975 0.98 0.985 0.99 0.995 1
0 0.5 1 1.5 2 2.5 3
Rate (bpp)
G-PCC Octree
V-PCC Intra
IT-DL-PCC-GC

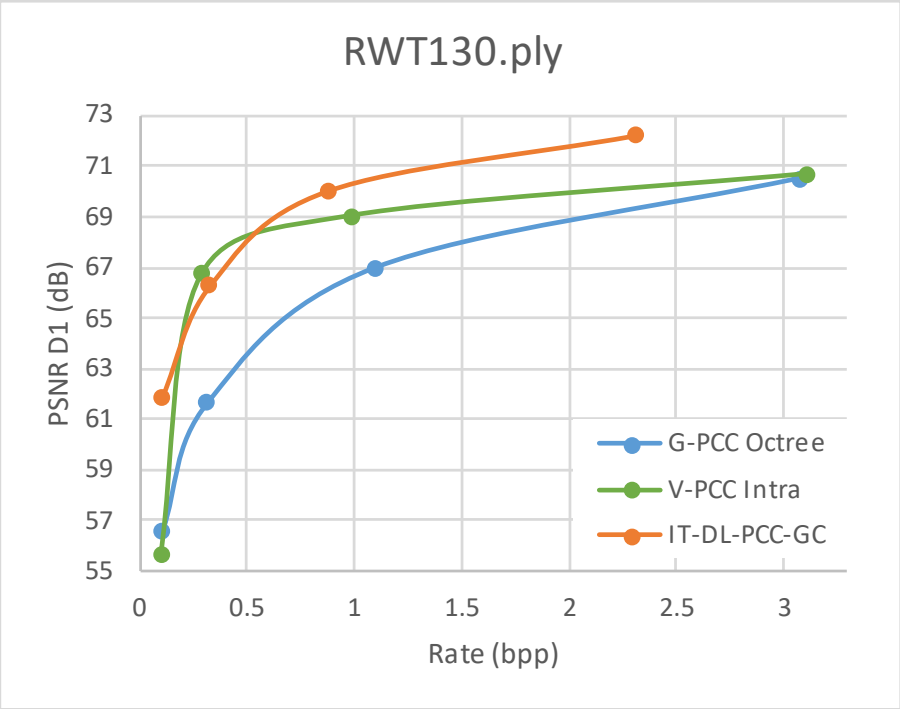

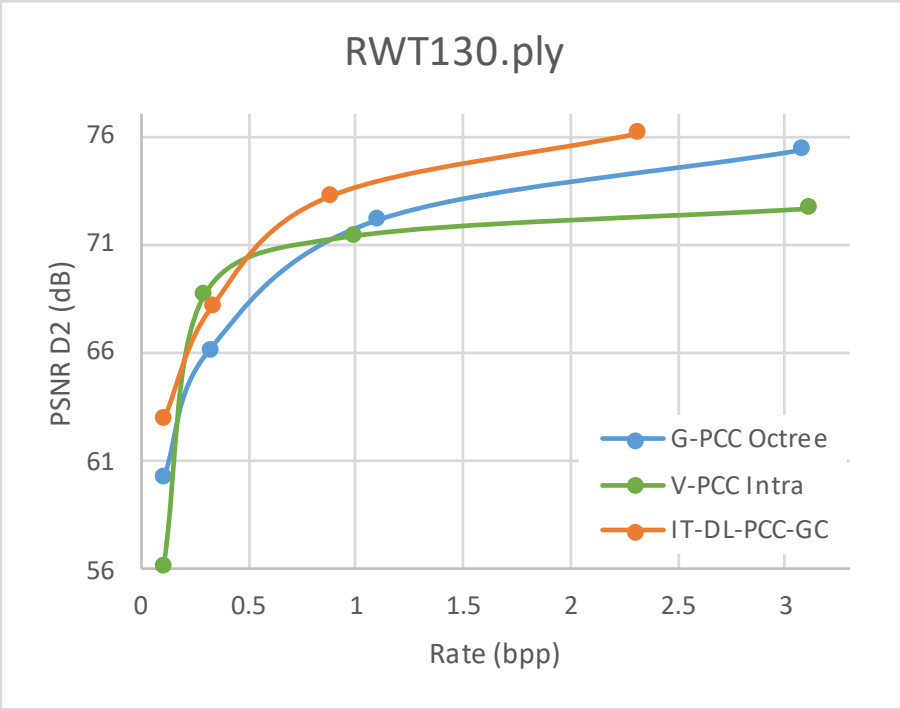

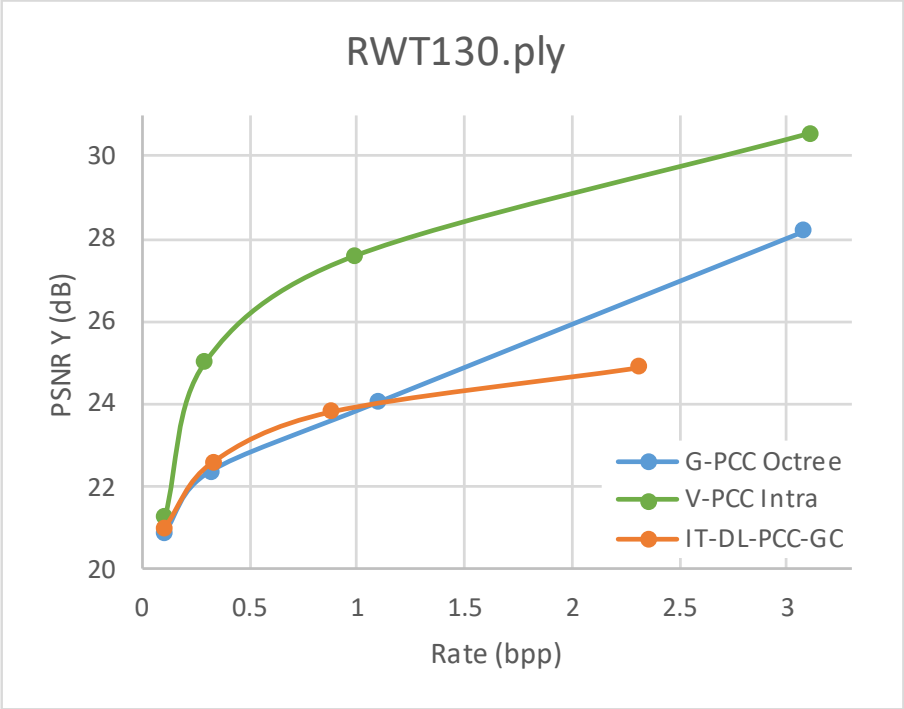

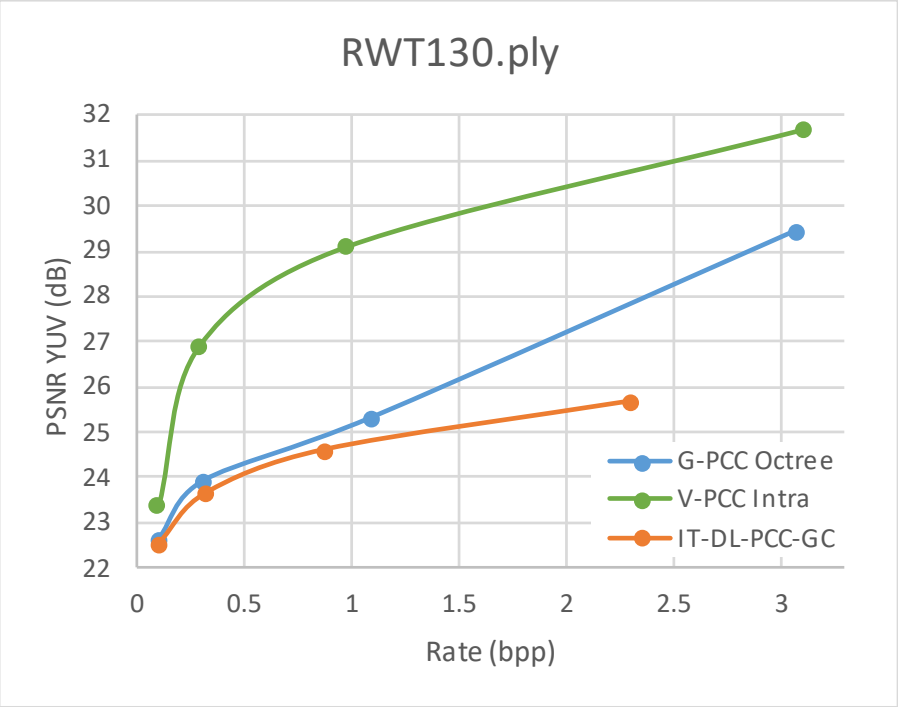

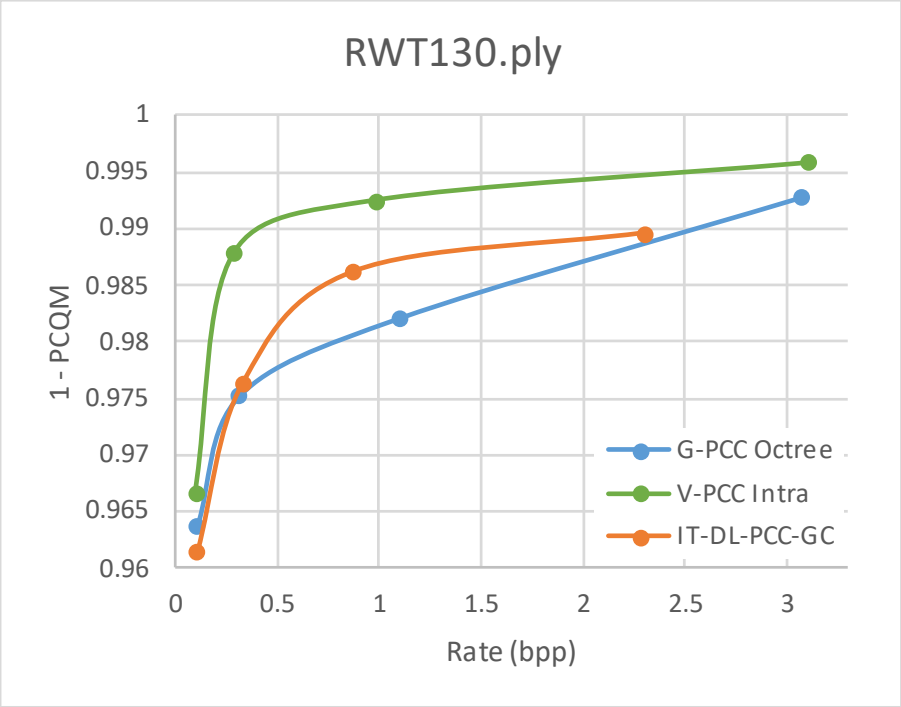

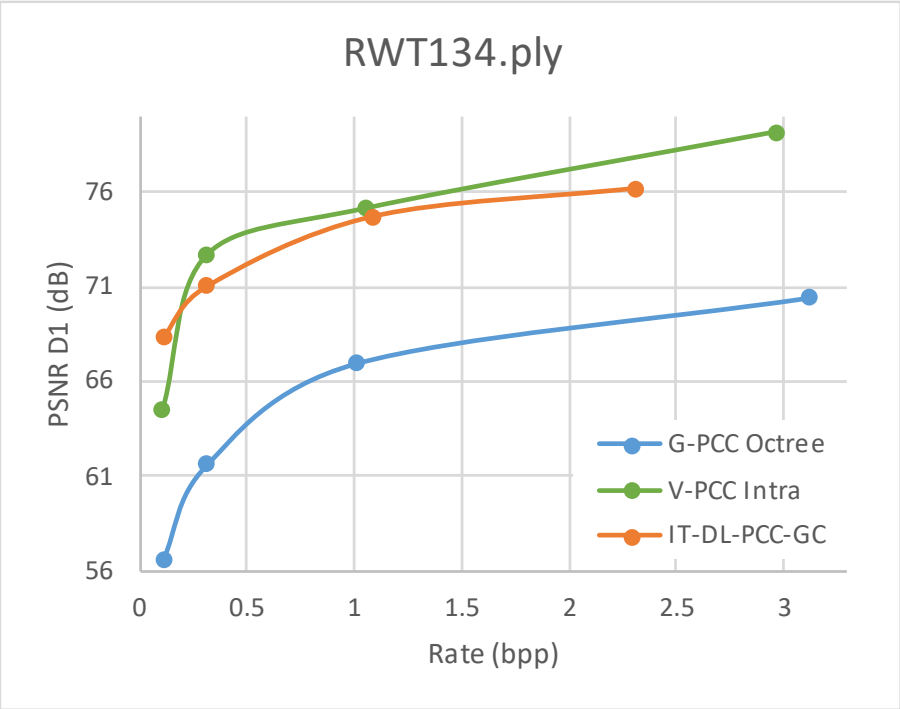

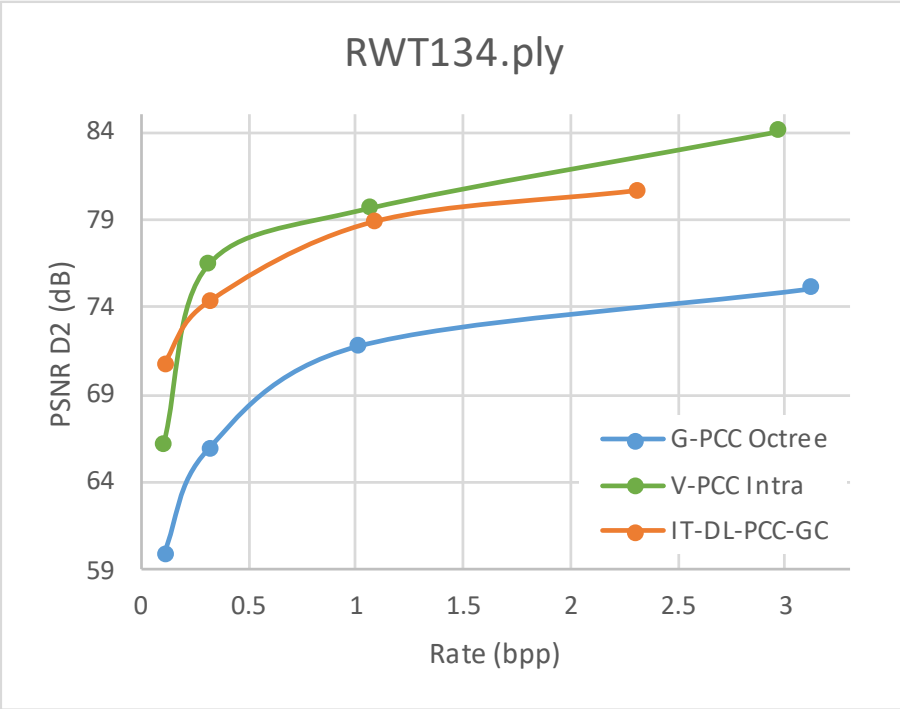

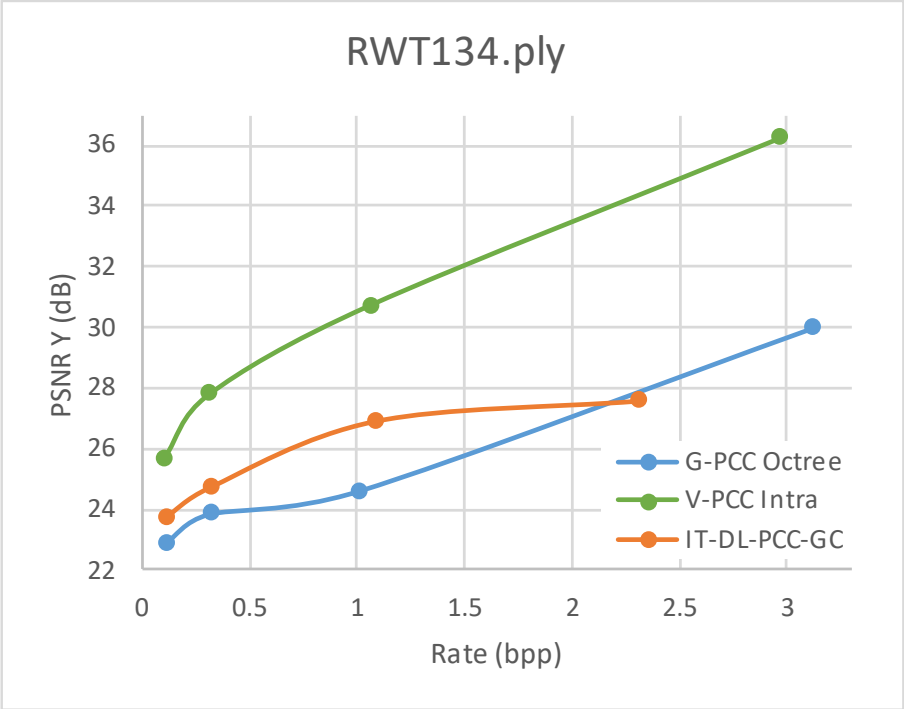

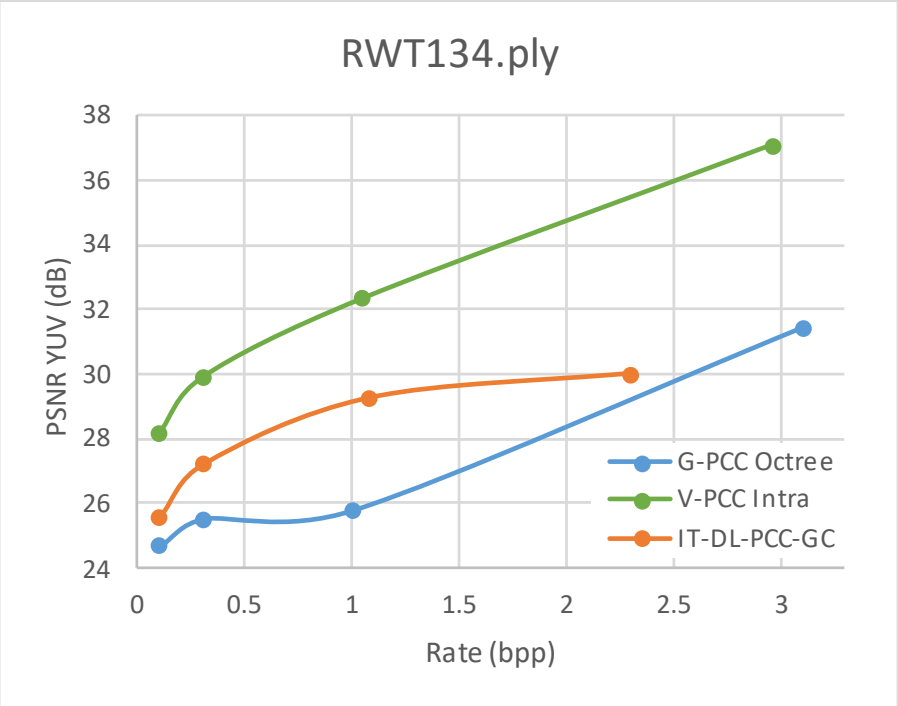

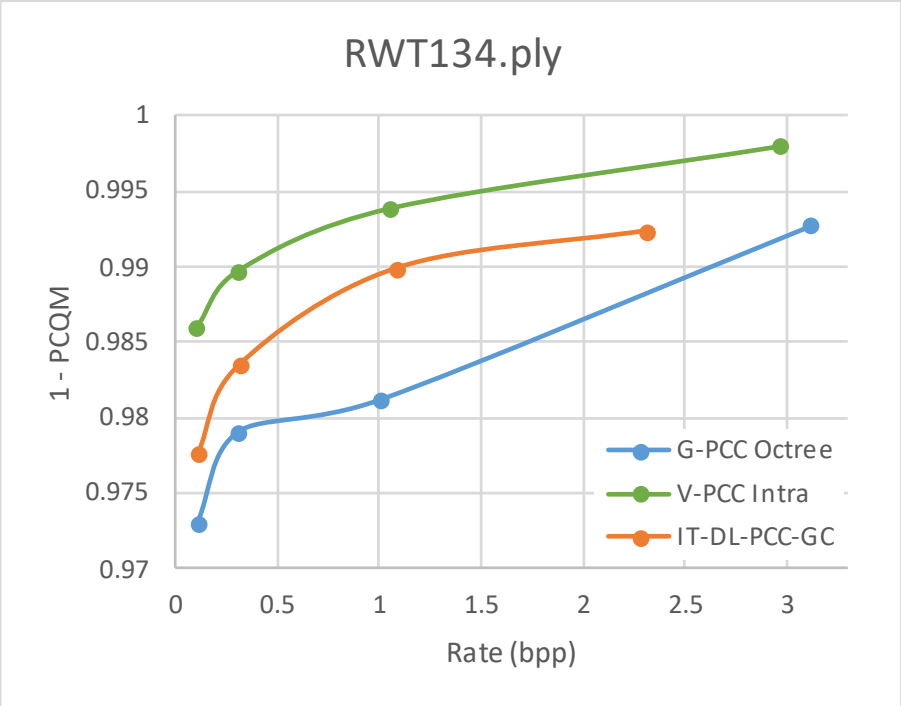

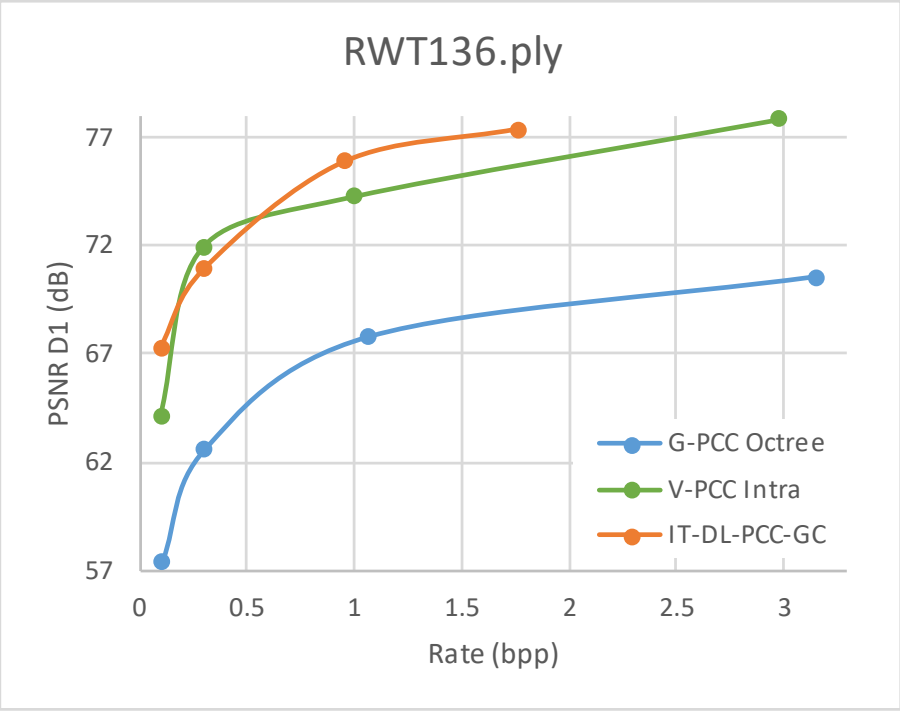

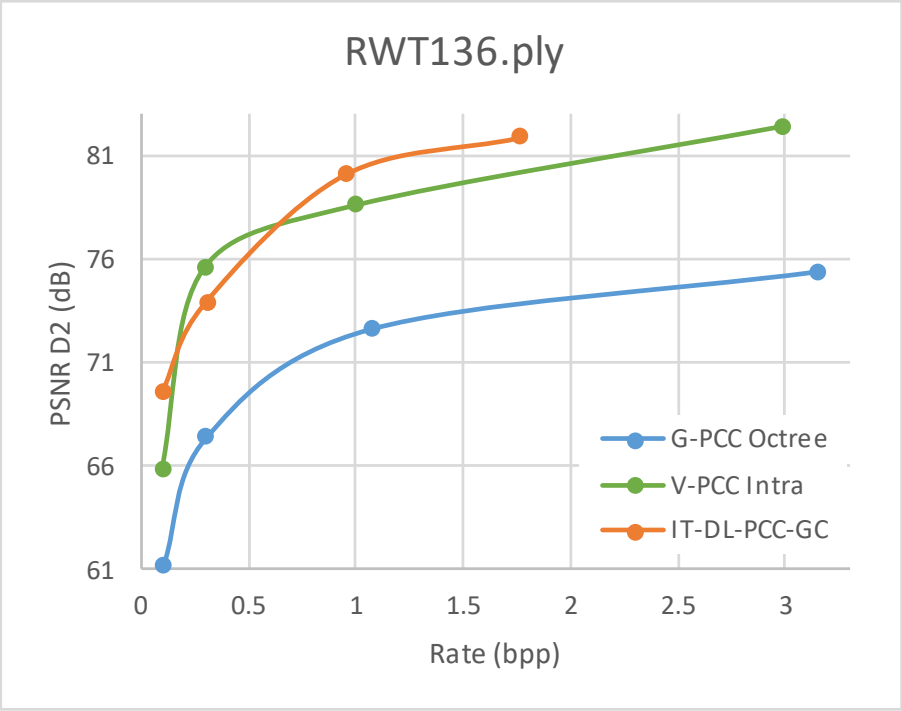

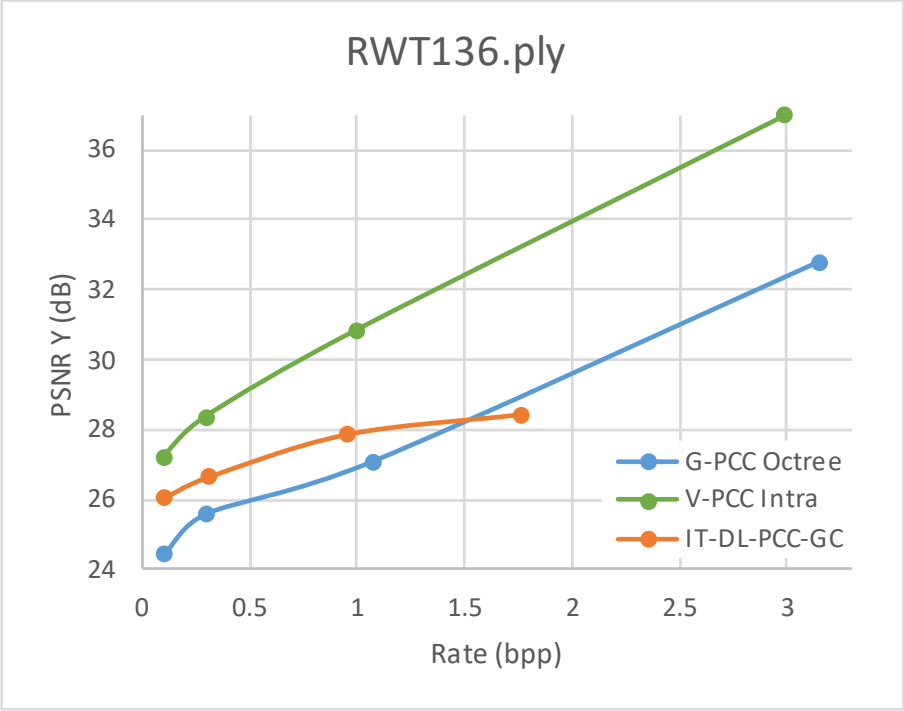

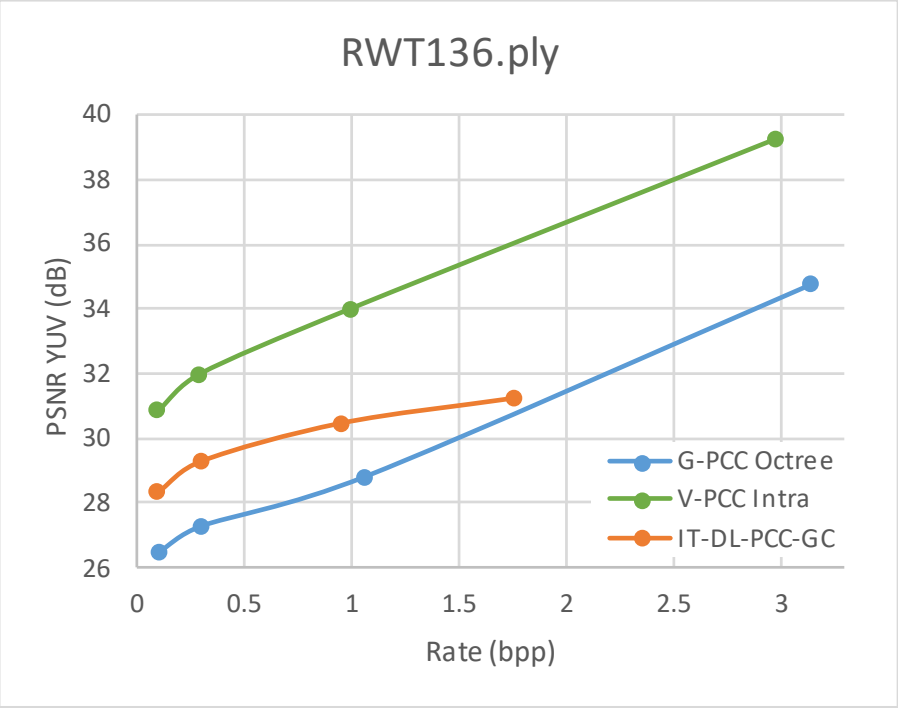

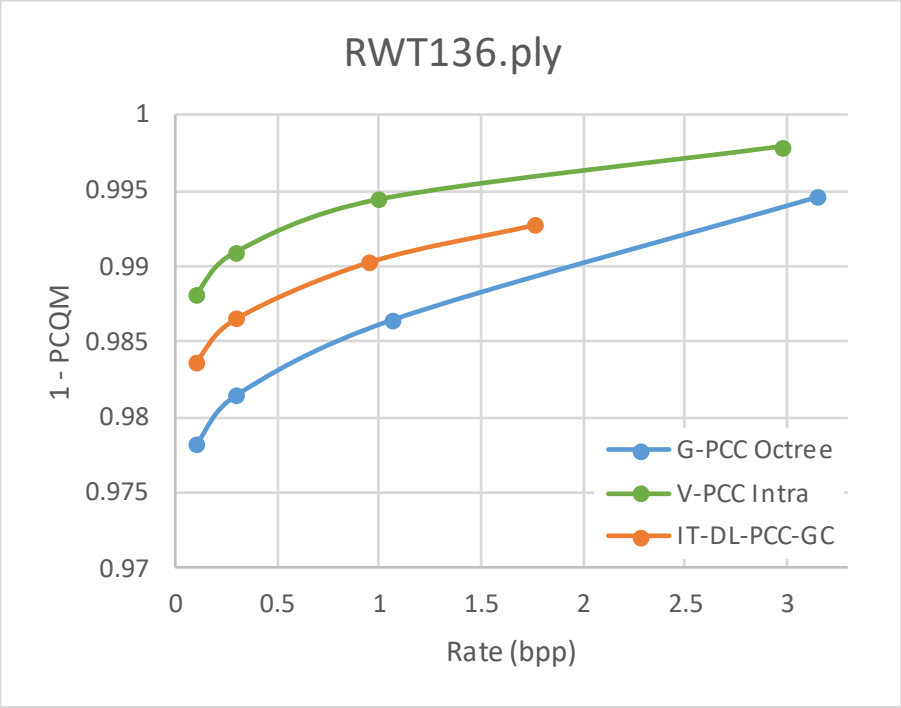

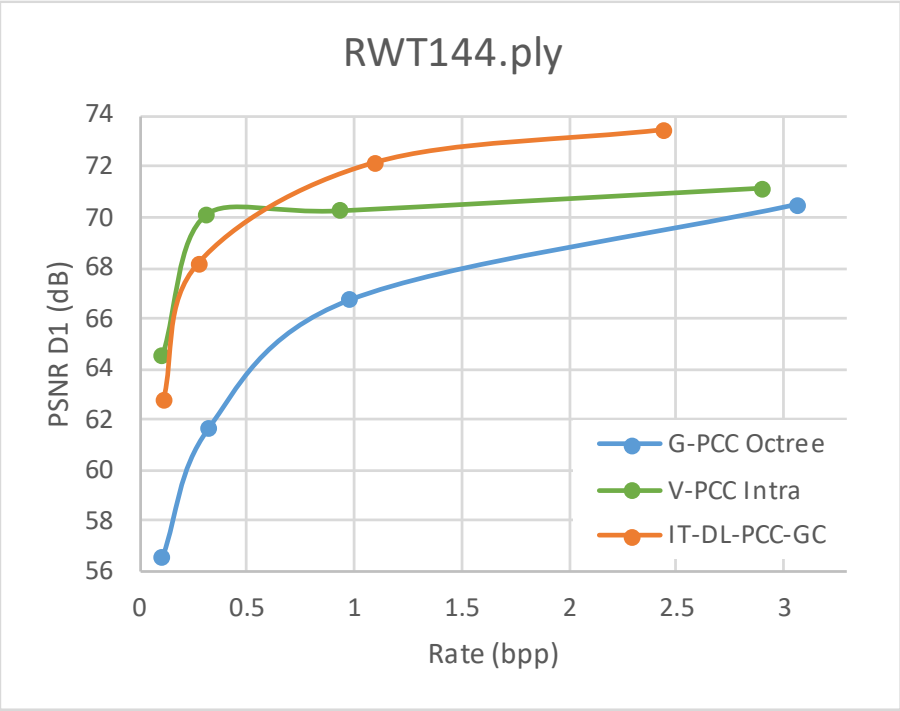

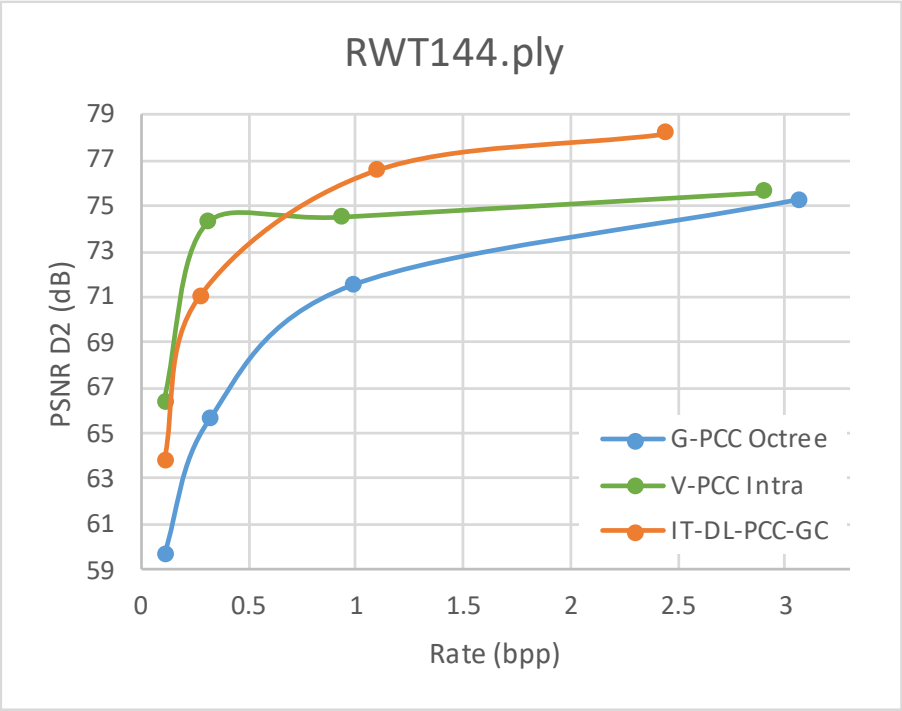

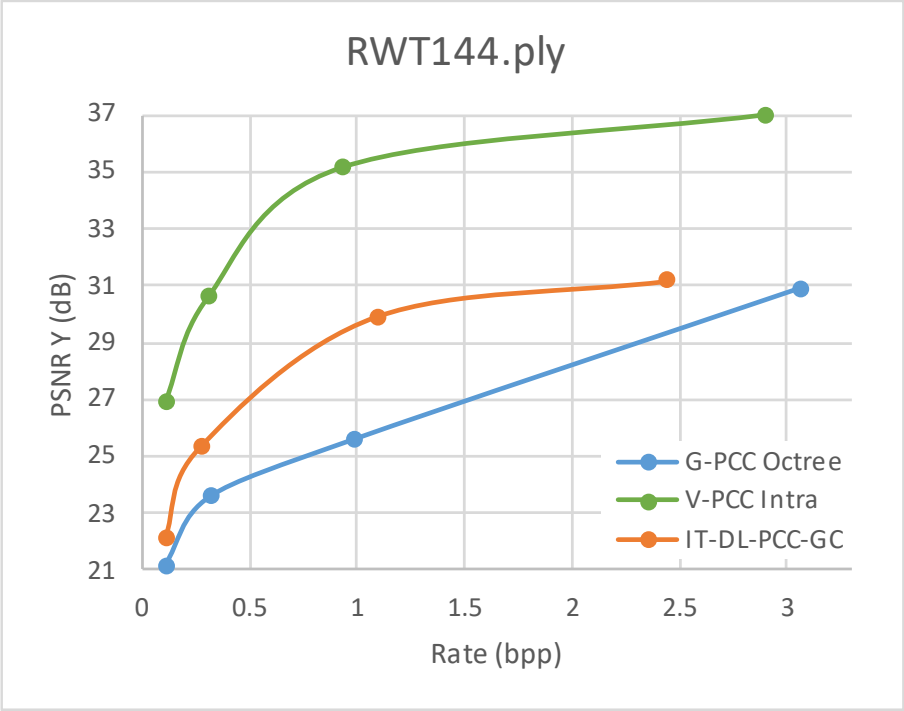

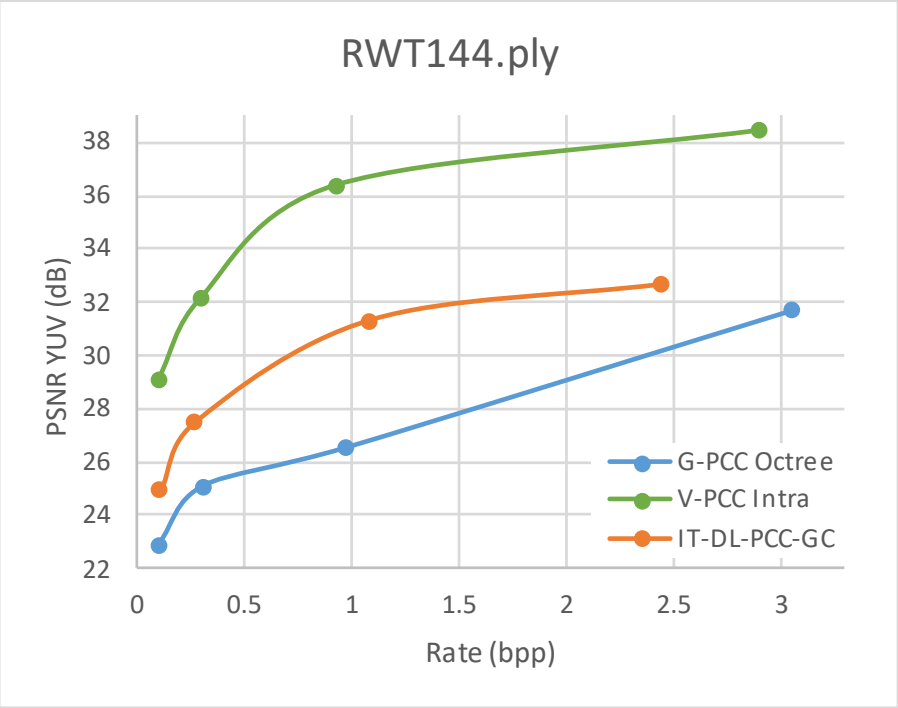

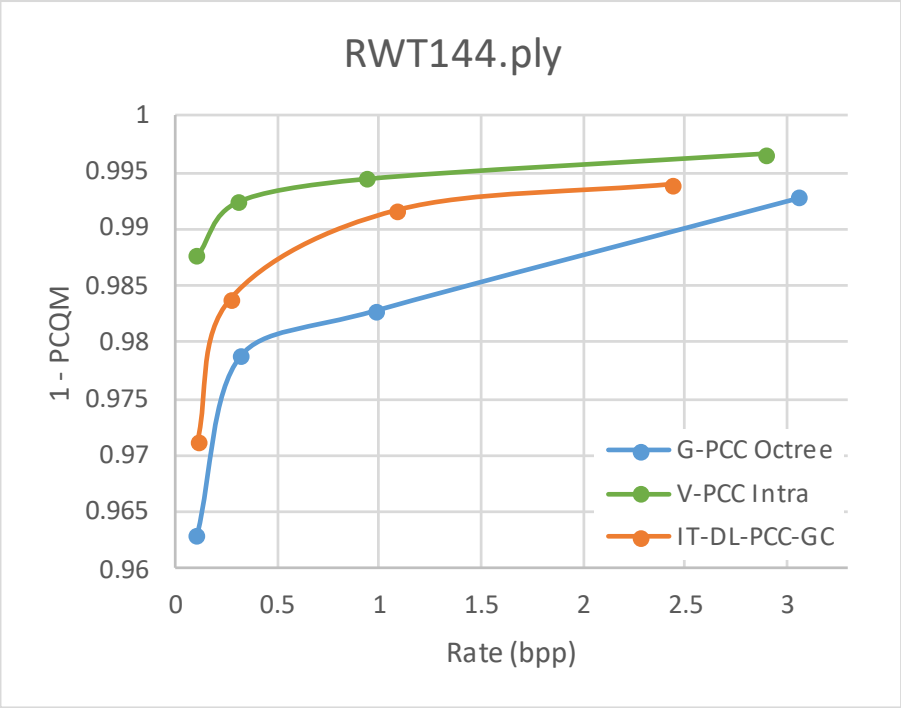

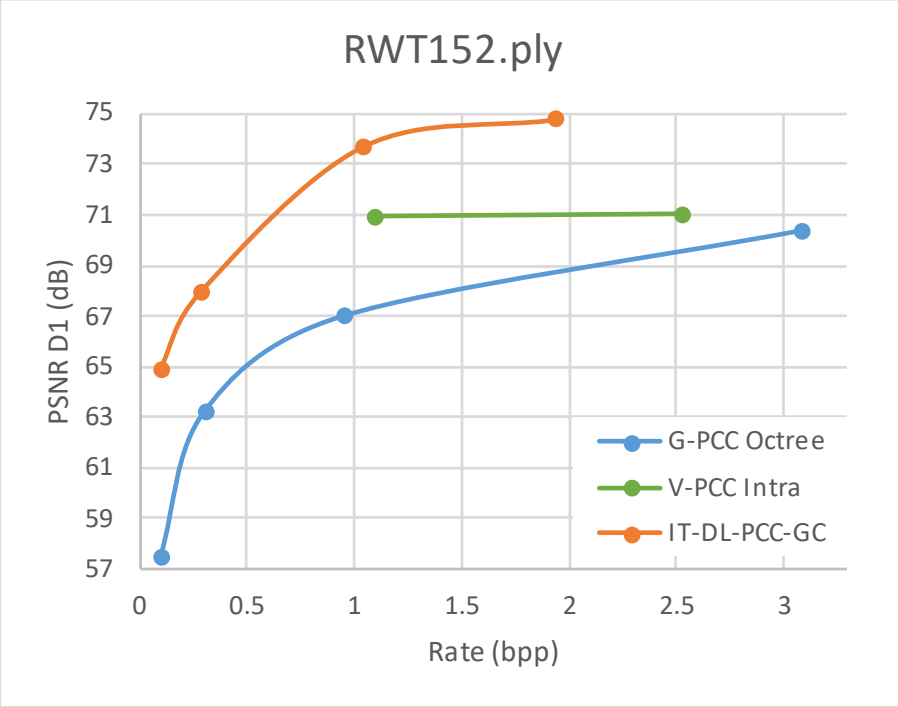

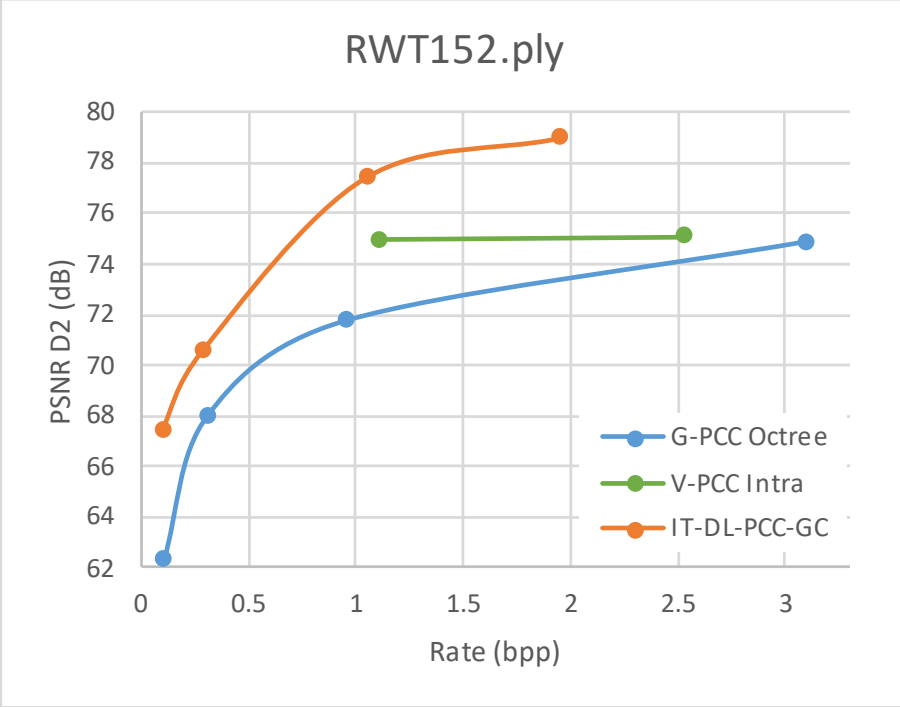

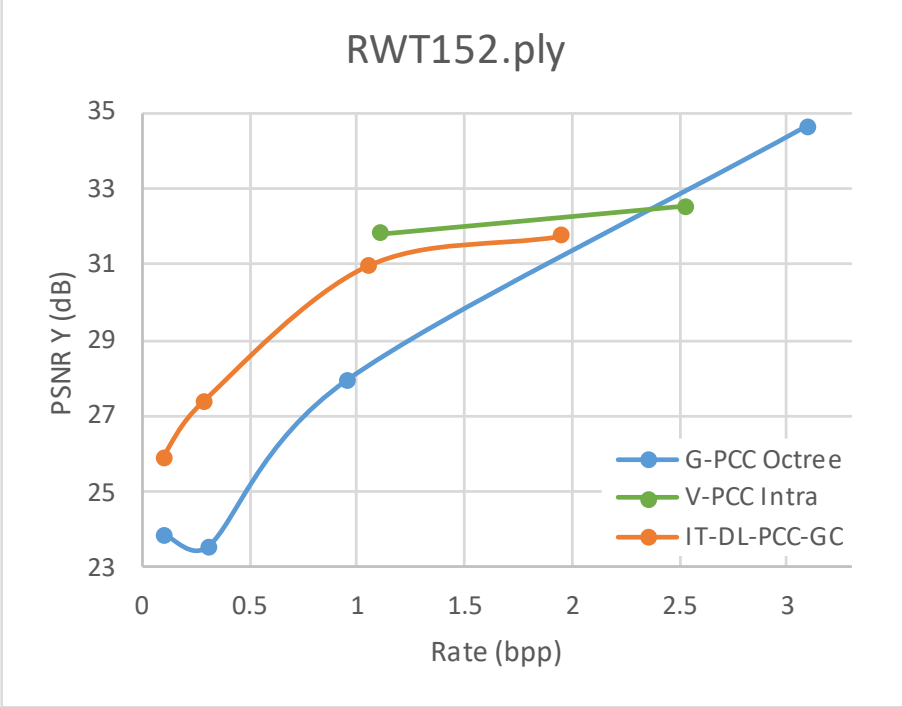

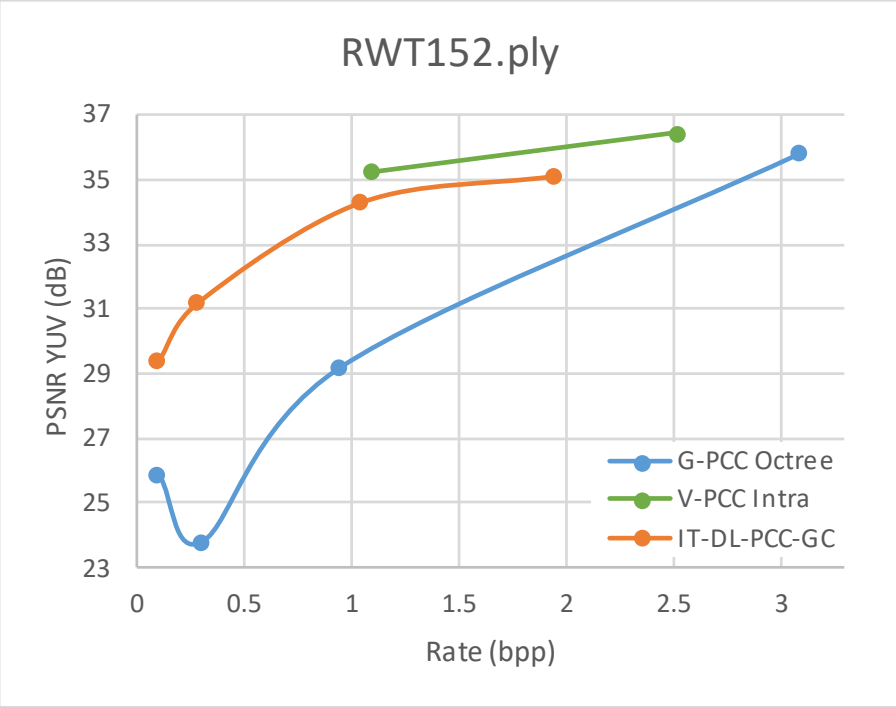

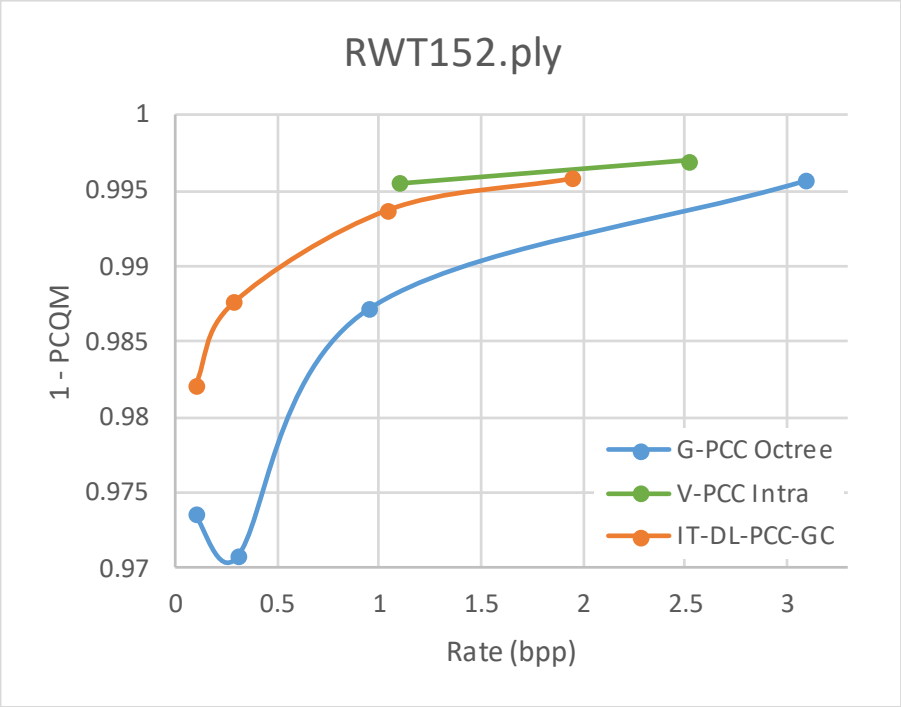

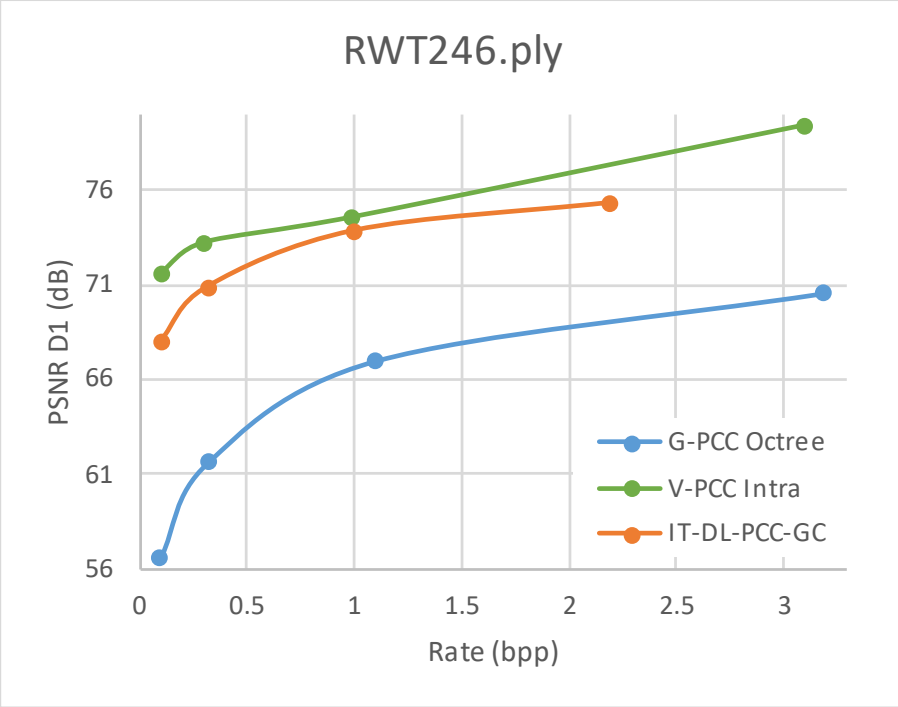

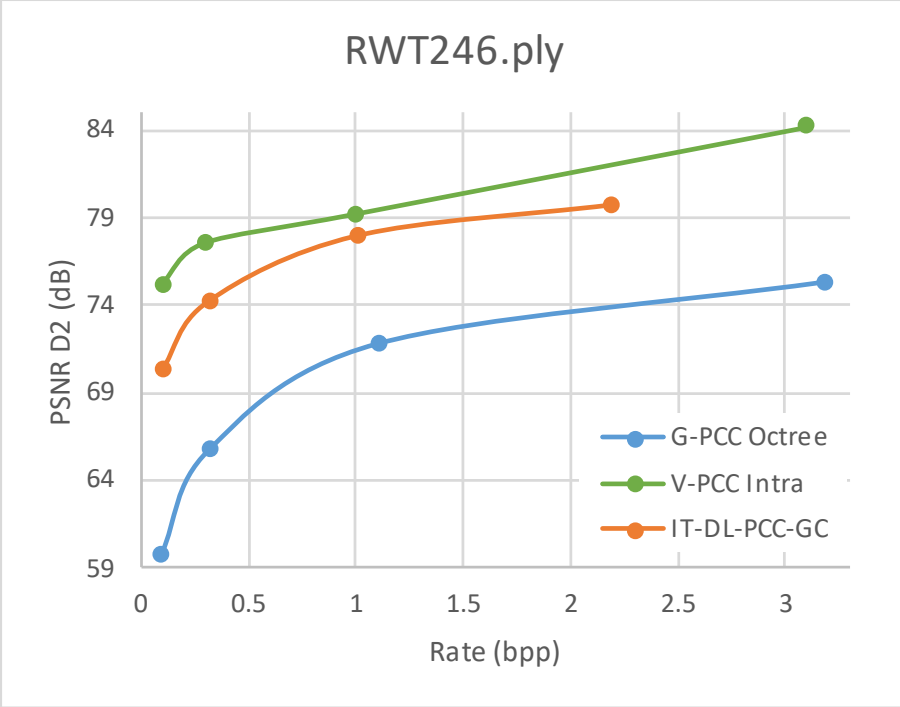

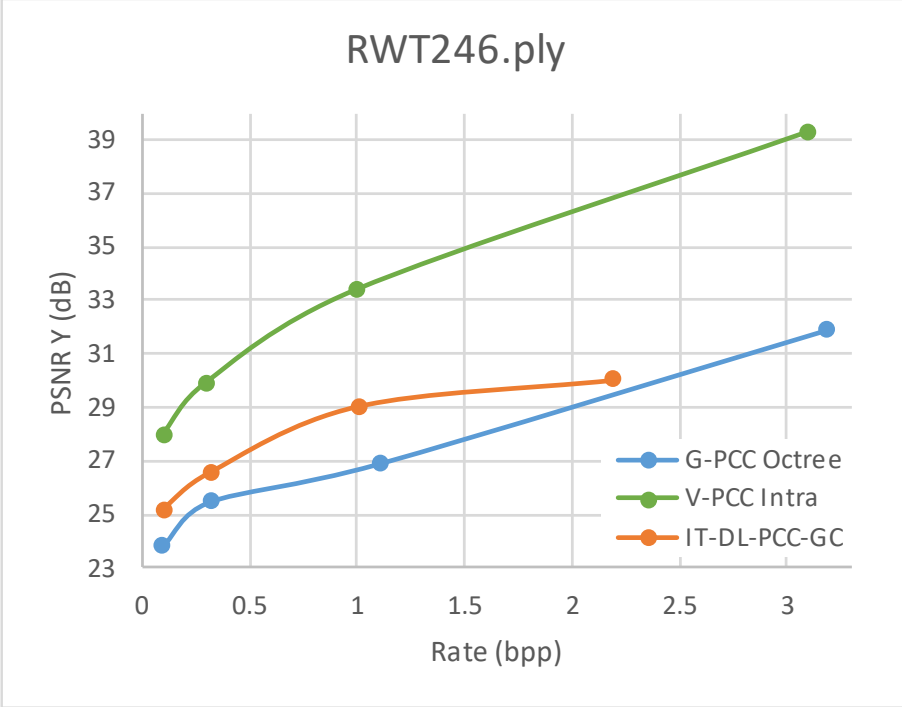

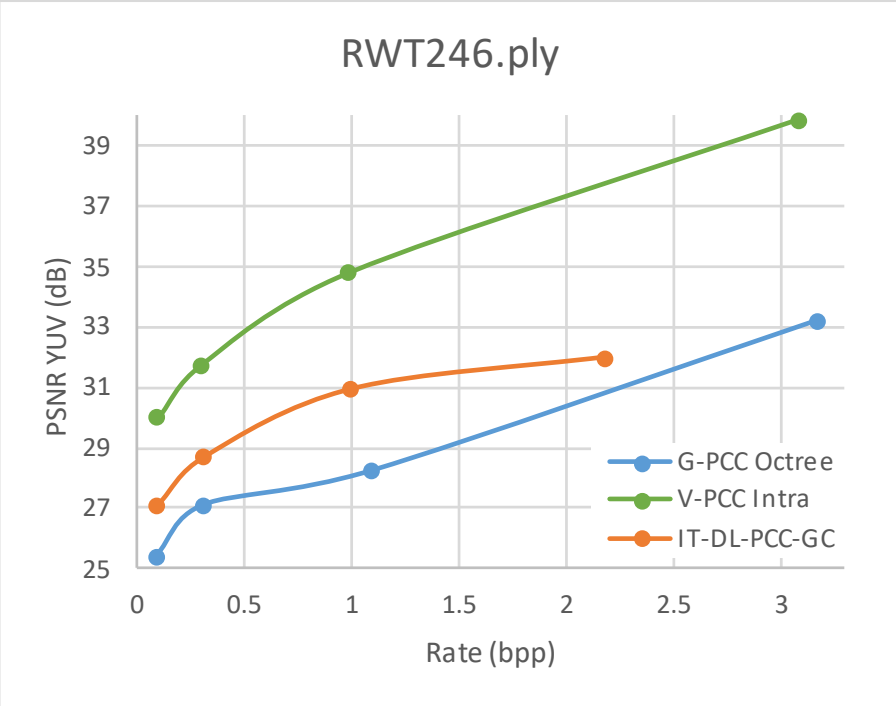

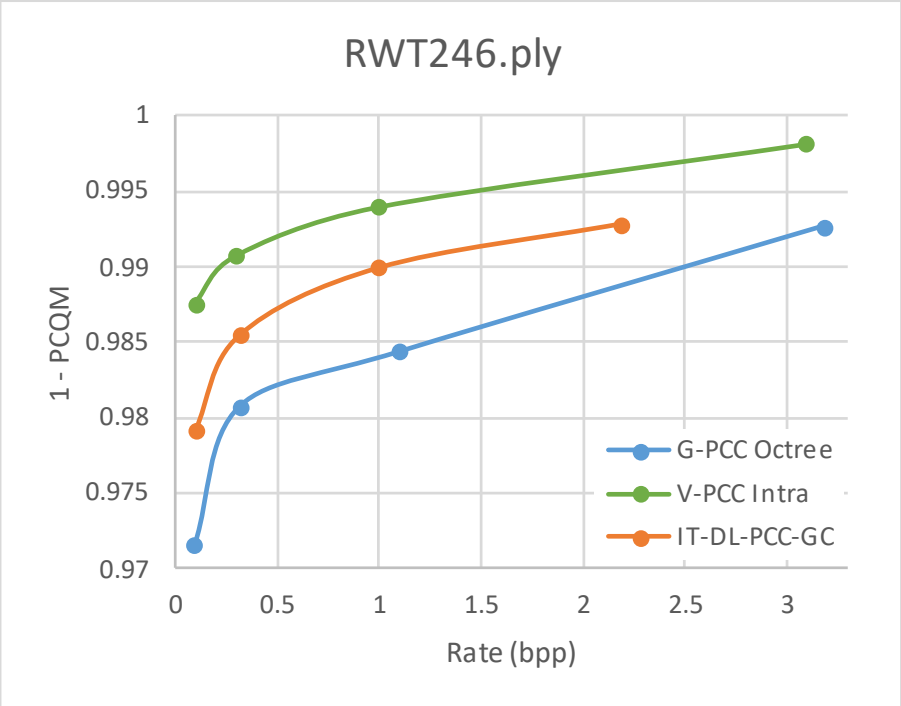

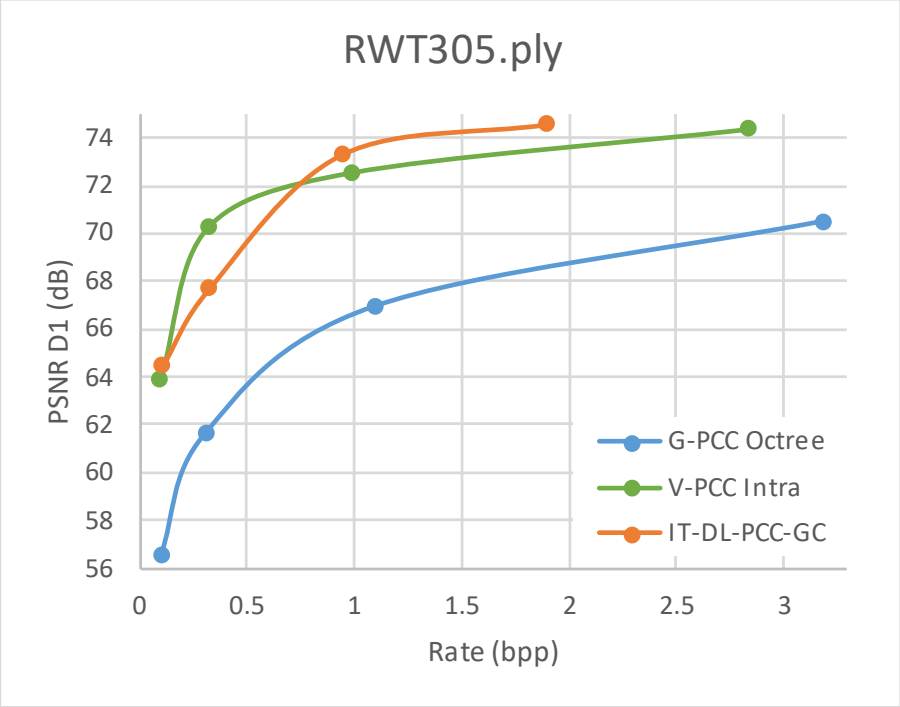
RWT305.ply
PSNR D1 (dB)
74 72 70 68 66 64 62 60 58 56
0 0.5 1 1.5 2 2.5 3
Rate (bpp)
G-PCC Octree
V-PCC Intra
IT-DL-PCC-GC

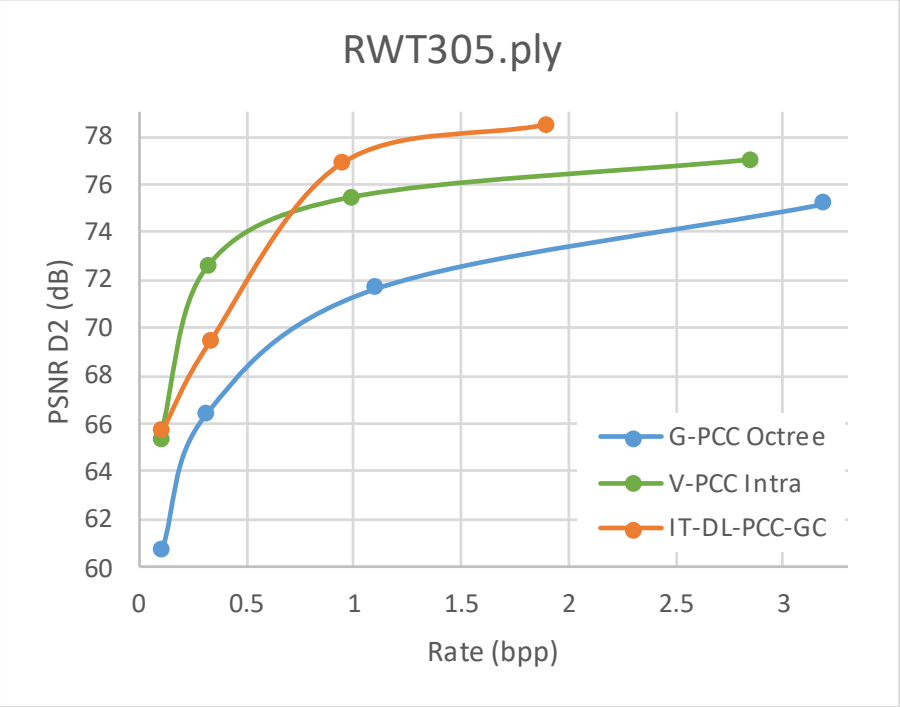
RWT305.ply
PSNR D2 (dB)
78 76 74 72 70 68 66 64 62 60
0 0.5 1 1.5 2 2.5 3
Rate (bpp)
G-PCC Octree
V-PCC Intra
IT-DL-PCC-GC

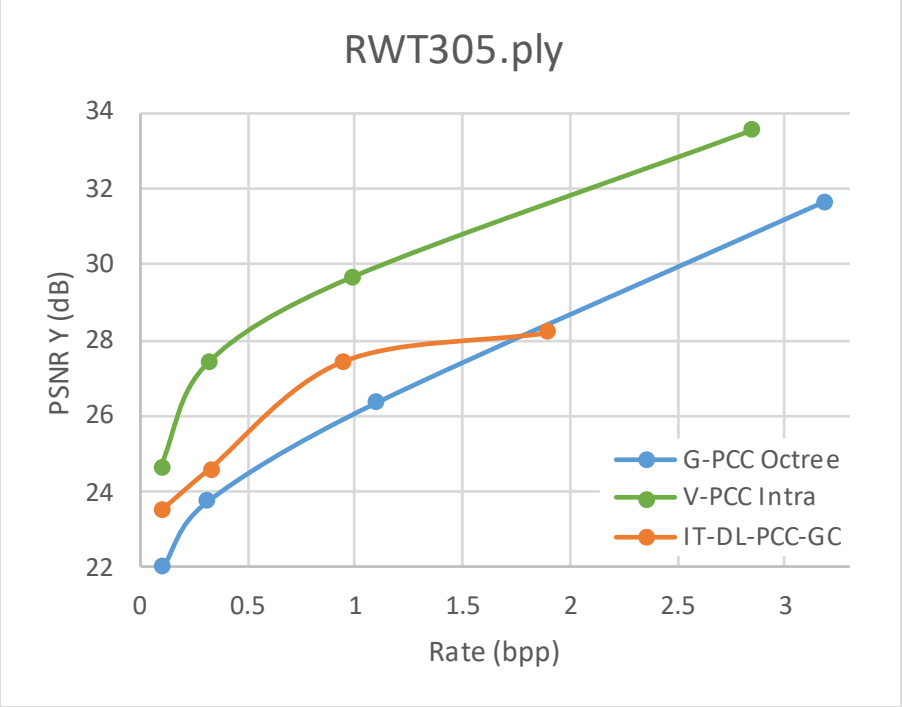
RWT305.ply
PSNR Y (dB)
34 32 30 28 26 24 22
0 0.5 1 1.5 2 2.5 3
Rate (bpp)
G-PCC Octree
V-PCC Intra
IT-DL-PCC-GC

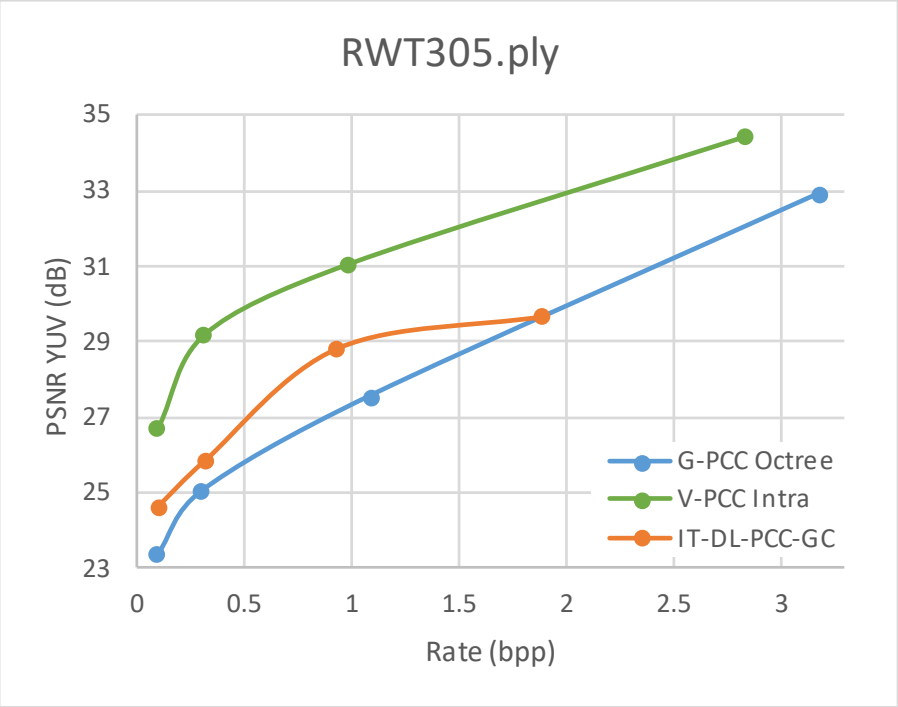
RWT305.ply
PSNR YUV (dB)
35 33 31 29 27 25 23
0 0.5 1 1.5 2 2.5 3
Rate (bpp)
G-PCC Octree
V-PCC Intra
IT-DL-PCC-GC

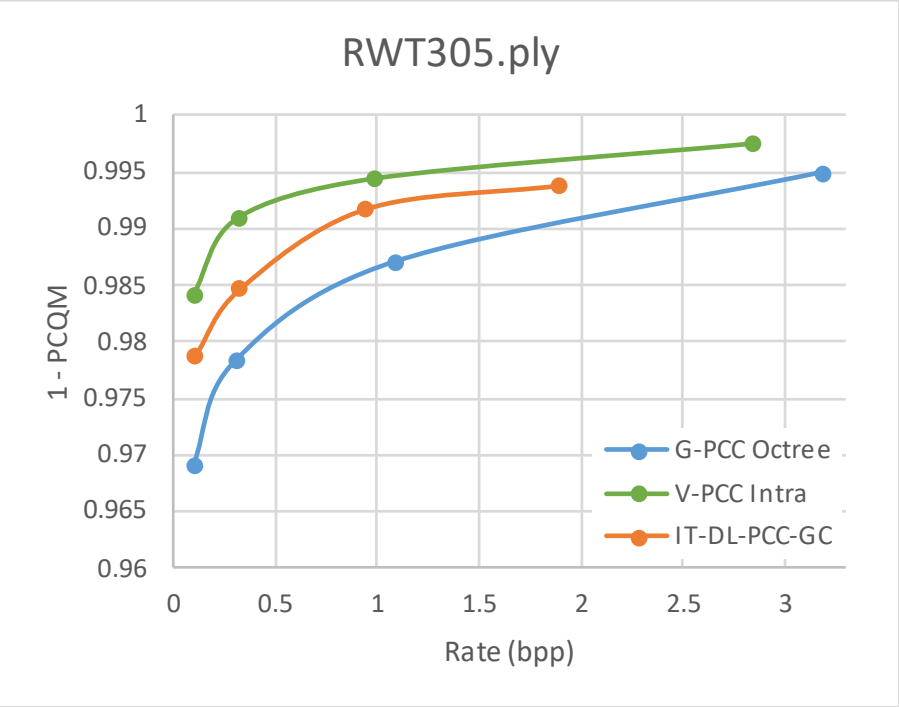
RWT305.ply
1 - PCQM
1 0.995 0.99 0.985 0.98 0.975 0.97 0.965 0.96
0 0.5 1 1.5 2 2.5 3
Rate (bpp)
G-PCC Octree
V-PCC Intra
IT-DL-PCC-GC

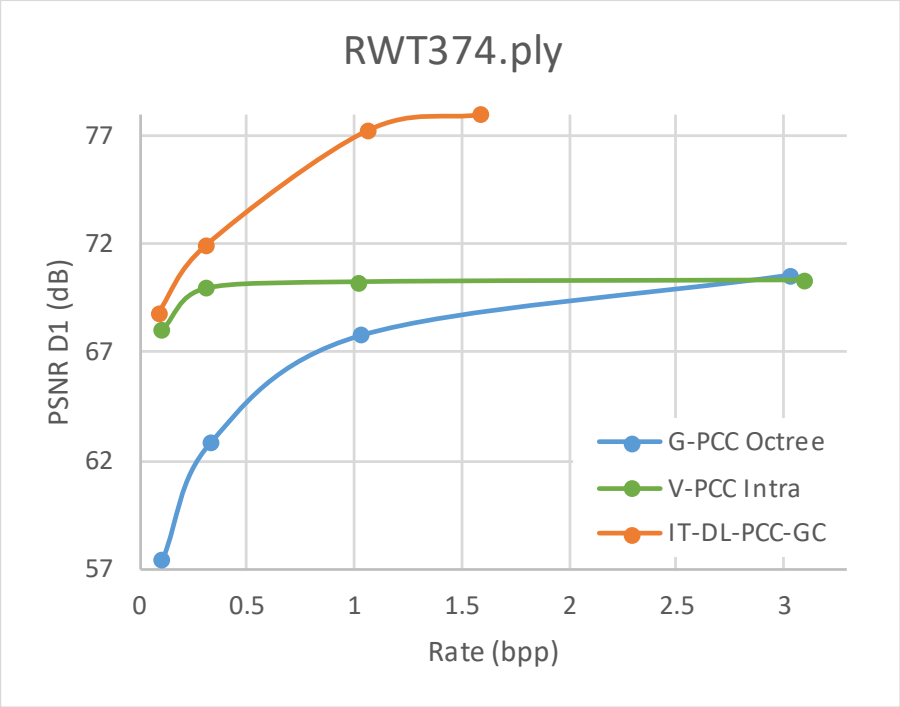
RWT374.ply
PSNR D1 (dB)
77 72 67 62 57
0 0.5 1 1.5 2 2.5 3
Rate (bpp)
G-PCC Octree
V-PCC Intra
IT-DL-PCC-GC

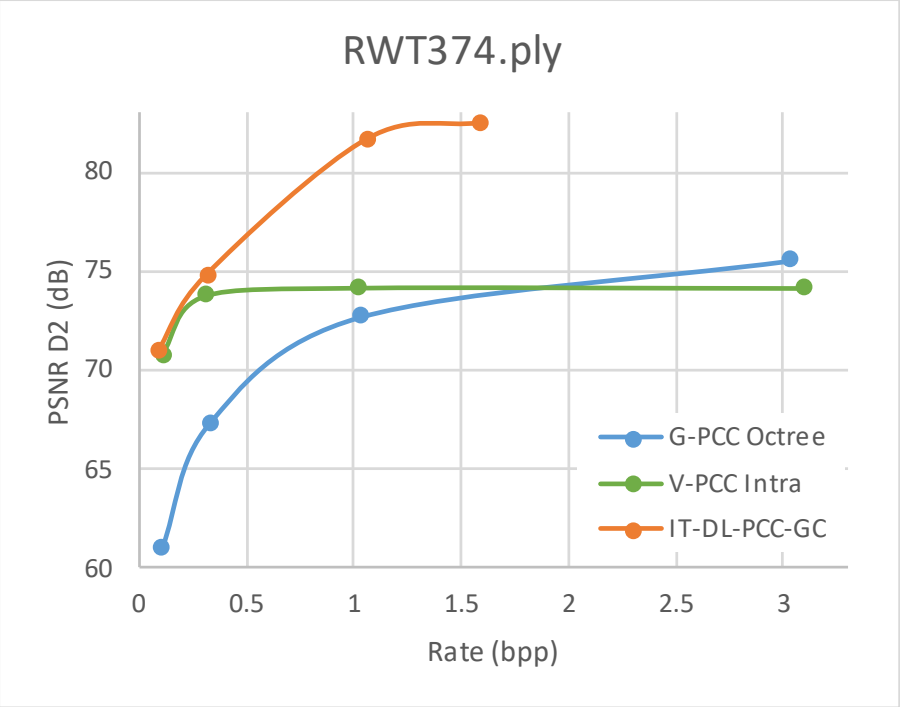
RWT374.ply
PSNR D2 (dB)
80 75 70 65 60
0 0.5 1 1.5 2 2.5 3
Rate (bpp)
G-PCC Octree
V-PCC Intra
IT-DL-PCC-GC

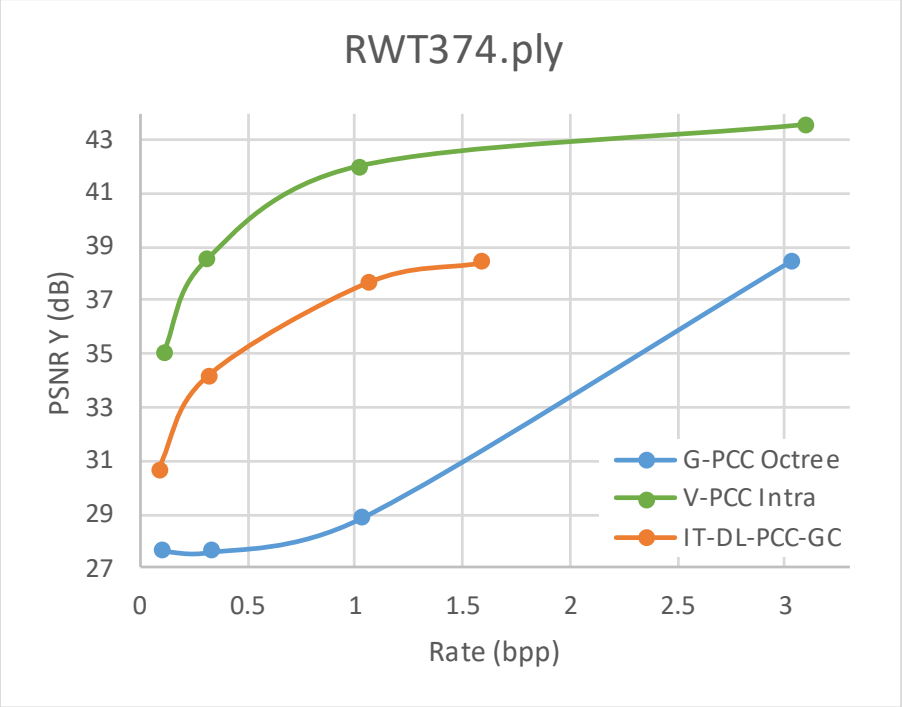
RWT374.ply
PSNR Y (dB)
43 41 39 37 35 33 31 29 27
0 0.5 1 1.5 2 2.5 3
Rate (bpp)
G-PCC Octree
V-PCC Intra
IT-DL-PCC-GC

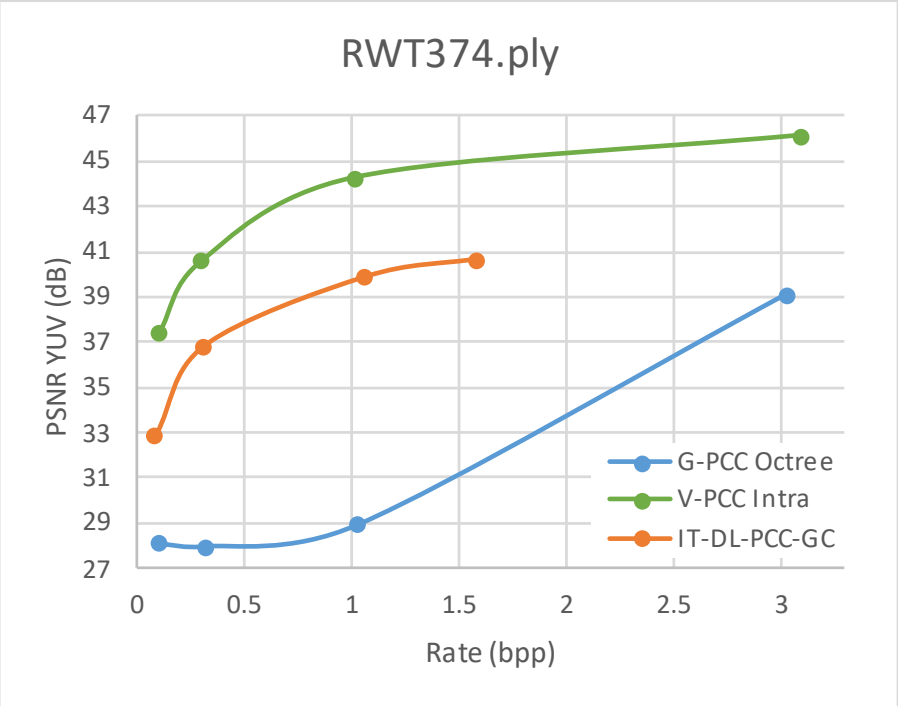
RWT374.ply
PSNR YUV (dB)
47 45 43 41 39 37 35 33 31 29 27
0 0.5 1 1.5 2 2.5 3
Rate (bpp)
G-PCC Octree
V-PCC Intra
IT-DL-PCC-GC

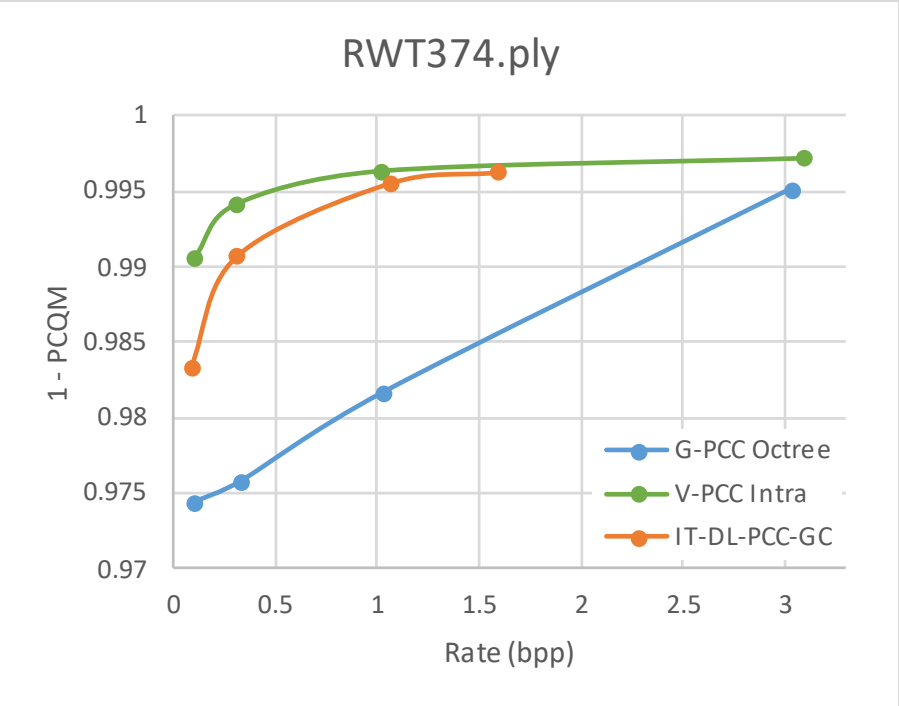
RWT374.ply
1 - PCQM
1 0.995 0.99 0.985 0.98 0.975 0.97
0 0.5 1 1.5 2 2.5 3
Rate (bpp)
G-PCC Octree
V-PCC Intra
IT-DL-PCC-GC

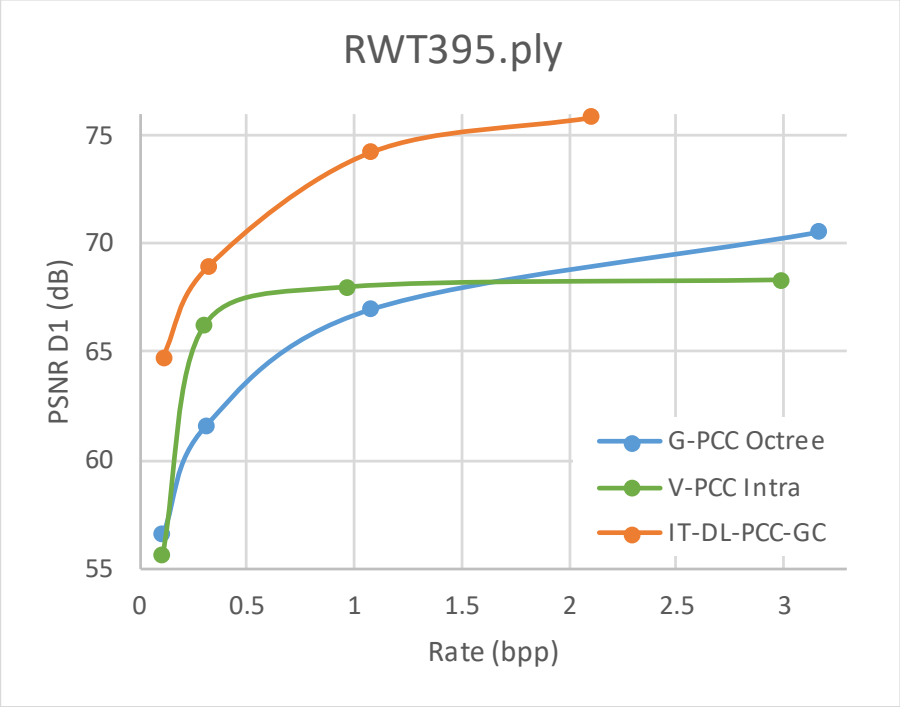

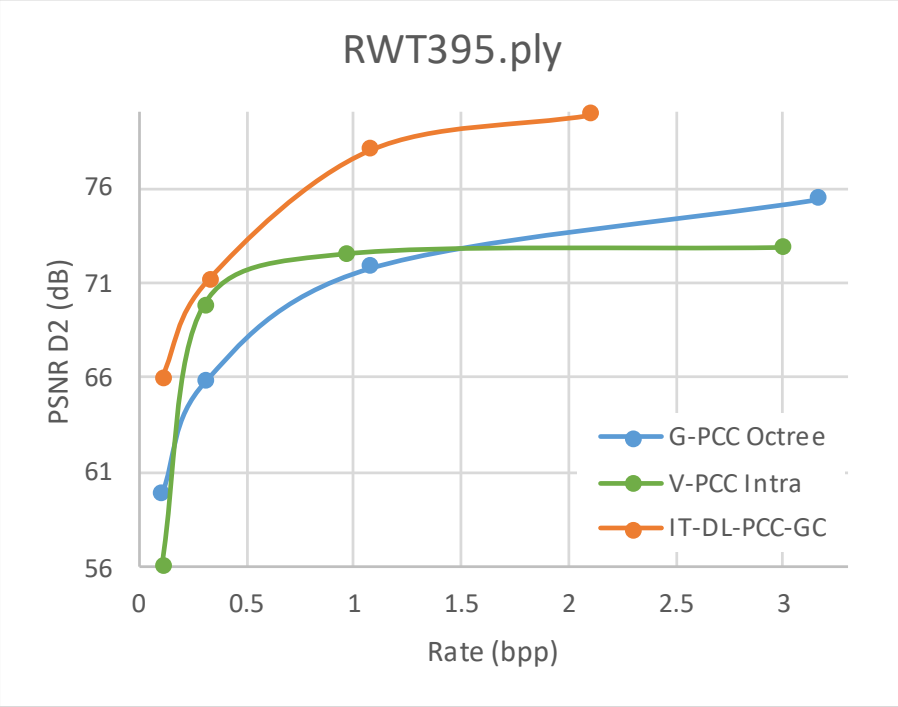

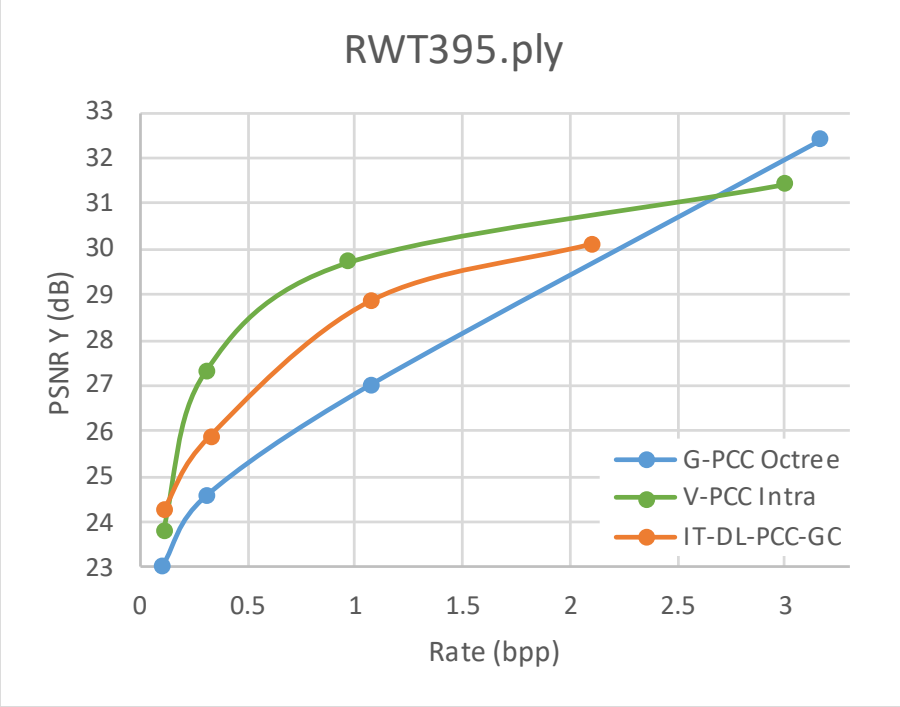
RWT395.ply
PSNR Y (dB)
33 32 31 30 29 28 27 26 25 24 23
0 0.5 1 1.5 2 2.5 3
Rate (bpp)
G-PCC Octree
V-PCC Intra
IT-DL-PCC-GC

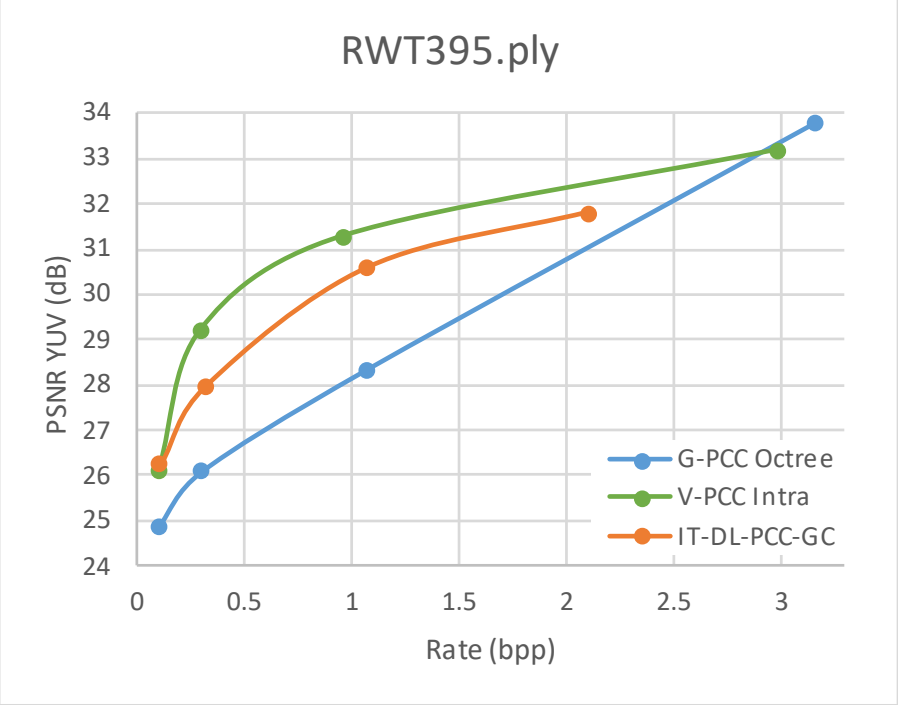
RWT395.ply
PSNR YUV (dB)
34 33 32 31 30 29 28 27 26 25 24
0 0.5 1 1.5 2 2.5 3
Rate (bpp)
G-PCC Octree
V-PCC Intra
IT-DL-PCC-GC

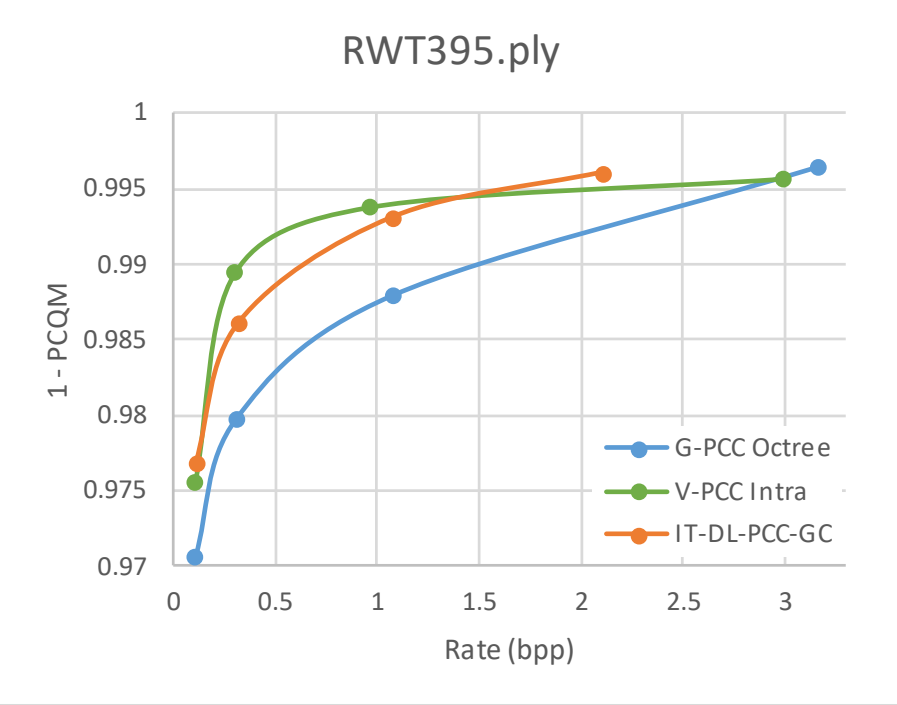
RWT395.ply
1 - PCQM
1 0.995 0.99 0.985 0.98 0.975 0.97
0 0.5 1 1.5 2 2.5 3
Rate (bpp)
G-PCC Octree
V-PCC Intra
IT-DL-PCC-GC

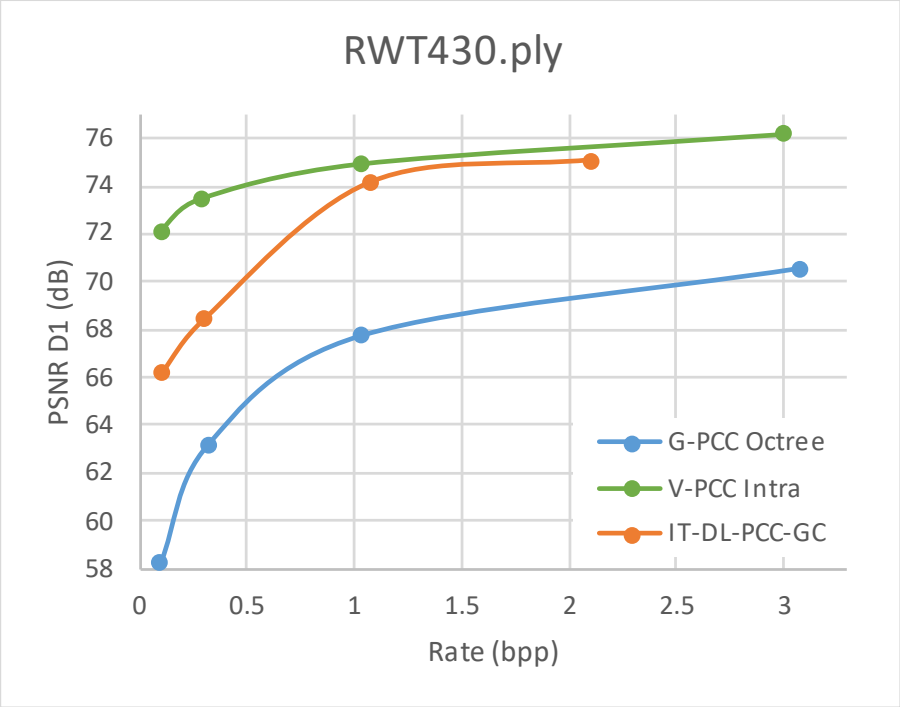
RWT430.ply
PSNR D1 (dB)
76 74 72 70 68 66 64 62 60 58
0 0.5 1 1.5 2 2.5 3
Rate (bpp)
G-PCC Octree
V-PCC Intra
IT-DL-PCC-GC

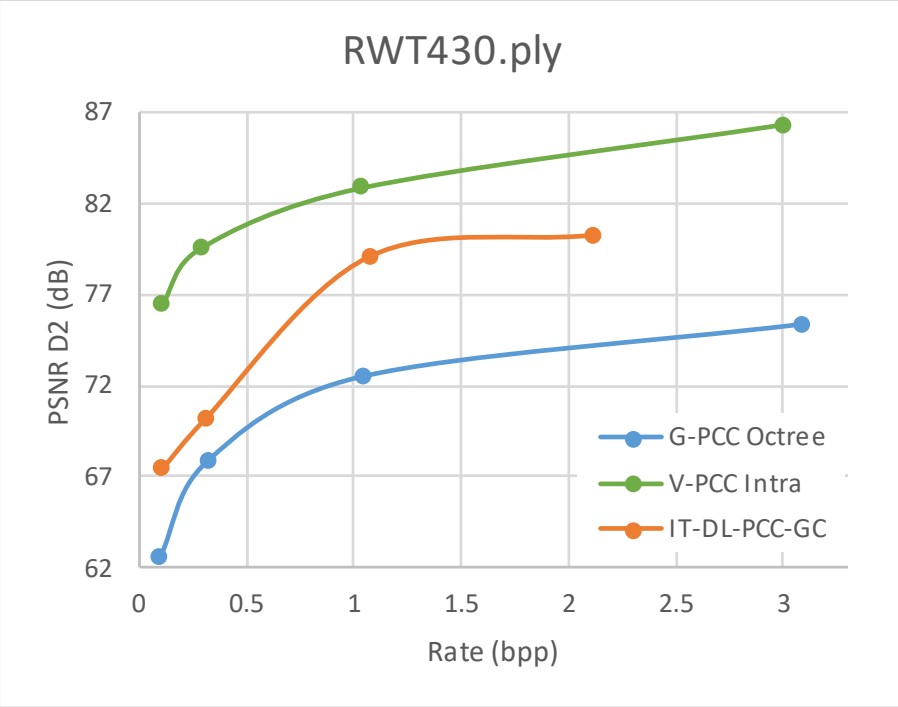
RWT430.ply
PSNR D2 (dB)
87 82 77 72 67 62
0 0.5 1 1.5 2 2.5 3
Rate (bpp)
G-PCC Octree
V-PCC Intra
IT-DL-PCC-GC

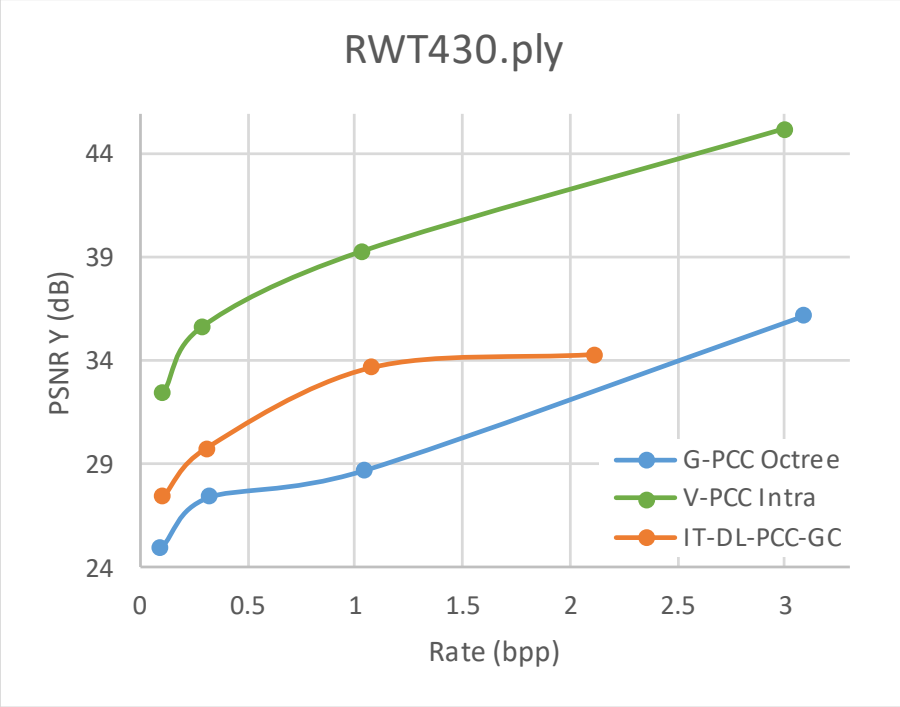
RWT430.ply
PSNR Y (dB)
44 39 34 29 24
0 0.5 1 1.5 2 2.5 3
Rate (bpp)
G-PCC Octree
V-PCC Intra
IT-DL-PCC-GC

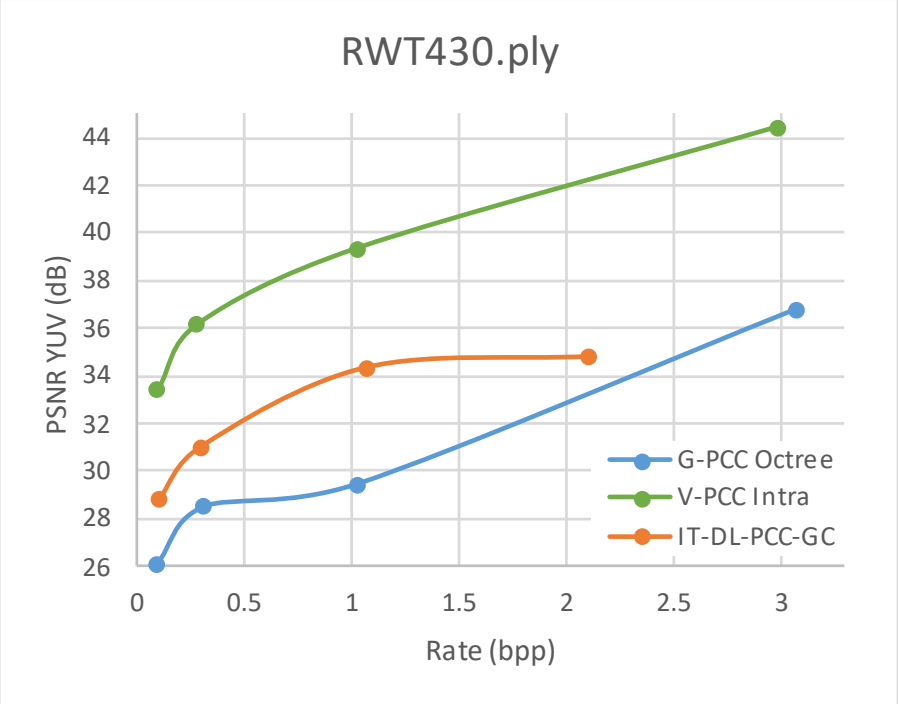
RWT430.ply
PSNR YUV (dB)
44 42 40 38 36 34 32 30 28 26
0 0.5 1 1.5 2 2.5 3
Rate (bpp)
G-PCC Octree
V-PCC Intra
IT-DL-PCC-GC

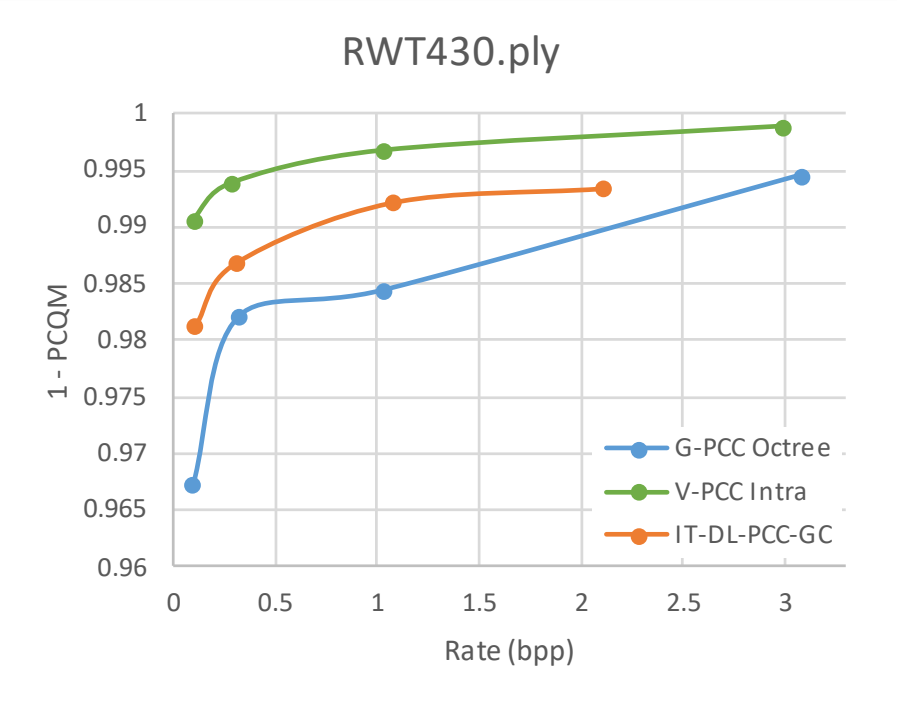
RWT430.ply
1 - PCQM
1 0.995 0.99 0.985 0.98 0.975 0.97 0.965 0.96
0 0.5 1 1.5 2 2.5 3
Rate (bpp)
G-PCC Octree
V-PCC Intra
IT-DL-PCC-GC

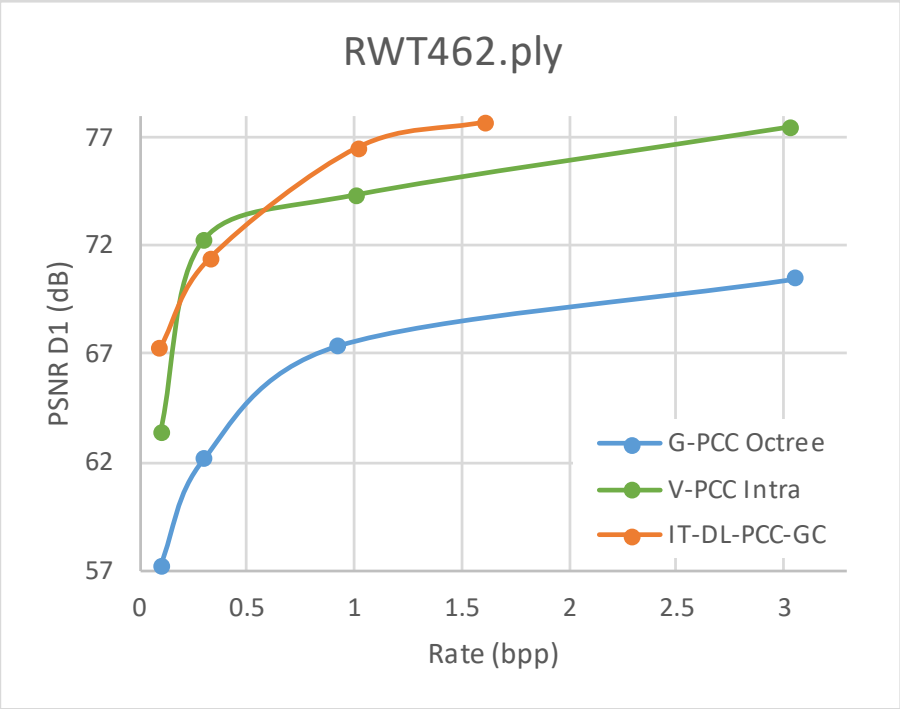

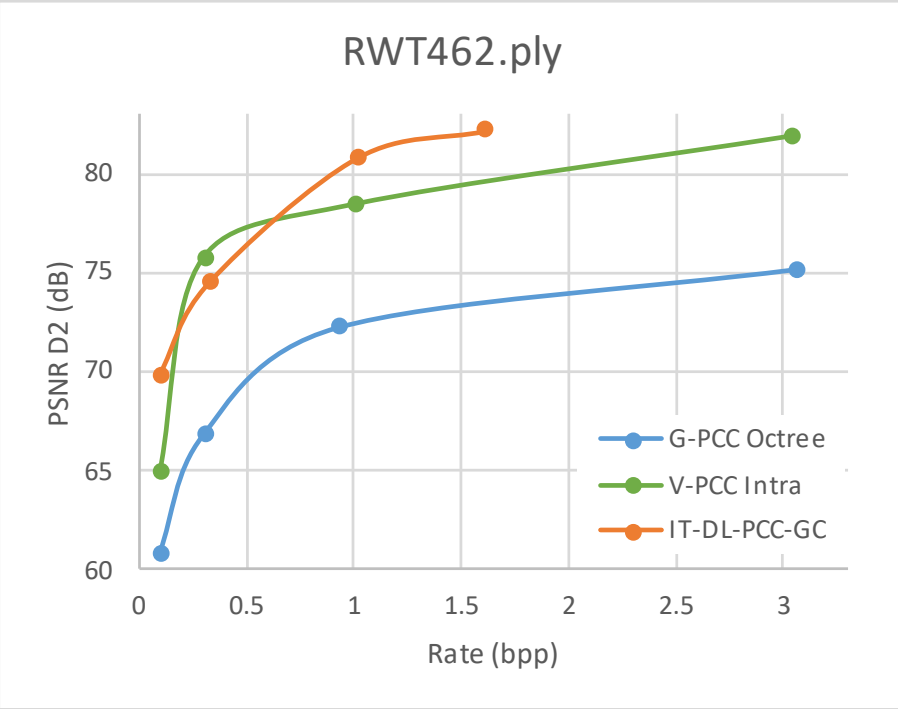

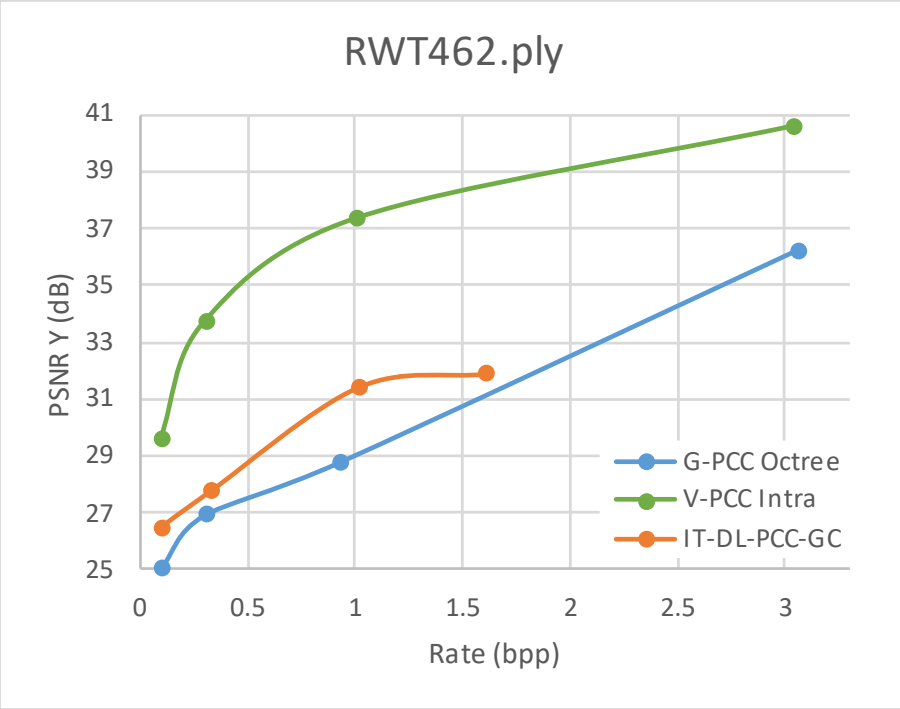

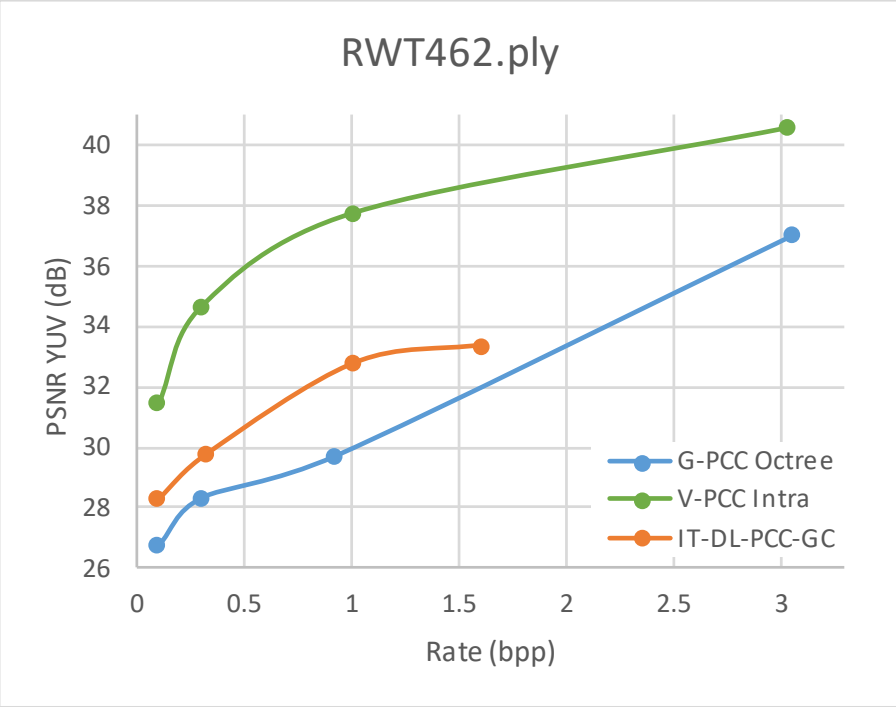

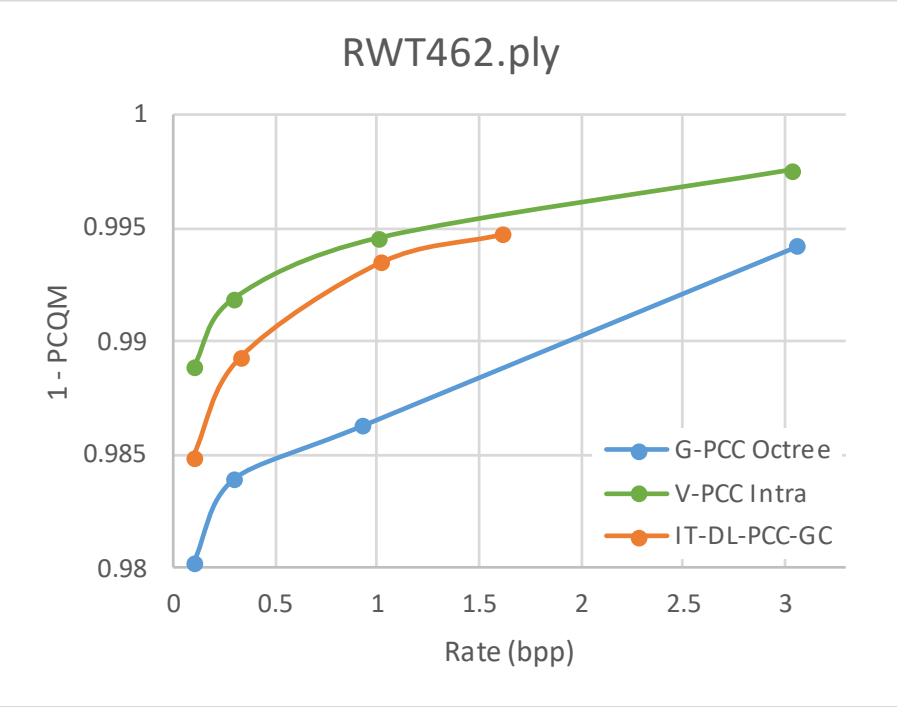

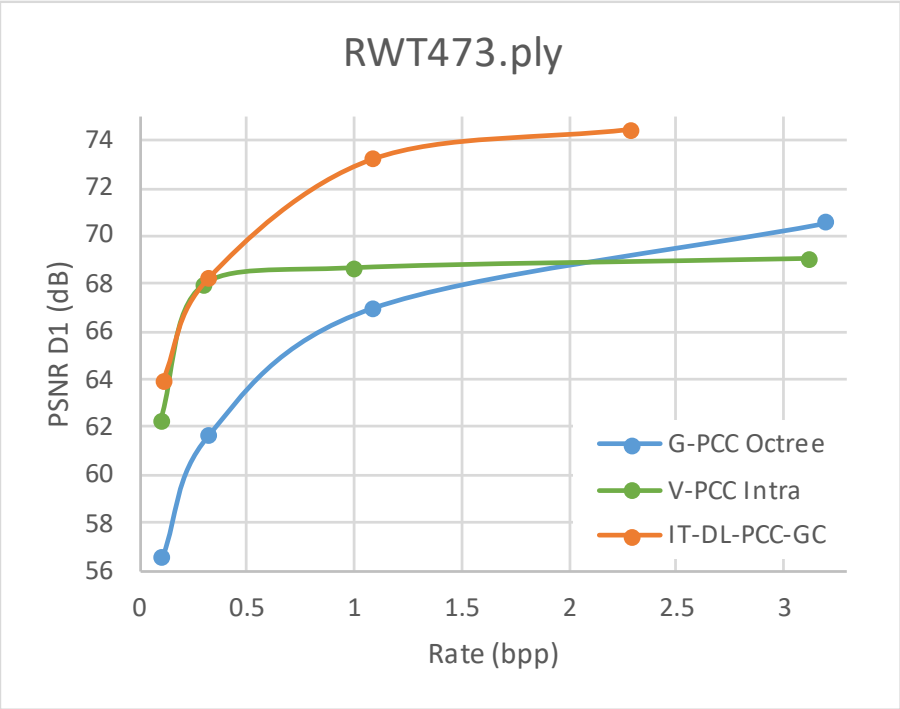

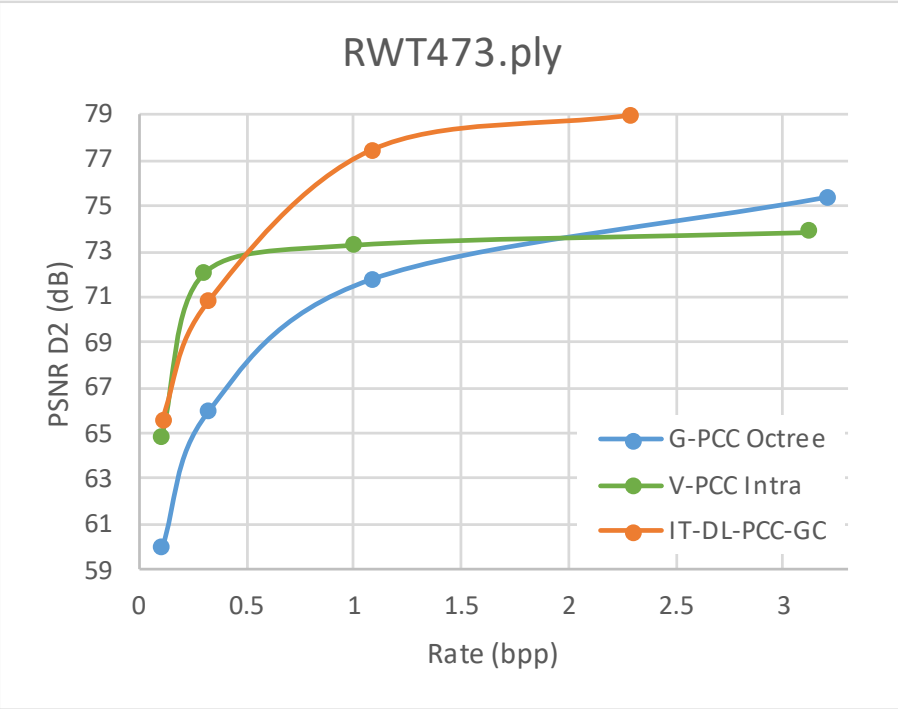

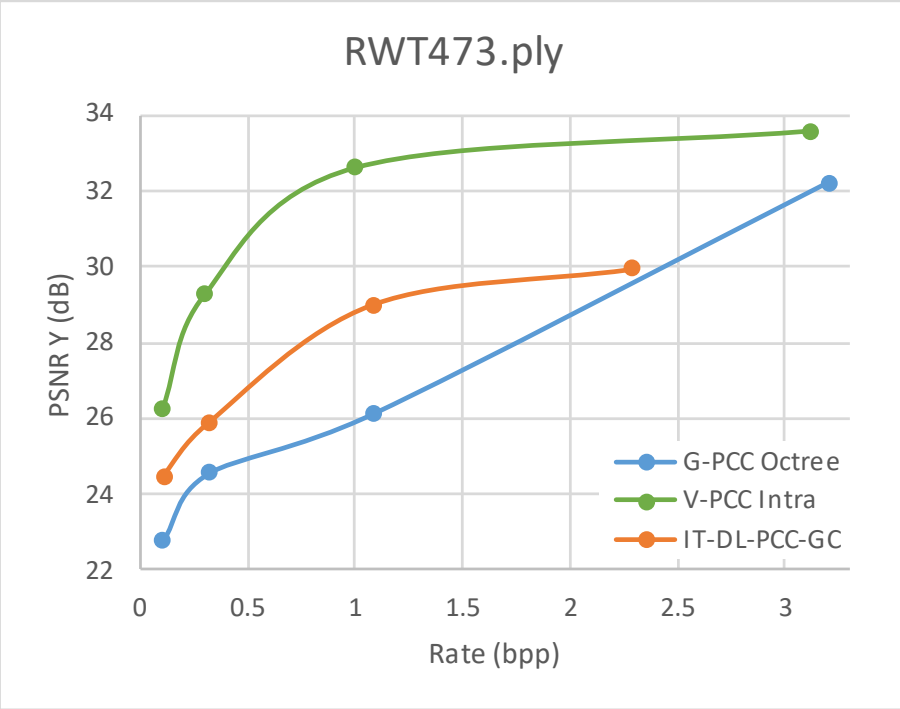

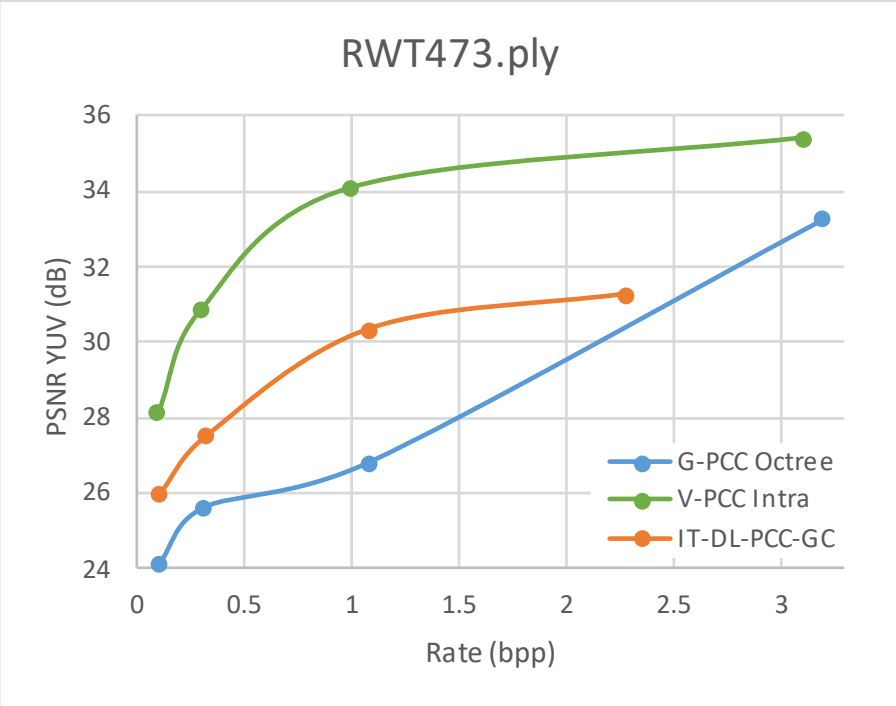

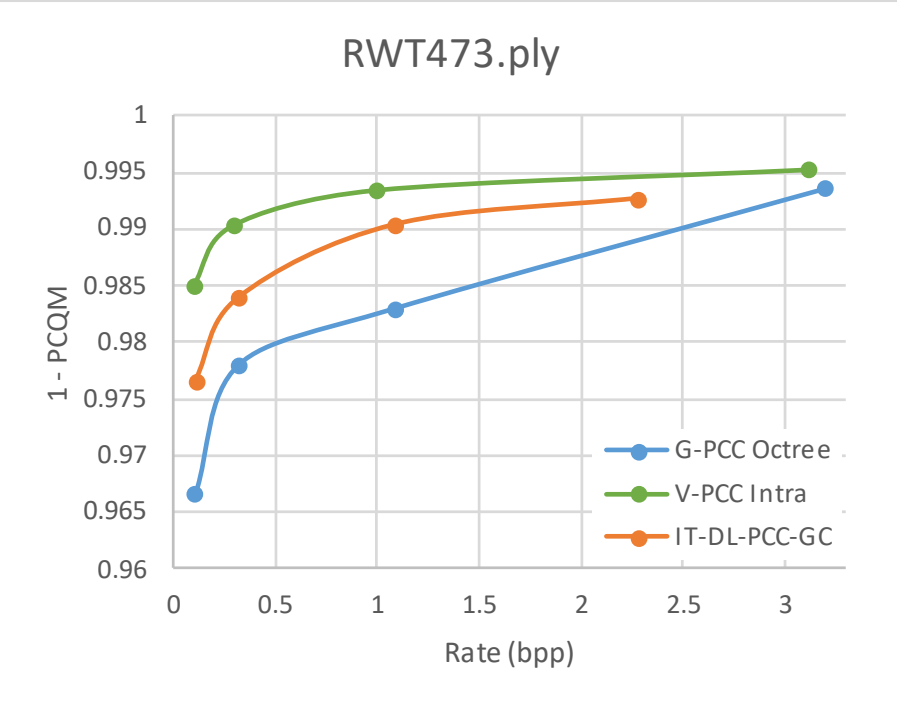

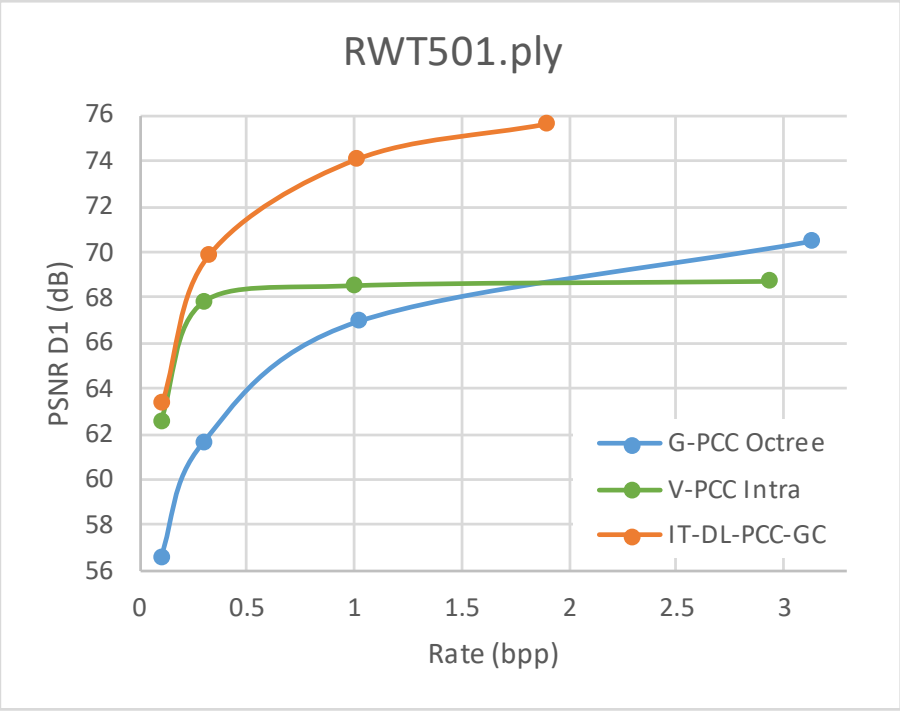

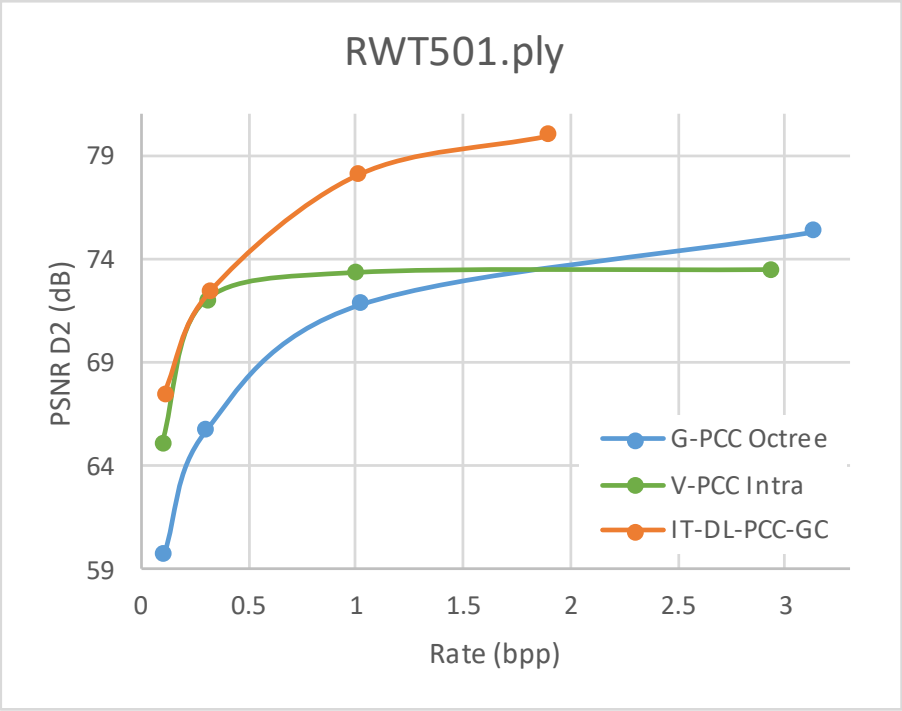

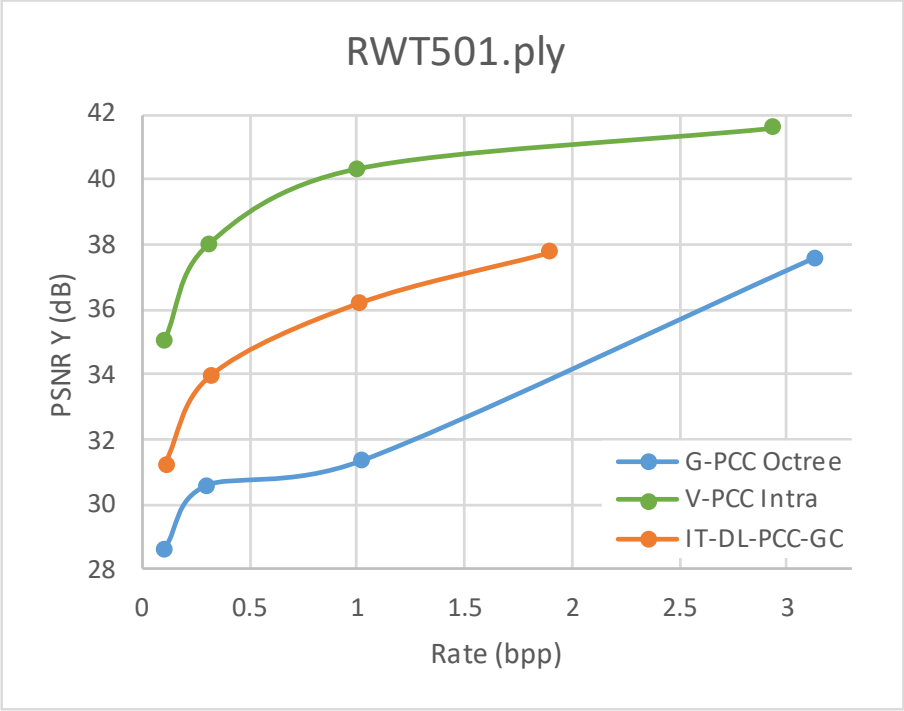

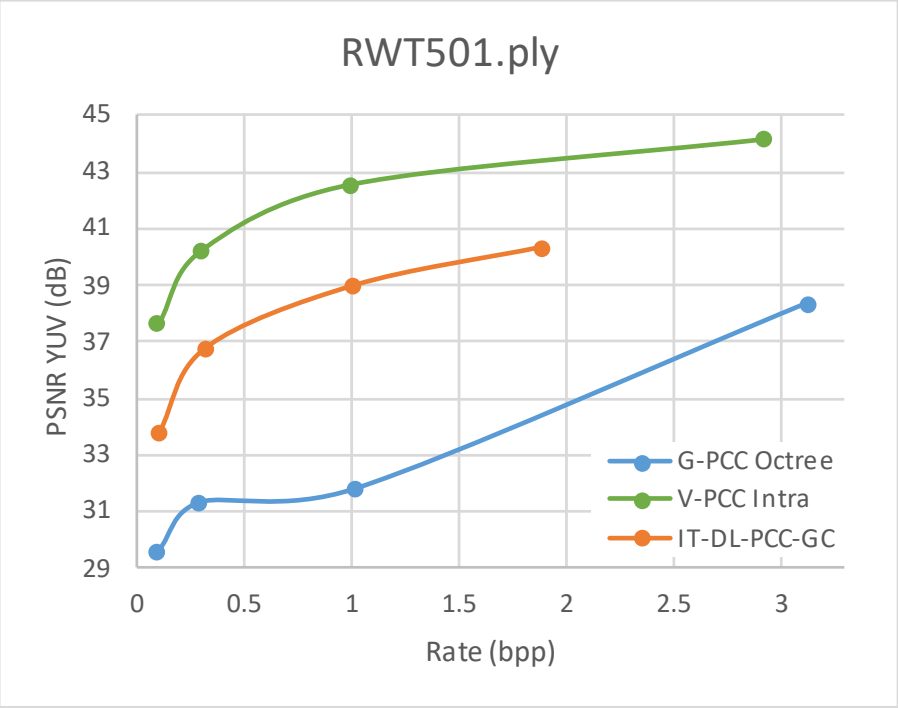

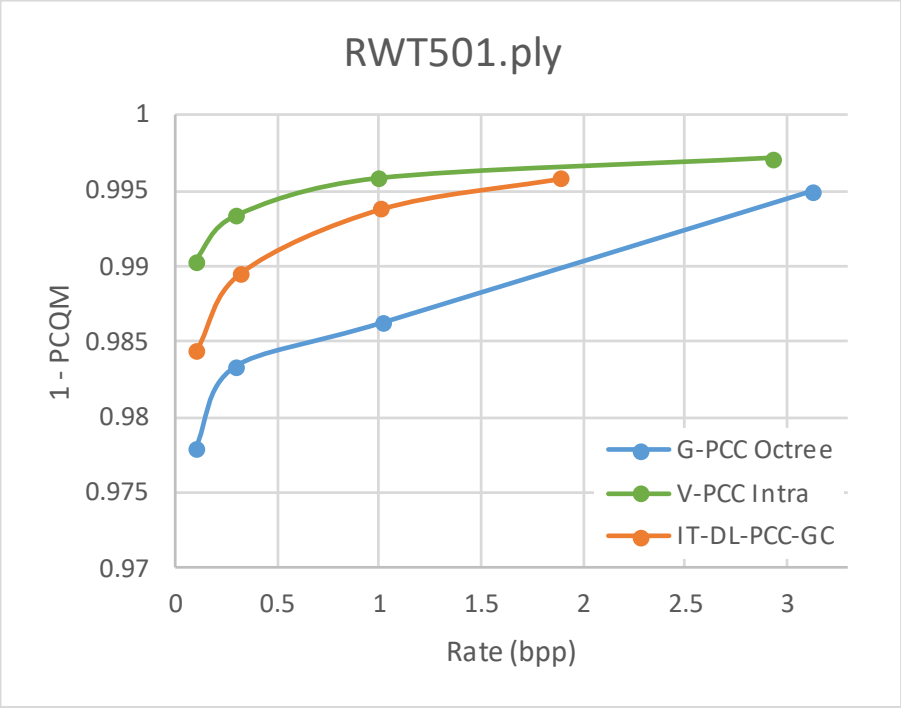

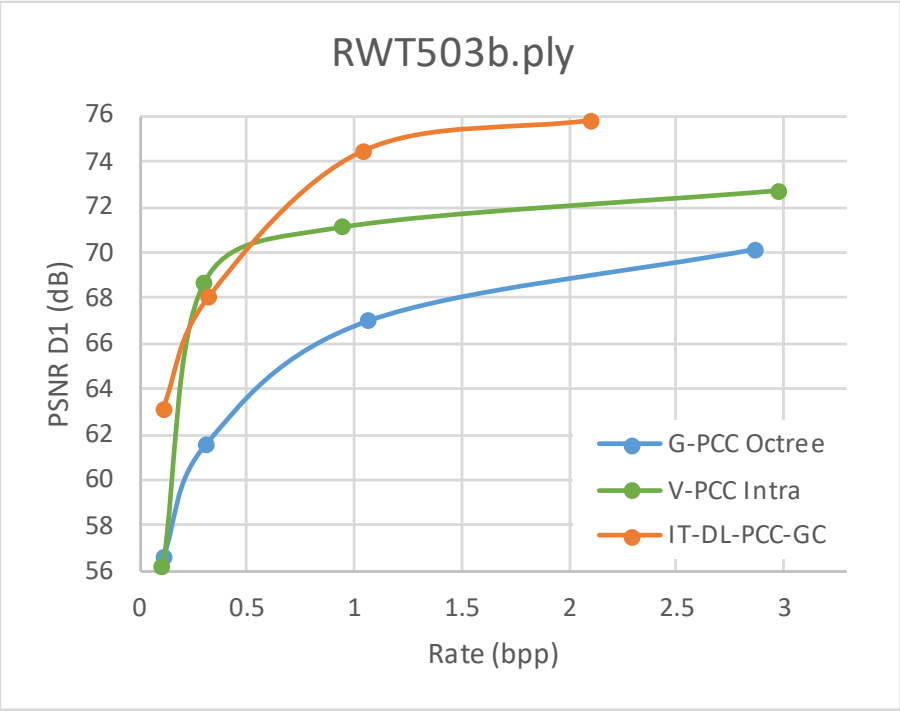

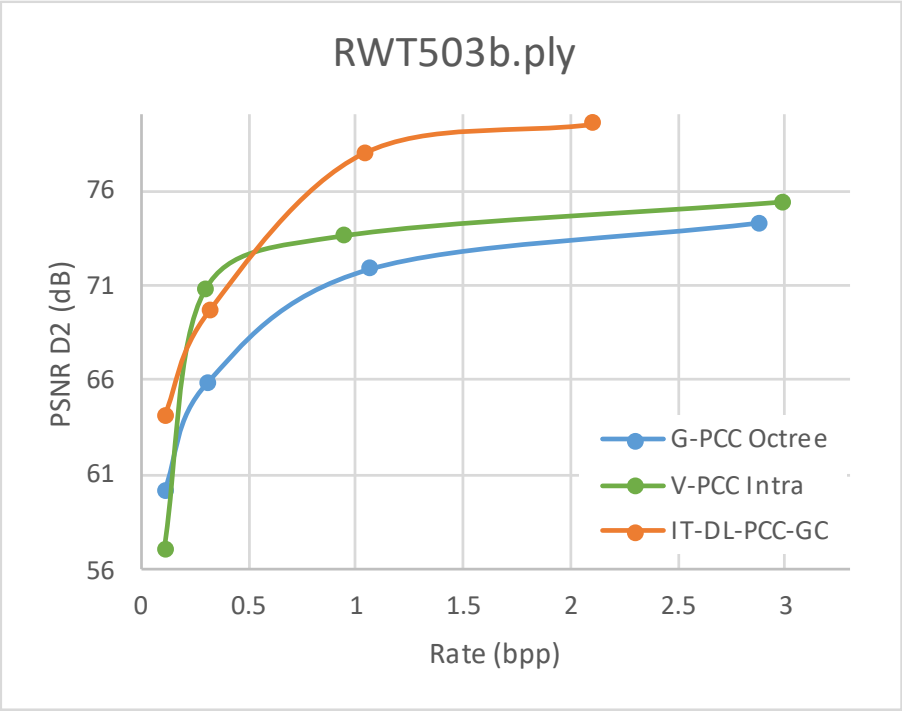

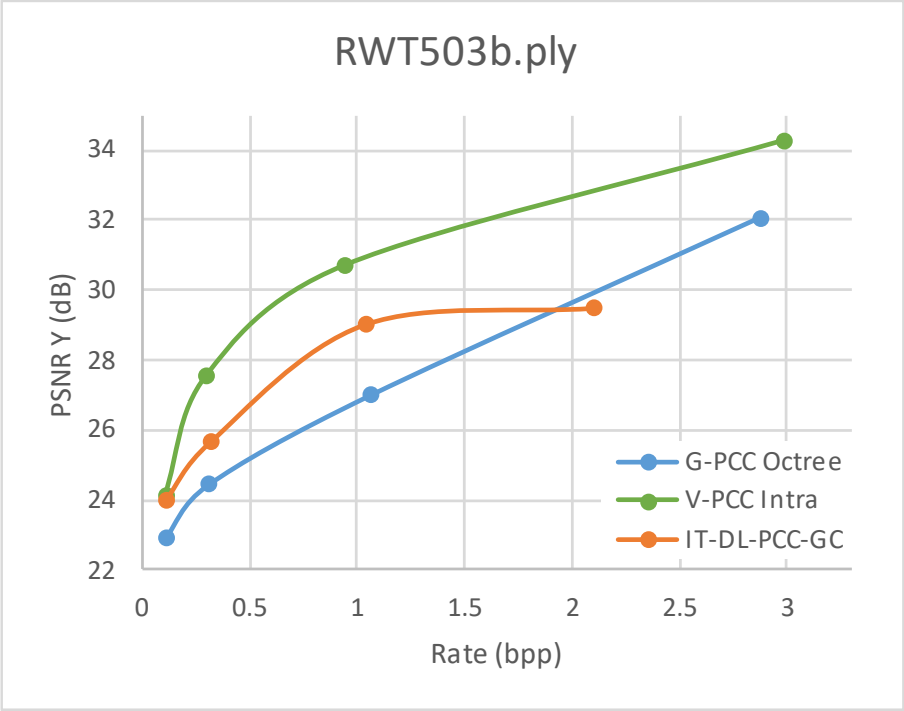

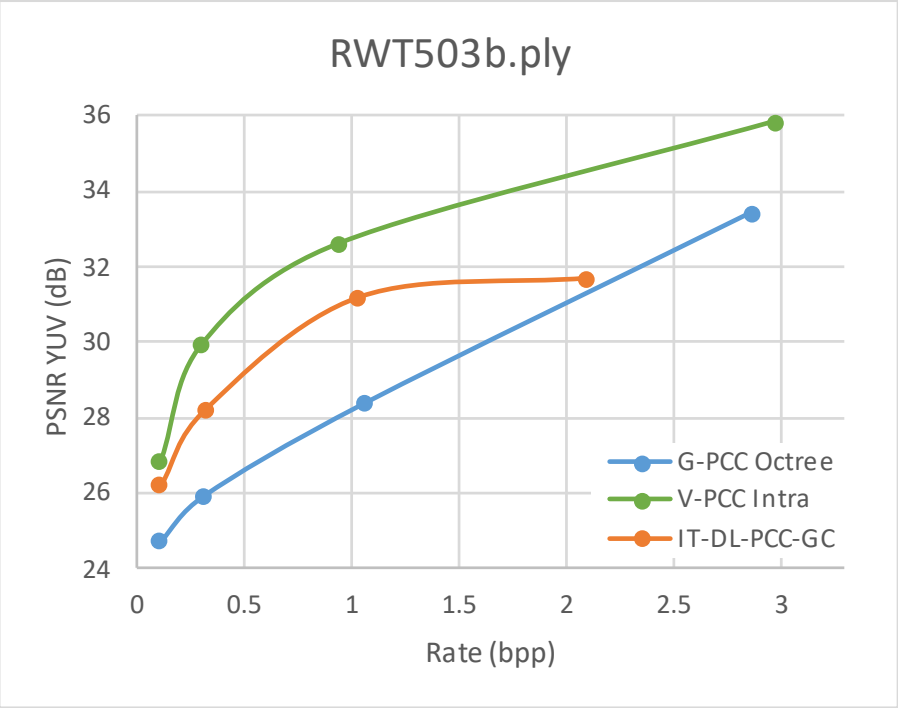

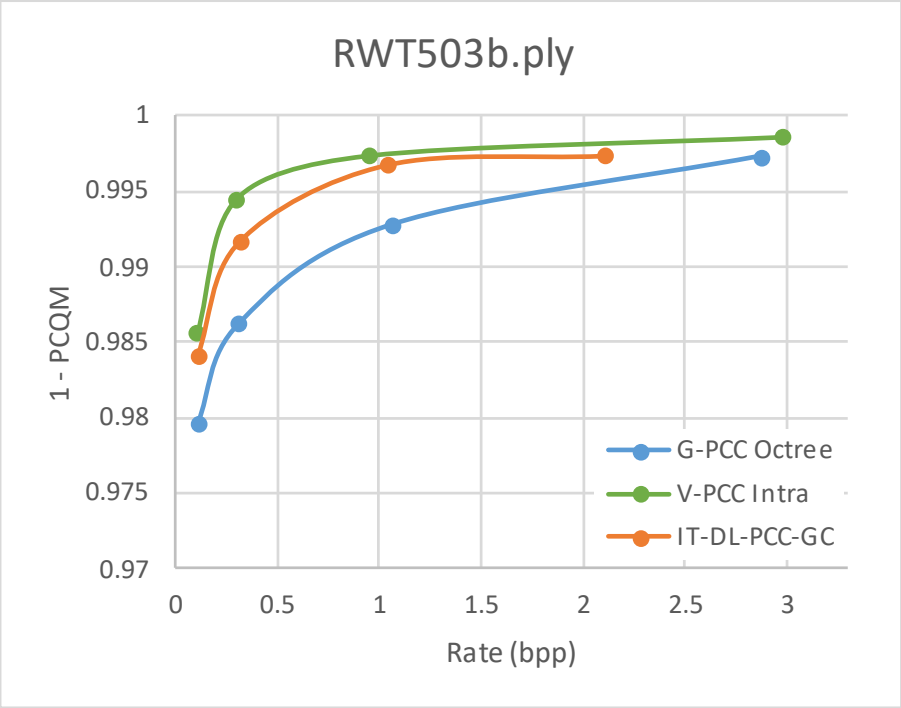

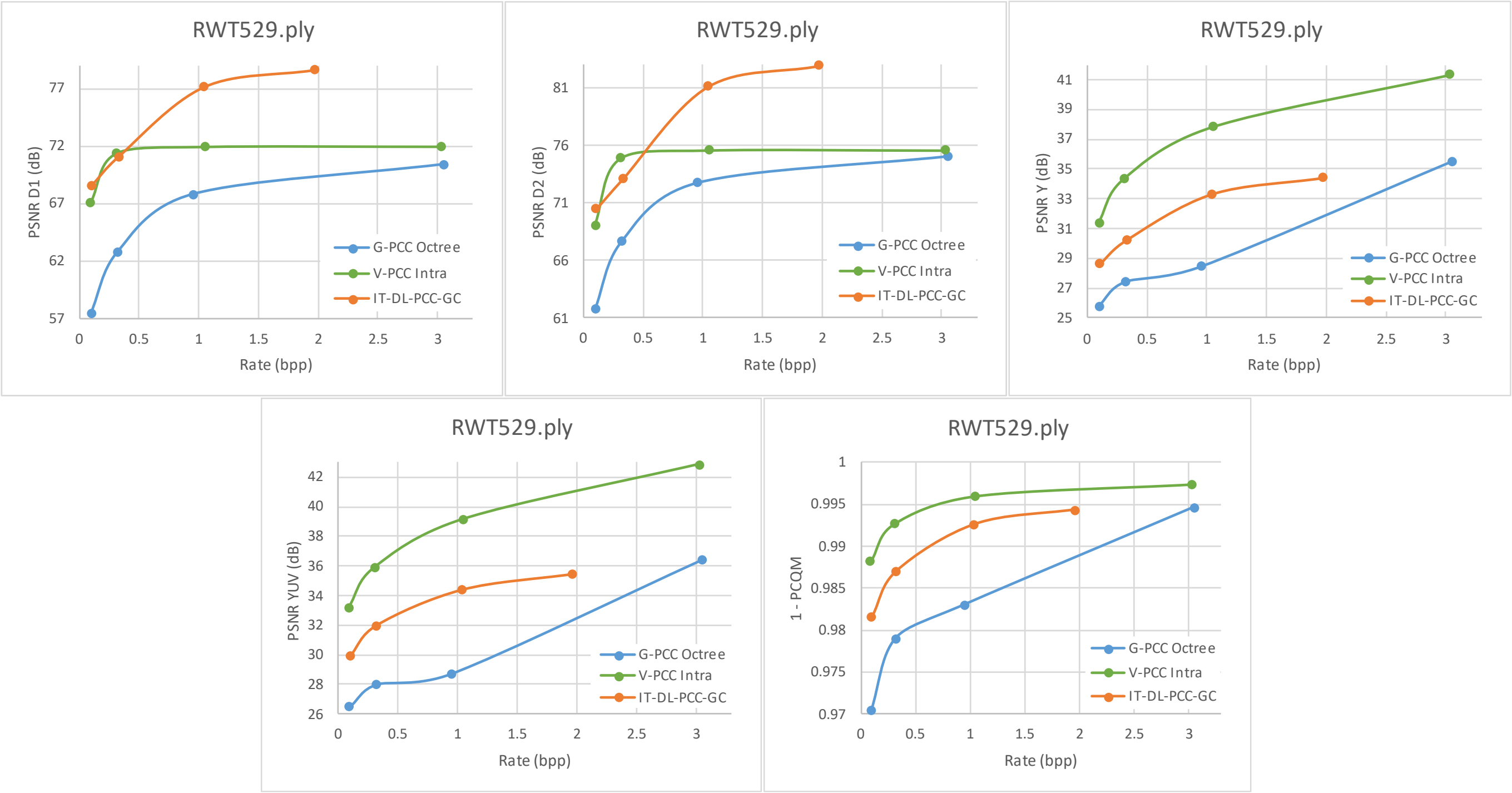

*Fig. 8. RD performance for the proposed IT-DL-PCC-GC solution, in comparison with the V-PCC and G-PCC anchors. Results are presented, for each test PC, for the following objective quality metrics: PSNR-D1, PSNR-D2, PSNR-Y, PSNR-YUV, and PCQM.*

*Table VIII. BD-PSNR and BD-Rate values for the proposed IT-DL-PCC-GC solution over the two anchors, for the geometry quality metrics PSNR-D1 and PSNR-D2.*

| | **G-PCC Octree Reference** | | | | **V-PCC Intra Reference** | | | |
|---|---|---|---|---|---|---|---|---|
| | PSNR-D1 | | PSNR-D2 | | PSNR-D1 | | PSNR-D2 | |
| | BD-Rate | BD-PSNR | BD-Rate | BD-PSNR | BD-Rate | BD-PSNR | BD-Rate | BD-PSNR |
| ***RWT2*** | -83.0% | 7.21 | -74.8% | 6.14 | -16.9% | 1.03 | -15.1% | 1.06 |
| ***RWT34*** | -90.8% | 9.30 | -82.6% | 8.36 | -31.9% | 1.33 | -23.1% | 1.24 |
| ***RWT53*** | -88.1% | 8.40 | -79.5% | 6.94 | -18.8% | 0.89 | -12.5% | 0.85 |
| ***RWT70*** | -80.0% | 6.61 | -68.9% | 5.18 | -26.1% | 3.02 | -20.3% | 2.81 |
| ***RWT120*** | -83.0% | 7.69 | -72.4% | 6.24 | 27.3% | 2.02 | 33.6% | 1.80 |
| ***RWT130*** | -66.4% | 4.20 | -35.8% | 2.00 | -13.4% | 1.04 | -10.7% | 1.51 |
| ***RWT134*** | -92.2% | 8.77 | -85.7% | 7.98 | 34.9% | -0.64 | 39.1% | -0.93 |
| ***RWT136*** | -89.9% | 8.63 | -80.8% | 7.36 | -13.3% | 0.64 | -3.3% | 0.36 |
| ***RWT144*** | -79.5% | 5.93 | -67.4% | 5.15 | 29.6% | 0.03 | 45.1% | -0.46 |
| ***RWT152*** | -79.9% | 5.94 | -64.1% | 4.23 | -68.3% | 3.40 | -62.3% | 3.36 |
| ***RWT246*** | -92.4% | 8.59 | -86.3% | 7.87 | 129.9% | -1.99 | 175.8% | -2.83 |
| ***RWT305*** | -78.9% | 6.51 | -59.6% | 4.24 | 29.9% | -0.80 | 32.6% | -0.82 |
| ***RWT374*** | -93.7% | 9.70 | -85.1% | 8.54 | -50.6% | 3.66 | -10.2% | 3.27 |
| ***RWT395*** | -82.7% | 7.25 | -67.6% | 5.58 | -57.1% | 4.94 | -31.3% | 4.02 |
| ***RWT430*** | -80.3% | 6.04 | -59.7% | 4.42 | 193.1% | -3.38 | 419.4% | -7.17 |
| ***RWT462*** | -90.1% | 9.08 | -82.0% | 7.94 | -17.9% | 0.82 | -13.5% | 0.81 |
| ***RWT473*** | -79.5% | 6.39 | -66.1% | 5.15 | -8.1% | 2.25 | 21.0% | 1.20 |
| ***RWT501*** | -82.8% | 7.38 | -74.7% | 6.47 | -17.9% | 3.22 | -8.1% | 2.42 |
| ***RWT503b*** | -77.1% | 6.83 | -56.2% | 4.78 | -17.8% | 1.85 | -13.8% | 2.09 |
| ***RWT529*** | -91.4% | 8.90 | -75.5% | 6.91 | 17.3% | 2.06 | 56.8% | 1.49 |
| **Average** | -84.1% | 7.47 | -71.2% | 6.07 | 5.2% | 1.27 | 30.0% | 0.80 |

*Table IX. BD-PSNR and BD-Rate values for the proposed IT-DL-PCC-GC solution over the two anchors, for the colour quality metrics PSNR-Y and PSNR-YUV.*

| | **G-PCC Octree Reference** | | | | **V-PCC Intra Reference** | | | |
|---|---|---|---|---|---|---|---|---|
| | PSNR-Y | | PSNR-YUV | | PSNR-Y | | PSNR-YUV | |
| | BD-Rate | BD-PSNR | BD-Rate | BD-PSNR | BD-Rate | BD-PSNR | BD-Rate | BD-PSNR |
| ***RWT2*** | -68.0% | 2.85 | -81.1% | 4.42 | 196.1% | -3.08 | 188.9% | -2.80 |
| ***RWT34*** | -71.5% | 3.24 | -84.7% | 5.22 | 420.2% | -5.40 | 382.8% | -4.78 |
| ***RWT53*** | -54.2% | 1.82 | -74.2% | 3.27 | 202.5% | -2.68 | 156.9% | -2.13 |
| ***RWT70*** | -48.6% | 1.43 | -67.2% | 2.35 | 62.0% | -0.82 | 71.6% | -0.92 |
| ***RWT120*** | -61.3% | 1.77 | -84.5% | 4.37 | 222.6% | -1.88 | 176.2% | -1.56 |
| ***RWT130*** | -6.1% | -0.04 | 36.0% | -0.49 | 214.8% | -2.90 | 513.3% | -3.63 |
| ***RWT134*** | -60.8% | 1.22 | -77.1% | 2.08 | 464.5% | -3.56 | 443.5% | -3.11 |
| ***RWT136*** | -50.6% | 0.99 | -66.9% | 1.87 | 415.2% | -2.23 | 1174.7% | -3.06 |
| ***RWT144*** | -62.8% | 2.68 | -76.2% | 3.32 | 326.7% | -5.26 | 382.9% | -4.88 |
| ***RWT152*** | -63.4% | 2.98 | -78.4% | 5.72 | 100.0% | -0.64 | 100.0% | -0.93 |
| ***RWT246*** | -58.5% | 1.49 | -68.9% | 1.99 | 459.7% | -3.96 | 470.3% | -3.62 |
| ***RWT305*** | -39.7% | 1.03 | -40.8% | 1.05 | 225.4% | -2.43 | 308.9% | -2.80 |
| ***RWT374*** | -85.8% | 6.75 | -91.3% | 8.96 | 326.3% | -4.35 | 308.6% | -4.06 |
| ***RWT395*** | -46.6% | 1.35 | -57.6% | 1.81 | 68.7% | -1.10 | 64.7% | -0.98 |
| ***RWT430*** | -73.7% | 3.18 | -78.1% | 3.26 | 627.2% | -5.98 | 678.1% | -5.30 |
| ***RWT462*** | -47.0% | 1.36 | -60.6% | 1.85 | 517.1% | -5.62 | 560.4% | -4.76 |
| ***RWT473*** | -59.6% | 1.79 | -69.9% | 2.33 | 305.6% | -3.46 | 356.1% | -3.57 |
| ***RWT501*** | -80.9% | 3.74 | -90.8% | 5.79 | 581.0% | -4.10 | 490.2% | -3.68 |
| ***RWT503b*** | -50.0% | 1.37 | -65.5% | 2.23 | 85.1% | -1.88 | 93.9% | -1.74 |
| ***RWT529*** | -74.7% | 3.36 | -80.5% | 4.28 | 423.5% | -4.22 | 553.6% | -4.28 |
| **Average** | -58.2% | 2.22 | -67.9% | 3.28 | 312.2% | -3.28 | 373.8% | -3.13 |

*Table X. BD-PSNR and BD-Rate values for the proposed IT-DL-PCC-GC solution over the two anchors, for the joint geometry and colour quality metric PCQM.*

| | **G-PCC Octree Reference** | | **V-PCC Intra Reference** | |
|---|---|---|---|---|
| | PCQM | | PCQM | |
| | BD-Rate | BD-PSNR | BD-Rate | BD-PSNR |
| ***RWT2*** | -72.9% | 0.007 | 83.9% | -0.002 |
| ***RWT34*** | -83.0% | 0.007 | 119.4% | -0.003 |
| ***RWT53*** | -64.1% | 0.007 | 106.0% | -0.003 |
| ***RWT70*** | -53.9% | 0.005 | 47.6% | -0.002 |
| ***RWT120*** | -71.7% | 0.008 | 127.9% | -0.003 |
| ***RWT130*** | -21.3% | 0.002 | 167.8% | -0.009 |
| ***RWT134*** | -70.0% | 0.006 | 268.1% | -0.006 |
| ***RWT136*** | -65.1% | 0.005 | 303.2% | -0.004 |
| ***RWT144*** | -62.7% | 0.007 | 368.7% | -0.007 |
| ***RWT152*** | -73.5% | 0.011 | 45.9% | -0.001 |
| ***RWT246*** | -67.4% | 0.005 | 321.6% | -0.005 |
| ***RWT305*** | -58.2% | 0.006 | 184.3% | -0.005 |
| ***RWT374*** | -88.9% | 0.014 | 135.5% | -0.003 |
| ***RWT395*** | -53.9% | 0.005 | 41.6% | -0.002 |
| ***RWT430*** | -74.3% | 0.007 | 603.0% | -0.006 |
| ***RWT462*** | -79.7% | 0.006 | 111.6% | -0.002 |
| ***RWT473*** | -66.5% | 0.007 | 277.3% | -0.005 |
| ***RWT501*** | -79.8% | 0.007 | 195.0% | -0.003 |
| ***RWT503b*** | -60.8% | 0.005 | 58.5% | -0.002 |
| ***RWT529*** | -77.9% | 0.009 | 283.9% | -0.005 |
| **Average** | -67.3% | 0.007 | 192.5% | -0.004 |

### 8.3 Analysis of the Results

From the full set of performance results included in the previous subsection, it is possible to observe that:

**IT-DL-PCC-GC versus G-PCC Octree**

- IT-DL-PCC-GC achieves significant overall compression gains when compared with G-PCC Octree for all rates.

- In terms of geometry, the proposed IT-DL-PCC-GC solution massively outperforms G-PCC Octree, with average rate savings of 84.1% and 71.2% and average BD-PSNR gains of 7.47 dB and 6.07 dB, for PSNR-D1 and PSNR-D2 metrics, respectively.

- In terms of colour, the improvement is not as pronounced as for geometry, but are still substantial gains for almost all PCs, with average rate savings of 58.2% and 67.9% and average BD-PSNR gains of 2.22 dB and 3.28 dB, for PSNR-Y and PSNR-YUV metrics, respectively.

- As for the joint metric PCQM, the results show the same tendency as for previous metrics, with average rate savings of 67.3%, confirming a clear advantage of IT-DL-PCC-GC over G-PCC Octree for both geometry and colour coding.

**IT-DL-PCC-GC versus V-PCC Intra**

- When compared with V-PCC Intra, the proposed IT-DL-PCC-GC presents less impressive results.

- Regarding geometry, IT-DL-PCC-GC achieves competitive results when compared with V-PCC Intra, outperforming it for most PCs, albeit with much smaller gains than for G-PCC Octree. For the *RWT246* (ArmChair) and *RWT430* (WoodenChest) PCs, V-PCC presents significantly advantage over the proposed IT-DL-PCC-GC. These PCs represent objects with simpler shapes, smooth surfaces, and without obstructions, which V-PCC Intra can code very efficiently.

- However, when evaluating the colour, IT-DL-PCC-GC falls behind V-PCC Intra significantly, for all PCs.

- In terms of PCQM, the colour appears to have a very significant impact, as once again V-PCC considerably outperforms the proposed solution.

- These results suggest that the proposed IT-DL-PCC-GC codec is not yet able to properly balance the relevance of colour and geometry in the joint coding approach and achieve competitive results for PC colour coding.

## 8.4 Complexity Evaluation

Regarding the complexity of the proposed IT-DL-PCC-GC solution, the resource footprint and the computational complexity are both reported according to the JPEG CTTC [17].

To measure the resource footprint, the number of parameters of the DL models is reported, as well as their precision. For each DL coding model, the total number of parameters is 6605664, represented with 32-bit floats. In addition, for each ABU model, the total number of parameters is 7291488, also represented with 32-bit floats.

As for the computational complexity, the encoding and decoding times are reported in Table XI and Table XII, with and without GPU. Tests were run on an Intel Core i5-12600k CPU @ 3.70 GHz computer with a NVidia GeForce RTX 3080 Ti 12 GB GPU and 32 GB of RAM, running Ubuntu 20.04. **Due to time constraints, only one complete run over the entire test dataset was performed (and not the average of 10 runs as requested by the JPEG CTTC)**.

*Table XI. Real encoding times of the IT-DL-PCC-GC codec, with and without GPU, for each target rate and each PC.*

| | CPU | | | | GPU | | | |
|---|---|---|---|---|---|---|---|---|
| | R1 | R2 | R3 | R4 | R1 | R2 | R3 | R4 |
| ***RWT2*** | 12m4.3s | 11m1.1s | 8m56.1s | 9m7.5s | 6m58.9s | 5m50.6s | 4m38.4s | 4m37.9s |
| ***RWT34*** | 10m5.3s | 9m38.8s | 8m3.1s | 8m5.6s | 5m33.3s | 5m4.8s | 4m14.8s | 4m13.0s |
| ***RWT53*** | 7m6.4s | 6m40.3s | 4m8.1s | 5m30.4s | 4m3.4s | 3m41.1s | 2m18.3s | 2m59.3s |
| ***RWT70*** | 9m58.2s | 8m59.3s | 7m20.3s | 7m19.1s | 5m31.1s | 4m37.7s | 3m39.6s | 3m38.6s |
| ***RWT120*** | 11m14.3 | 10m13.1s | 4m50.0s | 8m29.3s | 5m57.0s | 4m53.5s | 2m40.3s | 3m58.7s |
| ***RWT130*** | 17m51.7s | 14m26.5s | 11m21.8s | 10m54.7s | 11m36.9s | 8m19.9s | 6m16.3s | 5m43.6s |
| ***RWT134*** | 12m23.1s | 12m6.5s | 10m7.3s | 10m5.3s | 5m50.3s | 5m29.4s | 4m49.0s | 4m53.0s |
| ***RWT136*** | 3m38.4s | 3m32.8s | 2m53.5s | 2m54.9s | 2m11.2s | 2m6.2s | 1m44.2s | 1m43.1s |
| ***RWT144*** | 23m39.4s | 22m38.0s | 18m9.3s | 18m2.6s | 12m54.2s | 10m45.9s | 8m9.9s | 2m21.0s |
| ***RWT152*** | 16m43.9s | 15m59.2s | 12m41.7s | 12m39.1s | 9m43.3s | 8m49.8s | 7m40.8s | 6m37.7s |
| ***RWT246*** | 19m29.1s | 18m58.3s | 15m58.4s | 15m53.8s | 8m49.2s | 8m19.8s | 6m58.9s | 5m11.4s |
| ***RWT305*** | 7m18.3s | 6m32.0s | 5m1.7s | 4m59.2s | 5m2.8s | 4m19.8s | 3m18.4s | 3m12.6s |
| ***RWT374*** | 8m56.9s | 8m47.6s | 4m21.6s | 7m19.4s | 4m21.4s | 4m12.6s | 2m24.7s | 2m8.2s |
| ***RWT395*** | 3m21.9s | 3m9.5s | 2m37.1s | 2m37.0s | 1m55.8s | 1m42.4s | 1m26.6s | 1m26.8s |
| ***RWT430*** | 20m34.6s | 19m10.8s | 10m1.6s | 15m23.2s | 11m10.0s | 9m39.5s | 5m16.8s | 7m15.9s |
| ***RWT462*** | 3m12.5s | 3m7.8s | 2m33.1s | 2m32.7s | 1m55.4s | 1m51.6s | 1m32.4s | 1m32.4s |
| ***RWT473*** | 15m54.8s | 14m30.7s | 11m57.4s | 11m55.9s | 7m46.2s | 6m52.0s | 5m28.3s | 0m45.3s |
| ***RWT501*** | 14m46.9s | 14m1.7s | 11m46.8s | 11m45.2s | 7m4.6s | 6m26.6s | 5m19.2s | 0m51.9s |
| ***RWT503b*** | 2m14.9s | 2m7.2s | 0m49.6s | 1m43.5s | 1m11.0s | 1m3.4s | 0m31.4s | 0m53.2s |
| ***RWT529*** | 3m2.4s | 3m2.0s | 2m26.6s | 2m27.3s | 1m46.7s | 1m45.9s | 1m27.1s | 1m27.2s |

*Table XII. Real decoding times of the IT-DL-PCC-GC codec, with and without GPU, for each target rate and each PC.*

| | CPU | | | | GPU | | | |
|---|---|---|---|---|---|---|---|---|
| | R1 | R2 | R3 | R4 | R1 | R2 | R3 | R4 |
| ***RWT2*** | 7m5.9s | 7m6.4s | 5m34.1s | 5m36.6s | 2m1.8s | 2m1.6s | 1m44.3s | 1m43.1s |
| ***RWT34*** | 6m21.2s | 6m21.8s | 4m57.4s | 4m58.6s | 1m50.8s | 1m50.8s | 1m34.6s | 1m33.6s |
| ***RWT53*** | 4m20.0s | 4m20.4s | 2m25.8s | 3m22.3s | 1m21.1s | 1m21.3s | 0m45.1s | 1m8.4s |
| ***RWT70*** | 6m7.3s | 6m6.4s | 4m46.5s | 4m45.0s | 1m46.0s | 1m45.8s | 1m29.9s | 1m28.5s |
| ***RWT120*** | 7m21.3s | 7m21.2s | 2m49.5s | 5m45.3s | 2m4.8s | 2m4.5s | 0m51.2s | 1m44.9s |
| ***RWT130*** | 8m27.9s | 8m28.2s | 6m32.3s | 6m36.1s | 2m22.5s | 0m33.0s | 2m0.0s | 0m24.3s |
| ***RWT134*** | 8m57.4s | 8m57.9s | 7m2.7s | 7m0.5s | 2m28.8s | 2m29.1s | 2m7.5s | 2m2.4s |
| ***RWT136*** | 2m14.8s | 2m14.9s | 1m42.9s | 1m43.2s | 0m48.8s | 0m48.9s | 0m41.0s | 0m40.8s |
| ***RWT144*** | 14m42.8s | 15m58.6s | 12m37.0s | 12m32.1s | 3m59.5s | 4m13.7s | 3m41.1s | 3m33.1s |
| ***RWT152*** | 9m44.7s | 9m43.9s | 7m36.7s | 7m38.2s | 2m42.4s | 2m42.7s | 2m31.9s | 2m16.8s |
| ***RWT246*** | 14m21.0s | 14m21.8s | 11m19.3s | 11m16.5s | 3m50.5s | 3m51.1s | 3m19.5s | 3m1.5s |
| ***RWT305*** | 3m17.3s | 3m17.0s | 2m32.7s | 2m32.3s | 1m4.7s | 1m4.5s | 0m55.0s | 0m54.1s |
| ***RWT374*** | 6m20.0s | 6m20.6s | 2m32.5s | 4m57.5s | 1m49.0s | 1m49.3s | 0m46.2s | 0m46.8s |
| ***RWT395*** | 2m15.8s | 2m16.3s | 1m44.5s | 1m43.9s | 0m49.5s | 0m49.6s | 0m41.7s | 0m41.8s |
| ***RWT430*** | 12m49.1s | 12m49.6s | 5m58.6s | 10m4.5s | 3m29.0s | 3m29.1s | 1m44.6s | 3m4.0s |
| ***RWT462*** | 1m59.9s | 1m59.9s | 1m31.6s | 1m31.4s | 0m45.1s | 0m45.2s | 0m37.6s | 0m37.6s |
| ***RWT473*** | 10m29.7s | 10m29.8s | 8m15.9s | 8m14.2s | 3m1.4s | 2m52.1s | 2m27.4s | 0m21.1s |
| ***RWT501*** | 10m28.1s | 10m27.9s | 8m11.4s | 8m13.3s | 2m52.1s | 2m51.8s | 2m26.1s | 0m30.3s |
| ***RWT503b*** | 1m42.9s | 1m42.9s | 0m31.1s | 1m17.9s | 0m39.9s | 0m40.0s | 0m14.8s | 0m33.1s |
| ***RWT529*** | 1m58.5s | 1m58.6s | 1m30.3s | 1m30.2s | 0m44.1s | 0m44.0s | 0m36.8s | 0m36.6s |

## 9. JPEG Pleno PCC Requirements Checking

Considering the JPEG Pleno PCC requirements [18], the proposed IT-DL-PCC-GC solution offers the following properties:

- **Compression Efficiency**: IT-DL-PCC-GC offers a considerably better RD performance than the G-PCC Octree standard adopted as one of the anchors.
- **Random Access**: In IT-DL-PCC-GC, the PC is divided into blocks which are coded independently. Since the positions of the coded blocks are transmitted to the decoder, this allows the user to choose decoding only selected regions of the PC, thus providing spatial random access. The granularity of the random access may be controlled by selecting the block size for coding.

## 10. Submitted Proposal Materials

This section lists the materials that were submitted in the context of this proposal, notably:

1. JPEG document with textual description and JPEG document for presentation:
    - team1/Documents/
2. Source code/scripts of the encoder and decoder for the proposed codecs (IT-DL-PCC-G and IT-DL-PCC-GC):
    - team1/encoder/
    - team1/decoder/
3. Trained DL models for the proposed codecs (IT-DL-PCC-G and IT-DL-PCC-GC):
    - team1/models/
4. Configuration files for the target rates for both codecs (IT-DL-PCC-G and IT-DL-PCC-GC):

- team1/cfgs/

5. Bitstream files of the proposed codecs (IT-DL-PCC-G and IT-DL-PCC-GC) for the four target rates:

    - team1/testbit/

6. Decoded PCs obtained with the proposed codecs (IT-DL-PCC-G and IT-DL-PCC-GC) for the four target rates:

    - team1/testrec/

# PART III – Software Description

This part describes the software implementing the codecs described in Part I and Part II. The two codecs are integrated in the same software framework and running one or the other codec only depends on one configuration parameter which serves precisely to control if geometry only or joint geometry and colour coding is performed.

## 11. Software Contents

The IT-DL-PCC software made available through the submission platform contains the following material:

- **Source code**: The full source code to run both the IT-DL-PCC codecs is included in "*encoder*" and duplicated in "*decoder*". The same python script "*IT-DL-PCC.py*" is used to run both the encoder and decoder.

- **Trained DL coding models**: All the trained DL models (in a total of 16 models) used in the IT-DL-PCC codecs are included in "*models*", for both the geometry-only (IT-DL-PCC-G) and the joint geometry + color (IT-DL-PCC-GC) codecs, notably:

    - For the DL coding model, six were trained for different rate-distortion trade-offs (λ = 0.00025, 0.0005, 0.001, 0.0025, 0.005 and 0.01);

    - For the ABU model, there are two DL up-sampling models trained for different sampling factors (2 and 4).

Instructions on the software requirements and usage are also available and described in the following sections.

## 12. Software Requirements

The prerequisites to run the IT-DL-PCC software and DL models, in particular, are as follows:

- Python 3.9.7

- Tensorflow 2.8 with CUDA Version 11.2 and cuDNN 8.1.0, or compatible

- tensorflow-compression 2.8 (https://github.com/tensorflow/compression/tree/v2.8.0)

- Python packages:
    - numpy
    - scipy

- scikit-learn
- pandas
- pyntcloud

Using a Linux distribution (e.g. Ubuntu) is recommended, as other operating systems have not been tested.

## 13. Software Usage

The main script "*IT-DL-PCC.py*" is used to encode and decode a PC using the IT-DL-PCC codec.

**Running the script**:

```
python IT-DL-PCC.py [--with_color] {compress,decompress} [OPTIONS]
```

The flag "*--with_color*", used before the "*{compress,decompress}*" command, should be given to select the joint geometry + color codec (IT-DL-PCC-GC). Otherwise, the geometry-only codec (IT-DL-PCC-G) is used instead.

**Encoding a point cloud**:

```
usage: IT-DL-PCC.py [--with_color] compress [-h] [--helpfull] [--blk_size BLK_SIZE] [-
-q_step Q_STEP] [--scale SCALE] [--topk_metrics TOPK_METRICS] [--color_weight
COLOR_WEIGHT] [--use_fast_topk] [--max_topk MAX_TOPK] [--topk_patience TOPK_PATIENCE]
[--use_abu] [--abu_model_dir ABU_MODEL_DIR] [--abu_topk ABU_TOPK] [--abu_max_topk
ABU_MAX_TOPK] input_file model_dir output_dir

Reads a Point Cloud PLY file, compresses it, and writes the bitstream file.

positional arguments:
  input_file            Input Point Cloud filename (.ply).
  model_dir             Directory where to load model checkpoints. For compression, a
                        single directory should be provided: ../models/test
  output_dir            Directory where the compressed Point Cloud will be saved.

optional arguments:
  -h, --help            show this help message and exit
  --helpfull            show full help message and exit
  --blk_size BLK_SIZE   Size of the 3D coding block units. Should be a multiple of 64.
                        (default: 128)
  --q_step Q_STEP       Explicit quantization step. Can be any positive real value.
                        (default: 1)
  --scale SCALE         Down-sampling scale. If 'None', it is automatically
                        determined. Multiple comma separated values can be provided.
                        Can be any positive real value when not using ABU, otherwise
                        only an integer power of 2. (default: None)
  --topk_metrics TOPK_METRICS
                        Metrics to use for the optimized Top-k binarization.
                        Available: 'd1', 'd2', 'd1yuv', 'd2yuv', 'd1rgb', 'd2rgb'.
```

```
                        If coding geometry-only, only the geometry metric is used.
                        (default: d1yuv)
  --color_weight COLOR_WEIGHT
                        Weight of the colour metric in the joint top-k optimization,
                        in percentage. Between 0 and 1. (default: 0.5)
  --use_fast_topk       Use faster top-k optimization algorithm for the coding model.
                        (default: False)
  --max_topk MAX_TOPK   Define the maximum factor used by the Top-K optimization
                        algorithms. (default: 10)
  --topk_patience TOPK_PATIENCE
                        Define the patience for early stopping in the Top-K
                        optimization algorithms. (default: 5)
  --use_abu             Use Basic Up-sampling (False) or Advanced Block Up-sampling –
                        ABU (True). (default: False)
  --abu_model_dir ABU_MODEL_DIR
                        Directory where to load ABU model. A directory should be
                        provided for each down-sampling scale, separated by commas:
                        ../models/test1,../models/test2 (default: )
  --abu_topk ABU_TOPK   Type of Top-k optimization for the ABU at the encoder:
                          'none' - Use the same as the coding model.
                          'full' - Use the regular algorithm (default).
                          'fast' - Use the faster algorithm.
  --abu_max_topk ABU_MAX_TOPK
                        Define the maximum factor used by the Top-K optimization
                        algorithms. (default: 10)
```

Usage examples:

```
python IT-DL-PCC.py compress "../test_data_path/longdress.ply" "../models/Geometry-
only/Codec/0.00025" "../results/G/0.00025"
```

```
python IT-DL-PCC.py --with_color compress "../test_data_path/longdress.ply"
"../models/Geometry-Color/Codec/0.00025" "../results/GC/0.00025"
```

```
python IT-DL-PCC.py --with_color compress "../test_data_path/longdress.ply"
"../models/Geometry-Color/Codec/0.00025" "../results/GC/0.00025" --blk_size 64 --scale
2 --topk_metrics d2rgb
```

```
python IT-DL-PCC.py compress "../test_data_path/longdress.ply" "../models/Geometry-
only/Codec/0.0025" "../results/G/0.0025" --scale 4 --use_abu --abu_model_dir
"../models/Geometry-only/ABU/sfactor4"
```

**Decoding a point cloud**:

```
usage: IT-DL-PCC.py decompress [-h] [--helpfull] [--abu_model_dir ABU_MODEL_DIR]
input_file model_dir

Reads a bitstream file, decodes the voxel blocks, and reconstructs the Point Cloud.
```

```
positional arguments:
  input_file            Input bitstream filename (.gz).
  model_dir             Directory where to load model checkpoints. For decompression,
                        a single directory should be provided: ../models/test

optional arguments:
  -h, --help            show this help message and exit
  --helpfull            show full help message and exit
  --abu_model_dir ABU_MODEL_DIR
                        Directory where to load ABU model. A directory should be
                        provided for each down-sampling scale, separated by commas:
                        ../models/test1,../models/test2 (default: )
```

Usage examples:

```
python IT-DL-PCC.py decompress "../results/G/0.00025/longdress/longdress.pkl.gz"
"../models/Geometry-only/Codec/0.00025"
```

```
python IT-DL-PCC.py --with_color decompress
"../results/GC/0.00025/longdress/longdress.pkl.gz" "../models/Geometry-
Color/Codec/0.00025"
```

```
python IT-DL-PCC.py decompress "../results/G/0.0025/longdress/longdress.pkl.gz"
"../models/Geometry-only/Codec/0.0025" --abu_model_dir "../models/Geometry-
only/ABU/sfactor4"
```